%% file: main.tex
\documentclass{lmcs}
\pdfoutput=1

\usepackage{lastpage}
\lmcsdoi{15}{2}{3}
\lmcsheading{}{\pageref{LastPage}}{}{}%
{Apr.~25,~2017}{Apr.~26,~2019}{}

\input{packages}
\input{macros}
\pdfoutput=1

\def\cf{{\em cf.}}
\theoremstyle{plain}\newtheorem{rmrk}[thm]{Remark}

\begin{document}

\title[Generalised Mermin-type non-locality arguments]{Generalised Mermin-type non-locality arguments}

\author[S. Gogioso]{Stefano Gogioso}
\address{Quantum Group, University of Oxford, UK}	
\email{stefano.gogioso@cs.ox.ac.uk} 	

\author[W. Zeng]{William Zeng} 
\address{Rigetti Computing, Berkeley, CA} 
\email{zeng.will@gmail.com}  

\keywords{Mermin non-locality, categorical quantum mechanics, strongly complementary observables, All-vs-Nothing arguments, quantum secret sharing.}
\titlecomment{An extended abstract for a small part of this work has appeared in the Proceedings of QPL 2015, and can be found in EPTCS 195, 2015, pages 228-246, doi:\href{https://dx.doi.org/10.4204/EPTCS.195.17}{10.4204/EPTCS.195.17}}

\begin{abstract}
\noindent We broadly generalise Mermin-type arguments on GHZ states, and we provide exact group-theoretic conditions for non-locality to be achieved. Our results are of interest in quantum foundations, where they yield a new hierarchy of quantum-realisable All-vs-Nothing arguments. They are also of interest to quantum protocols, where they find immediate application to a non-trivial extension of the hybrid quantum-classical secret sharing scheme of Hillery, Bu\v{z}ek and Berthiaume (HBB). Our proofs are carried out in the graphical language of string diagrams for dagger compact categories, and their validity extends beyond quantum theory to any theory featuring the relevant algebraic structures.
\end{abstract}

\maketitle

\section*{Introduction}
\label{section_intro}

Non-locality is a defining feature of quantum mechanics, and its connection to the structure of phase groups is a key foundational question. A particularly crisp example of this connection is given by Mermin's argument for qubit GHZ states \cite{Mermin1990}, which finds practical application in the HBB quantum secret sharing protocol. 

In Mermin's argument, $N$ qubits are prepared in a GHZ state (with Pauli $Z$ as computational basis), then a controlled phase gate is applied to each, followed by measurement in the Pauli $X$ observable. Even though the $N$ outcomes (each valued in $\integersMod{2}$) are probabilistic, their parity turns out to satisfy certain deterministic equations. Mermin shows that the existence of a local hidden variable model would imply a joint solution for the equations, which however form an inconsistent system. Mermin concludes that the scenario is non-local.

Mermin's argument has sparked a number of lines of enquiry, and this work is concerned with two in particular: one leading to All-vs-Nothing arguments, and the other investigating the role played by the phase group. All-vs-Nothing arguments \cite{Abramsky2015} arise in the context of the sheaf-theoretic framework for non-locality and contextuality \cite{Abramsky2011}, and generalise the idea of a system of equations which is locally consistent but globally inconsistent. The second line of research is brought forward within the framework of categorical quantum mechanics \cite{Abramsky2009,Coecke2015,Coecke2016a}, and it focuses on the algebraic characterisation of phase gates and strongly complementary observables.

A detailed analysis of Mermin's argument shows that the special relationship between the Pauli $X$ and Pauli $Z$ observables, known as strong complementarity, is key to its success \cite{Coecke2012c}. A pair of complementary observables corresponds to mutually unbiased orthonormal bases: for example, both Pauli $X$ and Pauli $Y$ are complementary to Pauli $Z$. Strong complementarity \cite{Coecke2011,Duncan2016} amounts to a strictly stronger requirement: if one observable is taken as the computational basis, the other must correspond to the Fourier basis for some finite abelian group. Pauli $X$ fits the bill, for the abelian group $\integersMod{2}$, but Pauli $Y$ doesn't (Pauli $X$ is the only single-qubit observable strongly complementary to Pauli $Z$).

In \cite{Coecke2012c}, Mermin's argument is completely reformulated in terms of strongly complementary observables (using $\dagger$-Frobenius algebras) and phase gates. It can therefore be tested on theories different from quantum mechanics, to better understand the connection between non-locality and the structure of phase groups. A particularly insightful comparison is given by qubit stabiliser quantum mechanics \cite{Coecke2011,Backens2014} vs Spekkens' toy model \cite{Spekkens2007,Coecke2012a}: both theories sport very similar operational and algebraic features, but the difference in phase groups ($\integersMod{4}$ for the former vs $\integersMod{2} \times \integersMod{2}$ for the latter) results in the former being non-local and the latter being local (both models have $\integersMod{2}$ as group of measurement outcomes, like Mermin's original argument). The picture arising from comparing qubit stabiliser quantum mechanics and Spekkens' toy model is iconic, and provides a first real glimpse into the connection between phase groups and non-locality \cite{Coecke2010a}. 

While presenting an extremely compelling case for stabiliser qubits and Spekkens' toy qubits, the work of Refs.~\cite{Coecke2012c,Coecke2010a} does not treat the general case (i.e. beyond $\integersMod{2}$ as group of measurement outcomes), nor does it provide a complete algebraic characterisation of the conditions guaranteeing non-locality. In this work, we fully generalise Mermin's argument from $\integersMod{2}$ to arbitrary finite abelian groups, in arbitrary theories and for arbitrary phase groups (we will refer to these as \textbf{generalised Mermin-type arguments}). We also provide exact algebraic conditions for non-locality to be exhibited by our generalised Mermin-type arguments, thus bringing this line of investigation to a satisfactory conclusion.

We proceed to make contact with the All-vs-Nothing line of enquiry \cite{Abramsky2015}, showing that the non-local generalised Mermin-type arguments yield a new hierarchy of quantum-realisable All-vs-Nothing empirical models (and hence they are strongly contextual). As a corollary, we manage to show that the hierarchy of quantum-realisable All-vs-Nothing models over finite fields does not collapse. 

Mermin's argument for the qubit GHZ states also finds practical application in the quantum-classical secret sharing scheme of Hillery, Bu\v{z}ek and Berthiaume \cite{Hillery1999}. We extend the scheme to our generalised Mermin-type arguments, and use strong contextuality to provide some device-independent security guarantees (which apply to the original HBB scheme as a special case).

\subsection*{Synopsis of the paper}

In Section \ref{section_background}, we provide a brief recap of the technical background for this paper. We quickly review dagger compact categories, the CPM construction, Frobenius algebras, the CP* construction and the sheaf-theoretic framework for non-locality and contextuality. We introduce a new flavour of CP* categories, and prove some results connecting them to the sheaf-theoretic framework for non-locality and contextuality.

In Section \ref{section_merminOriginalArg}, we present Mermin's original argument in detail, deconstructing it in the interest of our upcoming generalisation.

In Sections \ref{section_strongCompl} and \ref{section_phaseGroup}, we introduce complementarity, strong complementarity and phase groups, and elaborate on the relationship that ties them together. Some of the simpler results are common knowledge in the field, and have appeared in similar form in other works (such as \cite{Coecke2012c,Kissinger2012,Coecke2016a}): they are re-proposed here to achieve a uniform, coherent presentation of the material.

In Section \ref{section_generalisedMerminArg}, we use strong complementarity and phase groups to formulate our generalised Mermin-type arguments, and we prove the exact algebraic conditions for the models to be non-local. In Section \ref{section_quantumRealisab}, we prove that our arguments are all quantum realisable. In Section \ref{section_AvN}, we connect with the All-vs-Nothing framework, showing that our arguments always result in All-vs-Nothing models whenever they are non-local.

In Section \ref{section_QSS}, finally, we present an extension of the HBB quantum-classical secret sharing scheme to our newly generalised Mermin-type arguments, and we provide some novel device-independent guarantees of security.

\section{Background}
\label{section_background}

\subsection*{Categories for quantum theory}

The framework of dagger compact categories captures some of the most fundamental structural features of pure-state quantum mechanics \cite{Abramsky2009,Coecke2015,Coecke2016a}: symmetric monoidal structure captures its characterisation as a theory of processes, composing sequentially and in parallel; the dagger captures the state-effect duality induced by the inner product; the compact closed structure captures operator-state duality. Dagger compact categories are commonplace in the practice of categorical quantum mechanics, and we will assume the reader is familiar with them. We recommend \cite{Selinger2009} for a detailed treatment of many subtle technicalities in the various constructions. The Hilbert space model of (finite-dimensional) pure-state quantum mechanics corresponds to the dagger compact category $\fdHilbCategory$ of (finite-dimensional) complex Hilbert spaces and linear maps between them, while the operator model of mixed-state quantum mechanics corresponds to the CPM category $\CPMCategory{\fdHilbCategory}$.

Categories of completely positive maps, also known as \textbf{CPM categories}, can be constructed for all dagger compact categories, in a process which mimics the way in which the operator model of mixed-state quantum mechanics is constructed from the Hilbert space model of pure-state quantum mechanics \cite{Selinger2007}. Given a dagger compact $\CategoryC$, the corresponding CPM category $\CPMCategory{\CategoryC}$ is defined as the subcategory of $\CategoryC$ having morphisms in the following form: 
\begin{equation}\begin{multlined}\label{eqn_CPMmorphismPrev}
\input{pictures/CPMmorphismPrev.tikz}
\end{multlined}\end{equation}
where $f: A \rightarrow B \otimes E$ is a morphism of $\CategoryC$, and $f^\ast: A^\ast \rightarrow E^\ast \otimes B^\ast$ is its conjugate, obtained via the dagger compact structure. Processes in the CPM category are called \textbf{completely positive (CP) maps}. The system $E$ in Diagram \ref{eqn_CPMmorphismPrev} is often interpreted as an \textbf{environment} which is operationally inaccessible, and hence must be \inlineQuote{discarded} after the process has taken place. In the case of $\CPMCategory{\fdHilbCategory}$, i.e. in the operator model of mixed-state quantum mechanics, Diagram \ref{eqn_CPMmorphismPrev} can then be seen as an alternative formulation of Kraus decomposition.

Because diagrammatic reasoning about categories of completely positive maps often involves two distinct SMCs (the original category $\CategoryC$ and the CPM category $\CPMCategory{\CategoryC}$), a stylistic choice is adopted where systems and processes of the CPM category are denoted by thicker wires, boxes and decorations. For example, the \inlineQuote{doubled} version $f \otimes f^\ast$ of a process $f: A \rightarrow B \otimes E$ will be denoted as $f$ with thicker wires and box:  
\begin{equation}\begin{multlined}\label{eqn_CPMdoubledNotation}
\input{pictures/CPMdoubledNotation.tikz}
\end{multlined}\end{equation}
The caps from compact closed structure play a particularly important role in the definition of the CPM category, and are given their own decoration:
\begin{equation}\begin{multlined}\label{eqn_CPMdiscardingMaps}
\input{pictures/CPMdiscardingMaps.tikz}
\end{multlined}\end{equation}
The CP map $\CPMdoubled{f}$ defined by Equation \ref{eqn_CPMdoubledNotation} is called the \textbf{double} of process $f$, while the CP map $\trace{A}$ defined by Equation \ref{eqn_CPMdiscardingMaps} is called the \textbf{discarding map} on system $A$. In mixed-state quantum mechanics, the double of a linear map $f$ is the rank-1 CP map $\CPMdoubled{f}: \rho \mapsto f \circ \rho \circ f^\dagger$, while the discarding map $\trace{A}$ sends a positive state $\rho \in \CPstates{A}$ to its trace $\Trace{\rho} \in \CPstates{\complexs} \isom \reals^+$.

The discarding maps are also called an \textit{environment structure} in the literature \cite{Coecke2012env}, and are tightly related to \textit{causality}, an important feature arising at the interface between quantum theory and relativity \cite{Coecke2013a,Coecke2015,Chiribella2011}. We say that a process is \textbf{normalised} (sometimes also called \textbf{causal}) if performing it and then discarding the output is the same as discarding the input:
\begin{equation}\begin{multlined}\label{eqn_normalised}
\input{pictures/normalised.tikz}
\end{multlined}\end{equation}
In particular, discarding normalised states results in the scalar $1$.

CPM categories $\CPMCategory{\CategoryC}$ are dagger compact, and the rules of diagrammatic reasoning for dagger compact categories apply to them. The compact structure for $\CPMCategory{\CategoryC}$ is given by the doubles of the cups and caps of $\CategoryC$, while the adjoint of a process in the form of Diagram \ref{eqn_CPMmorphism} is given by first taking the adjoint in $\CategoryC$, and then using the following equation for the adjoint of the discarding map:
\begin{equation}\begin{multlined}\label{eqn_CPMdiscardingMapsAdjoints}
\input{pictures/CPMdiscardingMapsAdjoints.tikz}
\end{multlined}
\end{equation}
Because the doubled processes $\CPMdoubled{f}$ and the discarding maps $\trace{A}$ are well-defined CP maps, it is legitimate to rephrase the very definition of the CPM category by saying that its processes are exactly those in the following form:
\begin{equation}\begin{multlined}\label{eqn_CPMmorphism}
\input{pictures/CPMmorphism.tikz}
\end{multlined}\end{equation}
This means that doubled processes and discarding maps are enough to \textit{express} all CP maps, but to \textit{prove results about} CP maps we need a graphical axiom relating a generic CPM category $\CPMCategory{\CategoryC}$ to the corresponding original category $\CategoryC$. The required relationship is encoded by the following \textbf{CPM axiom}, which characterises the action of discarding maps in $\CPMCategory{\CategoryC}$ in terms of the dagger structure of $\CategoryC$:
\begin{equation}\begin{multlined}\label{eqn_CPMaxiom}
\input{pictures/CPMaxiom.tikz}
\end{multlined}\end{equation}

\subsection*{Frobenius algebras}

Frobenius algebras are a fundamental ingredient of quantum theory, where they are intimately related to the notion of observable. A \textbf{$\dagger$-Frobenius algebra} on an object $A$ of a dagger symmetric monoidal category (henceforth $\dagger$-SMC) is given by a monoid $(A,\!\hbox{\input{symbols/ZbwmultSym.tex}}\!\!,\!\hbox{\input{symbols/ZbwunitSym.tex}}\!\!)$ on $A$ (i.e. $\!\hbox{\input{symbols/ZbwmultSym.tex}}\!\!: A \otimes A \rightarrow A$ is associative and has $\!\hbox{\input{symbols/ZbwunitSym.tex}}\!\!: \tensorUnit \rightarrow A$ as its bilateral unit), and a corresponding comonoid $(A,\!\hbox{\input{symbols/ZbwcomultSym.tex}}\!\!,\!\hbox{\input{symbols/ZbwcounitSym.tex}}\!\!)$, which are related by the following Frobenius law:
\begin{equation}\begin{multlined}\label{eqn_FrobeniusLaw}
\input{pictures/FrobeniusLaw.tikz}
\end{multlined}\end{equation}
A $\dagger$-Frobenius algebra is said to be \textbf{special} if the comultiplication $\!\hbox{\input{symbols/ZbwcomultSym.tex}}\!\!$ is an isometry, and \textbf{commutative} if the monoid and comonoid are commutative: 
\begin{equation}\begin{multlined}\label{eqn_specialCommutativeLaws}
\input{pictures/specialCommutativeLaws.tikz}
\end{multlined}\end{equation}
More in general, a \textbf{quasi-special} $\dagger$-Frobenius algebra is one with comultiplication $\!\hbox{\input{symbols/ZbwcomultSym.tex}}\!\!$ which is an isometry up to a \textbf{normalisation factor} $N$, where $N$ is in the form $n^\dagger n$ for some invertible scalar $n$:
\begin{equation}\begin{multlined}\label{eqn_quasiSpecialLaw}
\input{pictures/quasiSpecialLaw.tikz}
\end{multlined}\end{equation}
Because a several combinations of these properties will play a role in this work, we now introduce a number of short-hands for $\dagger$-Frobenius algebras:
\begin{center}
\begin{tabular}{c | c | c}
\textbf{$\dagger$-Frobenius algebras} 	& commutative 		& arbitrary 		\\ \hline
special    								& $\dagger$-SCFA 	& $\dagger$-SFA		\\ \hline
quasi-special  							& $\dagger$-qSCFA	& $\dagger$-qSFA	\\ \hline
arbitrary 								& $\dagger$-CFA		& $\dagger$-FA		\\ \hline
\end{tabular}
\end{center}

The importance of $\dagger$-SCFAs in categorical quantum mechanics comes from the fact that they correspond to orthonormal bases, i.e. non-degenerate quantum observables. Key to this correspondence is the notion of \textbf{classical states} for a $\dagger$-FA $\hbox{\input{symbols/ZbwdotSym.tex}}\!\!$, those states $\psi$ which are copied/transposed/deleted by $\!\hbox{\input{symbols/ZbwcomultSym.tex}}\!\!$ in the following sense:
\begin{equation}\begin{multlined}\label{eqn_classicalState}
\hspace{-8mm}
\resizebox{\textwidth}{!}{\input{pictures/classicalState.tikz}}
\hspace{-18mm}
\end{multlined}\end{equation}
\begin{thmC}[\cite{Coecke2013b}]
We denote the set of classical states for a $\dagger$-FA $\hbox{\input{symbols/ZbwdotSym.tex}}\!\!$ by $\classicalStates{\hbox{\input{symbols/ZbwdotSym.tex}}\!\!}$. In $\fdHilbCategory$, the classical states for a $\dagger$-SCFA $\hbox{\input{symbols/ZbwdotSym.tex}}\!\!$ always form an orthonormal basis\footnote{The copy and delete conditions are sufficient to characterise classical states in the case of $\fdHilbCategory$.}. Furthermore, any orthonormal basis arises this way for a unique $\dagger$-SCFA. More in general, if $\hbox{\input{symbols/ZbwdotSym.tex}}\!\!$ is a $\dagger$-qSCFA, with normalisation factor $N$, then the classical states for $\hbox{\input{symbols/ZbwdotSym.tex}}\!\!$ form an orthogonal basis, each state having norm $\sqrt{N}$. Furthermore, any orthogonal basis where all states have the same norm $\sqrt{N}$ arises this way for a unique $\dagger$-qSCFA.
\end{thmC}
\noindent The concept of classical states forming a basis is generalised to arbitrary $\dagger$-SMCs by the notion of enough classical states. A $\dagger$-FA $\hbox{\input{symbols/ZbwdotSym.tex}}\!\!$ on an object $A$ is said to have \textbf{enough classical states} if its classical states separate morphisms from $A$, i.e. two morphisms $f,g: A \rightarrow B$ are equal whenever they satisfy $f \circ \psi = g \circ \psi$ for all classical states $\psi$ of $\hbox{\input{symbols/ZbwdotSym.tex}}\!\!$. Because of the copy condition, a $\dagger$-FA with enough classical states is always commutative.

This algebraic characterisation of quantum observables is not limited to the non-degenerate case of orthonormal bases, but can be extended to the more general case of complete families of orthogonal projectors. To do so, one considers \textbf{balanced-symmetric} $\dagger$-Frobenius algebras\footnote{Commutative $\dagger$-FAs are a special case of balanced-symmetric $\dagger$-FAs.}, i.e. those satisfying the following equation:
\begin{equation}\begin{multlined}\label{eqn_balancedSymmetric}
\input{pictures/balancedSymmetric.tikz}
\end{multlined}\end{equation}
\begin{thmC}[\cite{Vicary2011}]\label{thm_FrobeniusCStar}
In $\fdHilbCategory$, balanced-symmetric $\dagger$-SFAs are in bijective correspondence with C*-algebras, and hence with complete families of orthogonal projectors.
\end{thmC}
\noindent The correspondence between balanced-symmetric $\dagger$-SFAs and C*-algebras will not play a role in this work, but it helps to frame the broad role played by $\dagger$-Frobenius algebras in categorical quantum mechanics. 

Further to their correspondence with quantum observables, $\dagger$-Frobenius algebras find direct use as fundamental building blocks of quantum algorithms and protocols \cite{Vicary2012a,Boixo2012,Coecke2016a,Gogioso2017b}. When designing quantum protocols, classical data is often encoded into quantum systems using orthonormal bases. In this context, the four processes in a $\dagger$-SCFA can be seen as coherent versions of the basic data manipulation primitives:
\begin{enumerate}[(a)]
\itemsep0em 
	\item the comultiplication $\!\!\hbox{\input{symbols/ZbwcomultSym.tex}}\!\!\!= \ket{\psi_x} \mapsto \ket{\psi_x} \otimes \ket{\psi_x}$ is the coherent copy of classical data;
	\item the counit $\!\!\hbox{\input{symbols/ZbwcounitSym.tex}}\!\!\! = \ket{\psi_x} \mapsto 1$ is the coherent deletion of classical data;
	\item the multiplication $\!\!\hbox{\input{symbols/ZbwmultSym.tex}}\!\!\! = \ket{\psi_x} \otimes \ket{\psi_y} \mapsto \delta_{xy}\ket{\psi_x}$ is the coherent matching of classical data;
	\item the unit $\!\!\hbox{\input{symbols/ZbwunitSym.tex}}\!\!\! = \sum_x \ket{\psi_x}$ is the coherent superposition of classical data (up to normalisation).
\end{enumerate} 
In this sense, $\dagger$-SCFAs in general $\dagger$-SMC are often interpreted as modelling coherent copy/delete/matching operations on some kind of classical data (usually modelled by their classical states)\footnote{This extends straightforwardly to $\dagger$-qSCFA and their unnormalised classical states.}. More in general, balanced-symmetric $\dagger$-SFAs can be though of as coherent manipulation of data which carries some residual entanglement after being copied\footnote{In quantum mechanics, this is because non-demolition measurements in a degenerate observable only breaks entanglement between the subspaces associated with distinct projectors, but not within each subspace.}.

Because of their diagrammatic definition, $\dagger$-FA in a dagger compact category $\CategoryC$ give rise to $\dagger$-FA in the CPM category $\CPMCategory{\CategoryC}$ by doubling: the $\dagger$-FA in $\CPMCategory{\CategoryC}$ that arise this way are said to be \textbf{canonical}. In this work, we will only consider canonical $\dagger$-FA when working with CPM categories.

\subsection*{The CP* construction}

In the framework of mixed-state quantum mechanics, classical systems can be though of as quantum systems which are constantly undergoing decoherence in some basis. In $\CPMCategory{\fdHilbCategory}$, the \textbf{decoherence map} $\decoh{\hbox{\input{symbols/ZbwdotSym.tex}}\!\!}$ in an orthonormal basis $(\ket{x})_x$ is the one zeroing out all non-diagonal elements of a positive state:
\begin{equation}
\decoh{\hbox{\input{symbols/ZbwdotSym.tex}}\!\!} := \rho \mapsto \sum_{x} \ket{x}(\bra{x}\rho\ket{x})\bra{x} 
\end{equation}
The decoherence map can be written as follows in terms of the associated $\dagger$-SCFA $\hbox{\input{symbols/ZbwdotSym.tex}}\!\!$:
\begin{equation}\begin{multlined}\label{eqn_decoherence}
\input{pictures/decoherence.tikz}
\end{multlined}\end{equation}
This means that a decohered quantum system can be though of as having undergone a coherent copy operation, with one copy lost in the environment. We take Equation \ref{eqn_decoherence} to be the definition of decoherence maps for arbitrary canonical balanced-symmetric $\dagger$-SFAs in arbitrary CPM categories.

Starting from a CPM category $\CPMCategory{\CategoryC}$, we wish to construct a new category which includes some kind of classical systems, as defined by decoherence maps. This new category $\CPStarCategory{\CategoryC}$, known as the \textbf{CP* category}, can be defined as follows:
\begin{enumerate}[(i)]
\item the objects are pairs $(A,a)$ of an object $A$ of $\CPMCategory{\CategoryC}$ and a normalised self-adjoint idempotent process $a: A \rightarrow A$ taking either the form $a := \id{A}$ or the form $a := \decoh{\hbox{\input{symbols/ZbwdotSym.tex}}\!\!}$ for some balanced-symmetric $\dagger$-SFA $\hbox{\input{symbols/ZbwdotSym.tex}}\!\!$ on $A$;
\item the morphisms $(A,a) \rightarrow (B,b)$ in CP* are the morphisms $f: A \rightarrow B$ in $\CPMCategory{\CategoryC}$ in CPM which are invariant under the specified idempotents on $A$ and $B$, i.e. those which satisfy $b \circ f \circ a = f$.
\end{enumerate}
Note that this definition is a hybrid of the perspective of Ref. \cite{Selinger2008}, based on decoherence maps and the Karoubi envelope, and the perspective of Refs. \cite{Coecke2014a,Cunningham2015}, based on C* algebras and quantum logic. Because we have picked processes to be invariant under self-adjoint idempotents, $\CPStarCategory{\CategoryC}$ is always dagger compact. 

The CP* category contains $\CPMCategory{\CategoryC}$ as the full subcategory associated with objects in the form $(A,\id{A})$: we refer to the latter as \textbf{CPM objects}\footnote{We will keep using the doubled notation for them and morphisms between them.}, and we simply denote them by $A$ for simplicity. We refer to the CP* objects in the form $(A,\decoh{\hbox{\input{symbols/ZbwdotSym.tex}}\!\!})$ as \textbf{super-selected objects}, and often denote them by $(A,\hbox{\input{symbols/ZbwdotSym.tex}}\!\!)$ for simplicity. 

In the case of $\CPStarCategory{\fdHilbCategory}$, a balanced-symmetric $\dagger$-SFA $\hbox{\input{symbols/ZbwdotSym.tex}}\!\!$ on a finite-dimensional Hilbert space $A$ corresponds, by Theorem \ref{thm_FrobeniusCStar}, to a complete family $(P_j)_j$ of orthogonal projectors\footnote{I.e. $P_i \circ P_j = \delta_{ij} P_i$ and $\sum_i P_i = \id{A}$.}. The associated decoherence map $\decoh{\hbox{\input{symbols/ZbwdotSym.tex}}\!\!}$ takes the following concrete form:
\begin{equation}
\decoh{\hbox{\input{symbols/ZbwdotSym.tex}}\!\!} = \rho \mapsto \sum_{j} P_j \, \rho \, P_j^\dagger
\end{equation} 
This means that objects in $\CPStarCategory{\fdHilbCategory}$ truly are super-selected quantum systems, with super-selection sectors given by the domains of the projectors $(P_j)_j$. In particular, super-selected objects associated to $\dagger$-SCFAs (corresponding to non-degenerate observables, i.e. families of 1-dimensional projectors) behave as classical probabilistic systems.

If $\hbox{\input{symbols/ZbwdotSym.tex}}\!\!$ is a balanced-symmetric canonical $\dagger$-SFA on an object $A$, the decoherence map $\decoh{\hbox{\input{symbols/ZbwdotSym.tex}}\!\!}$ is always a process $\decoh{\hbox{\input{symbols/ZbwdotSym.tex}}\!\!}: A \rightarrow A$ in $\CPStarCategory{\CategoryC}$. Because of idempotence, however, it is also a process $A \rightarrow (A,\hbox{\input{symbols/ZbwdotSym.tex}}\!\!)$ and a process $(A,\hbox{\input{symbols/ZbwdotSym.tex}}\!\!) \rightarrow A$: we will refer to the former as the \textbf{measurement} in $\hbox{\input{symbols/ZbwdotSym.tex}}\!\!$, and the latter as the \textbf{preparation} in $\hbox{\input{symbols/ZbwdotSym.tex}}\!\!$. The single and doubled notation distinguish between the different cases:
\begin{equation}\begin{multlined}\label{diagram_preparationMeasurementCPStar}
\input{pictures/preparationMeasurementCPStar.tikz}
\end{multlined}\end{equation}
In $\CPStarCategory{\fdHilbCategory}$, preparations and measurements for a $\dagger$-SCFA $\hbox{\input{symbols/ZbwdotSym.tex}}\!\!$ associated to an orthonormal basis $(\ket{x})_{x \in X}$ of a finite-dimensional Hilbert space $A$ take the familiar form traditionally adopted by the literature:
\begin{equation}\begin{multlined}\label{eqns_preparationMeasurementQM}
\input{pictures/preparationMeasurementQM.tikz}
\end{multlined}\end{equation}
Demolition measurements are traditionally thought to result in some kind of classical (probabilistic) data. Unfortunately, our framework does not yet allow us to conclude anything of the sort, as we lack an appropriate definition of classical systems to work with.
Following the footsteps of the sheaf-theoretic framework for non-locality and contextuality \cite{Abramsky2011}, we generalise probabilities from $\reals^+$ to some arbitrary commutative semiring $R$. For a fixed commutative semiring $R$, we define our category of \textbf{classical $R$-probabilistic  systems} to be the category $\RMatCategory{R}$ of free, finite-dimensional $R$-semimodules and $R$-semilinear maps between them:
\begin{enumerate}[(a)]
\item the objects of $\RMatCategory{R}$ are in the form $R^X$ for all finite sets $X$;
\item the morphisms $R^X \rightarrow R^Y$ in $\RMatCategory{R}$ are $Y \otimes X$ matrices with values in $R$
\item $\RMatCategory{R}$ is a SMC, with $\fSetCategory$ (finite sets and functions) as a sub-SMC;
\item $\RMatCategory{R}$ inherits the discarding maps $\trace{X} := x \mapsto \star$ of $\fSetCategory$;
\item $\RMatCategory{R}$ is enriched in itself, with morphisms $R^{Y \otimes X}$ forming the free finite-dimensional $R$-semimodule $R^{Y \otimes X}$.
\end{enumerate}
If we generalise this definition to \textit{involutive} commutative semirings $R$, i.e. those coming with an involution $\dagger: R \rightarrow R$, the category $\RMatCategory{R}$ is in fact a dagger compact category. 

The traditional definition of \textbf{classical probabilistic systems} corresponds to working in $\RMatCategory{\reals^+}$: normalised states are probability distributions over finite sets, and normalised processes are stochastic maps (also, we always think of $\reals^+$ as coming with the trivial involution $\id{\reals^+}$). However, using arbitrary semirings opens the way to interesting generalisations: a prominent example is that of \textbf{classical possibilistic systems}, which are associated to the semiring $R=\mathbb{B}$ of the booleans and play a large role in the sheaf-theoretic framework for non-locality and contextuality.

\begin{defi}
We say that a SMC $\CategoryD$ is \textbf{distributively $\CMonCategory$-enriched} if the following conditions hold:
\begin{enumerate} 
\item the category is $\CMonCategory$-enriched, i.e. morphisms $A \rightarrow B$ form a commutative monoid $\Big(\Hom{\CategoryD}{A}{B},+,0\Big)$ for any fixed objects $A,B$;
\item the tensor product $\otimes$, associators and unitors are all linear.
\end{enumerate}
The definition can be extended to a $\dagger$-SMC (or dagger compact category) $\CategoryD$ by asking that the dagger also be linear.
\end{defi}
The scalars of SMCs which are distributively $\CMonCategory$-enriched always form a commutative semiring $R$ (which is furthermore involutive in the case of $\dagger$-SMCs), and all homsets automatically inherit the structure of $R$-semimodules. We use this observation to define classical systems within the context of CP* categories. 
\begin{defi}
We say that a $\CPStarCategory{\CategoryC}$ is an \textbf{$R$-probabilistic CP* category}, or an \textbf{$R$-probabilistic theory}, if it satisfies the following conditions.
\begin{enumerate}
	\item[(i)] The dagger compact category $\CPStarCategory{\CategoryC}$ is distributively $\CMonCategory$-enriched, with $R$ as its involutive semiring of scalars\footnote{Equivalently, we can ask for $\CPMCategory{\CategoryC}$ to be enriched, as the two categories mutually inherit enrichment, scalars and discarding maps. }.
	\item[(ii)] For each $n \in \naturals$, there is some super-selected system $(A,\hbox{\input{symbols/ZbwdotSym.tex}}\!\!)$ in $\CPStarCategory{\CategoryC}$ such that:
		\begin{enumerate}
			\item[(a)] $\hbox{\input{symbols/ZbwdotSym.tex}}\!\!$ is a $\dagger$-SCFA with enough classical states;
			\item[(b)] the classical states of $\hbox{\input{symbols/ZbwdotSym.tex}}\!\!$ are mutually orthogonal.
			\item[(c)] $\hbox{\input{symbols/ZbwdotSym.tex}}\!\!$ has exactly $n$ classical states;
		\end{enumerate}
\end{enumerate}
In this context, we refer to super-selected systems $(A,\hbox{\input{symbols/ZbwdotSym.tex}}\!\!)$ satisfying conditions (a) and (b) above as \textbf{classical systems}. Hence requirement (ii) above can be rephrased to say that for every $n \in \naturals$ there is a classical system with $n$ classical states.
\end{defi}
\begin{thm}
The full sub-SMC of an $R$-probabilistic CP* category spanned by the classical systems is equivalent to $\RMatCategory{R}$.
\end{thm}
\proof
All we really need to show is that processes $(A,\hbox{\input{symbols/ZbwdotSym.tex}}\!\!) \rightarrow (B,\hbox{\input{symbols/YbwdotSym.tex}}\!\!)$ between two classical systems in the CP* category form an $R$-module which is isomorphic to the $R$-module of processes $R^{\classicalStates{\hbox{\input{symbols/YbwdotSym.tex}}\!\!}} \rightarrow R^{\classicalStates{\hbox{\input{symbols/ZbwdotSym.tex}}\!\!}}$ in the category $\RMatCategory{R}$ of classical $R$-probabilistic systems. Firstly, every process $f: (A,\hbox{\input{symbols/ZbwdotSym.tex}}\!\!) \rightarrow (B,\hbox{\input{symbols/YbwdotSym.tex}}\!\!)$ is determined by the $R$-valued matrix obtained by testing against classical states of the two $\dagger$-SCFAs:
	\begin{equation}\begin{multlined}\label{eqns_CPStarProcessesRModule1}
	\resizebox{\textwidth}{!}{\input{pictures/CPStarProcessesRModule1.tikz}}
	\end{multlined}\end{equation}
Secondly, every matrix $\big(F_x^y\big)_{x \in \classicalStates{\hbox{\input{symbols/ZbwdotSym.tex}}\!\!}}^{y \in \classicalStates{\hbox{\input{symbols/YbwdotSym.tex}}\!\!}}$ corresponds to a unique process $(A,\hbox{\input{symbols/ZbwdotSym.tex}}\!\!) \rightarrow (B,\hbox{\input{symbols/YbwdotSym.tex}}\!\!)$:
	\begin{equation}\begin{multlined}\label{eqns_CPStarProcessesRModule2}
	\input{pictures/CPStarProcessesRModule2.tikz}
	\end{multlined}\end{equation}
Because we have required that for each $n \in \naturals$ there be a classical system with $n$ classical states, the bijective correspondence above establishes the desired equivalence of categories.
\qed

As a special case, classical systems in CP* categories with $\reals^+$ as their semiring of scalars behave exactly like classical probabilistic systems, and the entire toolbox of probability theory becomes available: we will refer to these as \textbf{probabilistic theories}, and they will come into play in the context of our very last result. Since the time this work was first written, a stand-alone framework for probabilistic theories capturing the constructions above has been developed by one of the authors, and can be found in Ref. \cite{Gogioso2017a}. Examples of notable quantum-like theories captured by this framework can be found in Ref. \cite{Gogioso2017FQT}.

\subsection*{Non-locality and contextuality}

Consider the abstract setup of a Bell-type scenario:
\begin{enumerate}[(i)] 
\item $N$ parties are given devices $B_1,...,B_N$ which might share some global state $\rho$;
\item each device $B_j$ takes an input, the \textbf{measurement choice}, freely chosen by party $j$ from some finite set $M_j$;
\item upon receiving input $m_j \in M_j$, the device $B_j$ produces some output $o_j$ in some finite set $O_j$, the \textbf{measurement outcome};
\item no signalling is possible between the devices from before the first input is given to after the last outputs has been produced. 
\end{enumerate}
The sheaf-theoretic framework for non-locality and contextuality \cite{Abramsky2011} characterises the distribution of joint outputs conditional to joint inputs from the point of view of sheaf theory, showing that non-locality and contextuality are related to the (non-) existence of global sections for a particular presheaf. The framework does not rely on any concrete description of the state $\rho$ or the devices $B_1,...,B_N$, focusing instead on the distributional properties of joint outputs/measurement outcomes $\underline{o_j} := (o_1,...,o_N)$ conditional to the choice $\underline{M} := (m_1,...,m_N)$ of joint inputs/measurement choices.

The framework begins by identifying a finite set $\mathcal{X}$ of inputs, which in the Bell-type scenario setup above (the one used in this work) would be $\mathcal{X} = \sqcup_{j=1}^N M_j$. The disjoint union preserves information about which party each measurement is associated to, so we will adopt the notation $m_j$ for generic elements of $\mathcal{X}$, where $m$ is the measurement and $j$ is the party. For each subset $U \subseteq \mathcal{X}$, the family of all potential\footnote{Not all subsets of measurements need be compatible in each concrete scenario: see below for the definition of measurement contexts.} \textbf{joint outcomes} takes the following form:
\begin{equation}
	\sheafOfEvents{U} := \prod_{m_j \in U} O_j
\end{equation}
The powerset $\Powerset{\mathcal{X}}$ is a poset (hence a poset category) under inclusion $V \subseteq U$ of subsets. We can define a functor $\sheafOfEventsSym: \OpCategory{\Powerset{\mathcal{X}}} \rightarrow \SetCategory$, i.e. a \textbf{presheaf}, by setting:
\begin{enumerate}[(i)]
	\item if $U \in  \Powerset{\mathcal{X}} $, then we define $ \sheafOfEvents{U} := \prod_{m_j \in U} O_j$ as above
	\item if $V \subseteq U $, then we define $\sheafOfEvents{V \subseteq U}:=\restrictionMap{U}{V}$ to be the \textbf{restriction map} $U \stackrel{\SetCategory}{\longrightarrow} V$:
	\begin{equation}
		\restrictionMap{U}{V} = s \mapsto \restrict{s}{V}
	\end{equation}
\end{enumerate}
A \textbf{section $s$ over $U$} is a $U$-indexed family of outcomes in the following form:
\begin{equation}
	s = \suchthat{(m_j,s(m_j))}{m_j \in U} \in  \prod_{m_j \in U} O_j
\end{equation}
The restriction map then sends a section $s$ over $U$ to its restriction over $V$:
\begin{equation}
	\restrict{s}{V} = \suchthat{(m_j,s(m_j))}{m_j \in V} \in  \prod_{m_j \in V} O_j
\end{equation}

The definition of the set of possible joint inputs requires further consideration: it is a fundamental feature of quantum mechanics that not all measurements on a system are compatible, and we should not expect different measurement choices in each $M_j$ to have a consistent assignment of outputs. Instead, the framework requires us to specify a set $\mathcal{M}$ of \textbf{measurement contexts}, subsets $C \subseteq \mathcal{X}$ of measurements which are mutually compatible (and therefore have a well-defined notion of joint outcome). Even though more general setups are allowed, we will assume that our measurement contexts all take the form $C = \{m_1,...,m_N\}$ for $m_j \in M_j$, which we will denote by $\underline{m}$: each party chooses exactly one input for their device, but we allow the possibility that not all combinations of inputs might be allowed/interesting. The only requirement is that $\cup_{C \in \mathcal{M}} C = \mathcal{X}$, i.e. that $\mathcal{M}$ be a \textbf{global cover} of $\mathcal{X}$ (each measurement choice for each player appears in at least one measurement context), which we assume to be endowed with the discrete topology. One can also define the \textbf{local covers} for any $U \subseteq \mathcal{X}$ as the families $(U_i)_{i \in I}$ such that $\cup_{i \in I} U_i = U$.

The choice of the discrete topology on $\mathcal{X}$ makes $\Powerset{\mathcal{X}}$ its locale of open subsets, and one can define a notion of \textbf{sheaf} on it. Because it is defined in terms of sections\footnote{Compatibility of local sections amounts to compatibility over the intersection of the domains, and hence compatible local sections can always be glued together by taking their union as relations.}, the presheaf $\sheafOfEventsSym$ is in fact a sheaf on the locale $\Powerset{\mathcal{X}}$, and we shall refer to it as the \textbf{sheaf of events}. The measurement cover and the sheaf of events are the two ingredients required to define a \textbf{measurement scenario} $(\sheafOfEventsSym,\mathcal{M})$: the former gives the compatible joint measurement choices, while the latter gives the joint measurement outcomes conditional on all possible measurement choices.

The next step in the framework sees the introduction of generalised notions of probabilities and distributions. In quantum mechanics, probabilities can be seen as taking values in the commutative semiring $R = (\reals^+,+,0,\cdot,1)$ of the non-negative reals (in fact they fall within the interval $[0,1]$, a consequence in the semiring $R$ of the normalisation condition requiring that probabilities add up to $1$). In other circumstances, one may be interested in the \textbf{possibilities} associated with events, living in the commutative semiring $\mathbb{B} = (\{0,1\},\vee,0,\wedge,1)$ of the booleans. In the sheaf-theoretic treatment of contextuality, one works with an arbitrary commutative semiring $R = (|R|,+,0,\cdot,1)$. 

Given a set $U$, an \textbf{$R$-distribution} on $U$ is a function $d: U \rightarrow R$ which has finite \textbf{support} $\support{d} := \suchthat{s \in U}{d(s) \neq 0}$ and such that $\sum_{s \in \support{d}} d(s) = 1$.  One can then define a functor $\distributionFunctorSym{R}: \SetCategory \rightarrow \SetCategory$ as follows:
\begin{enumerate}[(i)]
	\item for any set $U$, define $\distributionFunctor{R}{U}$ to be the set of $R$-distributions of $U$
	\item for any function $f: U \rightarrow V$, define $\distributionFunctor{R}{f} := d \mapsto \left[ t \mapsto \sum_{f(s) = t} d(s) \right]$.
\end{enumerate}
Composing this functor with the sheaf of events yields the \textbf{presheaf of distributions} $\presheafOfDistributionsSym{R}:\OpCategory{\Powerset{\mathcal{X}}} \rightarrow \SetCategory$, which captures the structure of $R$-distributions on joint measurement outcomes under marginalisation. The presheaf sends each set $U$ of measurements (the objects of the presheaf category $\Powerset{\mathcal{X}}$) to the set $\presheafOfDistributions{R}{U}$ of \textbf{$R$-distributions on $U$-sections}, and sends any inclusion $V \subseteq U$ (the morphisms of the presheaf category $\Powerset{\mathcal{X}}$) to the corresponding marginalisation of distributions:
\begin{equation}
	\presheafOfDistributions{R}{V \subseteq U} = d \mapsto \restrict{d}{V} := \left[ t \mapsto \sum_{\restrict{s}{V} = t} d(s)\right]
\end{equation}
We will refer to $\restrict{d}{V}$ as the \textbf{marginal} of $d$.

In quantum mechanics, if $C$ is a set of compatible measurements on some state $\ket{\psi}$, then there is a probability distribution $d \in \presheafOfDistributions{\reals^+}{C}$ on the joint outcomes of the measurements, and the typical contextuality argument involves showing that the probability distributions on different contexts cannot be obtained, in a no-signalling scenario, as marginals of some non-contextual hidden variable. In the sheaf-theoretic framework, a \textbf{(no-signalling) empirical model} is defined to be a compatible family of distributions $(\zeta_C)_{C \in \mathcal{M}}$ for the global cover $\mathcal{M}$ of measurement contexts; the usual no-signalling property is shown in \cite{Abramsky2011} to be a special case of the compatibility condition. In other literature (usually treating probabilistic models), empirical models for Bell-type scenarios are usually given explicitly as conditional (probability) distributions, in a format akin to the following:
\begin{equation}
\zeta_{\,\underline{m}}\big(\underline{o}\big) := \mathbb{P}\big[\,\underline{o}\,\big\vert\, \underline{m}\,\big]
\end{equation}
where $\underline{m} = (m_1,...,m_N) \in \mathcal{M}$ are the measurement contexts used by the scenario and $\underline{o} \in \prod_{j} O_j$ are the joint outcomes. This is the format we will use in the last section of this work. In the probabilistic case, empirical models for a fixed scenario form a polytope. However, this need not be the same as the no-signalling polytope which usually studied in quantum information theory, because the set of measurement contexts need not include all possible combinations of all possible measurements for each party (i.e. it need not be the case that $\mathcal{M} = \prod_j M_j$, although it is the case that $\mathcal{M} \subseteq \prod_j M_j$). 

A \textbf{global section} for an empirical model\footnote{From now on, no-signalling is implicitly assumed.} $(\zeta_C)_{C \in \mathcal{M}}$ is a distribution $d \in \presheafOfDistributions{R}{\mathcal{X}}$ over the joint outcomes of all measurements which marginalises to the distributions specified by the empirical model:
\begin{equation}
	\restrict{d}{C}=\zeta_C \text{ for all } C \in \mathcal{M}
\end{equation} 
\noindent The fundamental observation behind the sheaf-theoretic framework is that the existence of a global section for an empirical model is equivalent to the existence of a \textbf{non-contextual hidden variable model} (also known as a \textbf{local hidden variable model}). Concretely, the existence of a global section $d$ means that there is a finite set $\Lambda$, an $R$-distribution $q(\lambda): \Lambda \rightarrow R$ and a family of functions $f_j^{\lambda}: M_j \rightarrow O_j$ such that:
\begin{equation}
\zeta_{\,\underline{m}}\big(\underline{o}\big) = \sum_{\lambda \in \Lambda} q(\lambda) \prod_{j} \delta_{f_j^{\lambda}(m_j) = o_j}
\end{equation}
We will say that an empirical model $(\zeta_C)_{C \in \mathcal{M}}$ is \textbf{contextual} (or \textbf{non-local}) if it doesn't admit a global section.

\subsection*{Strong contextuality}

Contextuality of probabilistic models is interesting in itself, but more refined notions can be obtained by relating $\reals^+$ to two other semirings: the reals, modelling signed probabilities, and the booleans, modelling possibilities. Observe that the construction $\distributionFunctorSym{R}$ is functorial in $R$, so that for any morphism of semirings $r: R \rightarrow R'$ we can define the following:
\begin{equation}
	\distributionFunctor{r}{U} = \left[d:U \rightarrow R\right] \mapsto \left[r \circ d: U \rightarrow R'\right]
\end{equation}
\noindent In particular, there is an injective morphism of semirings $i^+: \reals^+ \inject R$ sending $x \in \reals^+$ to $+x \in \reals$, as well as a surjective morphism of semirings $p: \reals^+ \rightarrow \mathbb{B}$ sending $0 \mapsto 0$ and $x \neq 0 \mapsto 1$ (the latter is well defined for all positive semirings, not just for $\reals^+$).  

If $(\zeta_C)_{C \in \mathcal{M}}$ is a probabilistic empirical model, i.e. one in the semiring $\reals^+$, then $(\zeta_C)_{C \in \mathcal{M}}$ can be seen as an empirical model $(i^+ \circ \zeta_C)_{C \in \mathcal{M}}$ in the semiring $\reals$: regardless of whether $(\zeta_C)_{C \in \mathcal{M}}$ was contextual or not over $\reals^+$, it can be shown \cite{Abramsky2011} that over the reals it always admits a global section. On the other hand, any probabilistic empirical model $(\zeta_C)_{C \in \mathcal{M}}$ can be assigned a corresponding possibilistic empirical model $(p\circ \zeta_C)_{C \in \mathcal{M}}$ in the semiring $\mathbb{B}$ of the booleans (and each boolean function $p \circ \zeta_C$ can equivalently be seen as the characteristic function of the subset $\support{\zeta_C} \subseteq \sheafOfEvents{C}$). 

Note that contextuality is a contravariant property with respect to change of semiring: if $(\zeta_C)_{C \in \mathcal{M}}$ is an empirical model in a semiring $R$ and $r: R \rightarrow R'$ is a morphism of semiring, then contextuality of $(r \circ \zeta_C)_{C \in \mathcal{M}}$ implies contextuality of $(\zeta_C)_{C \in \mathcal{M}}$ (because a global section $d$ of the latter is mapped to a global section $r \circ d$ of the former). We will say that a probabilistic empirical model $(\zeta_C)_{C \in \mathcal{M}}$ is \textbf{possibilistically contextual} if the corresponding possibilistic model $(p \circ \zeta_C)_{C \in \mathcal{M}}$ is contextual (as opposed to \textbf{probabilistically contextual}, which we use to say that $(\zeta_C)_{C \in \mathcal{M}}$ is contextual over $\reals^+$). Because of contravariance, possibilistic contextuality implies probabilistic contextuality, but the opposite is not true: the Bell model given in \cite{Abramsky2011} is probabilistically contextual but not possibilistically contextual.

Seeing distributions $d \in \presheafOfDistributions{\mathbb{B}}{U}$ as indicator functions of the subsets $\support{d} \subseteq \sheafOfEvents{U}$ endows them with a partial order:
\begin{equation}
	\label{eqn_SCimpliesC}
	d' \preceq d \text{ if and only if } \support{d'} \subseteq \support{d}
\end{equation}
\noindent The existence of a global section $d \in \presheafOfDistributions{\mathbb{B}}{U}$ for a possibilistic empirical model $(\zeta_C)_{C \in \mathcal{M}}$ implies that:
\begin{equation}
	\label{eqn_StrongContextualityCondition}
	\restrict{d}{C} \preceq \zeta_C \text{ for all } C \in \mathcal{M}
\end{equation}
We say that a possibilistic empirical model $(\zeta_C)_{C \in \mathcal{M}}$ is \textbf{strongly contextual} if there is no distribution $d \in \presheafOfDistributions{\mathbb{B}}{\mathcal{X}}$ such that Equation \ref{eqn_StrongContextualityCondition} holds. In particular, the GHZ model given in \cite{Abramsky2011}, corresponding to Mermin's original non-locality argument, is strongly contextual. Because of Equation \ref{eqn_SCimpliesC}, strong contextuality implies contextuality, but the opposite is not true: the possibilistic Hardy model give in \cite{Abramsky2011} is contextual, but not strongly contextual. We will say that a probabilistic empirical model is strongly contextual if the associated possibilistic empirical model is strongly contextual, yielding the following strict hierarchy of notions of contextuality for probabilistic empirical models:
\begin{equation}
	\text{probabilistically contextual } \Leftarrow \text{ possibilistically contextual } \Leftarrow \text{ strongly contextual}
\end{equation}

\subsection*{Empirical models within the CP* construction}

The relevance of the sheaf-theoretic framework to this work stems from the following result: in any $R$-probabilistic CP* category, all Bell-type scenarios give rise to an empirical model.
\begin{defi}\label{def_BellTest}
Consider an $R$-probabilistic CP* category. A \textbf{Bell-type scenario} is a process $\Phi: (A_1,\hbox{\input{symbols/YbwdotSym.tex}}\!\!_1) \otimes ... \otimes (A_N,\hbox{\input{symbols/YbwdotSym.tex}}\!\!_N) \rightarrow (B_1,\hbox{\input{symbols/DdotSym.tex}}\!\!_1) \otimes ... \otimes (B_N,\hbox{\input{symbols/DdotSym.tex}}\!\!_N)$, where all $(A_j,\hbox{\input{symbols/YbwdotSym.tex}}\!\!_j)$ and all $(B_j,\hbox{\input{symbols/DdotSym.tex}}\!\!_j)$ are classical systems, which takes the following form, for some normalised state $\rho$ and some normalised processes $B_1,...,B_N$:
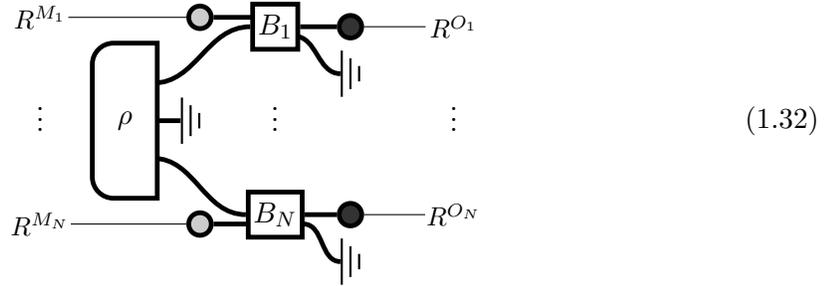
\begin{equation}\label{eqn_BellTest}
	\input{pictures/BellTest.tikz}
\end{equation}
\end{defi}
\begin{thm}
Consider a Bell-type scenario $\Phi$ in the form given by Definition \ref{def_BellTest}. Let $M_j$ be the finite set of classical states for $\hbox{\input{symbols/YbwdotSym.tex}}\!\!_j$, and $O_j$ be the finite set of classical states for $\hbox{\input{symbols/DdotSym.tex}}\!\!_j$. Then the process $\Phi$ gives rise to a no-signalling empirical model $(\zeta_{\,\underline{m}})_{\underline{m} \in \mathcal{M}}$ as follows, for any cover $\mathcal{M}$:
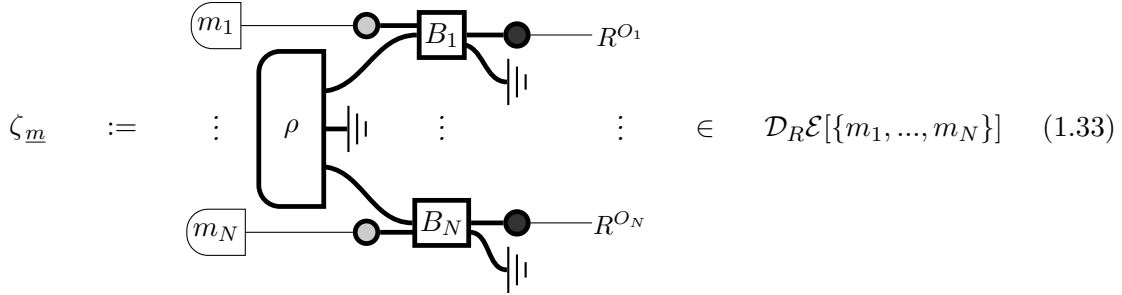
\begin{equation}\label{eqn_BellTestEM}
	\input{pictures/BellTestEM.tikz}
\end{equation}
\end{thm}
\proof 

We need to show that the states in Equation \ref{eqn_BellTestEM} (indexed by the measurement contexts $\underline{m} \in \mathcal{M}$) satisfy no-signalling and are normalised (i.e. are $R$-distributions). To do so, we (i) marginalise over party $j$, (ii) use the fact that the discarding map on the classical systems $(B_j,\hbox{\input{symbols/DdotSym.tex}}\!\!_j)$ can be written as $\trace{\,(B_j,\hbox{\input{symbols/DdotSym.tex}}\!\!_j)} = \sum_{o_j} \bra{o_j}$, and (iii) use the fact that the measurement on $\hbox{\input{symbols/DdotSym.tex}}\!\!_j$ and the process $B_j$ are both normalised to show that the resulting state is independent of $m_j$:
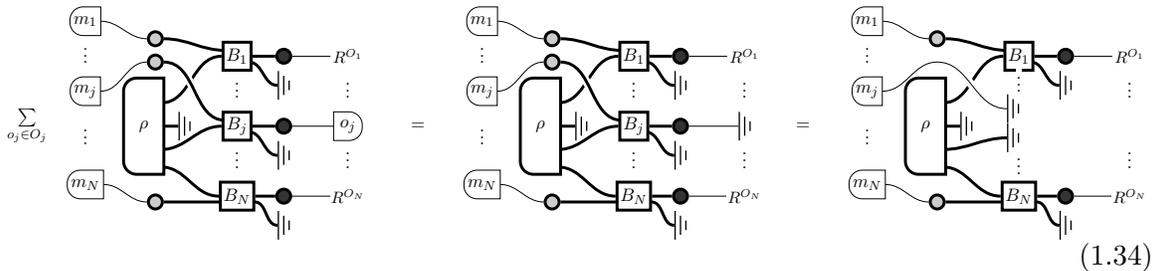
\begin{equation}\label{diagram_BellTestEMproof}
	\resizebox{\textwidth}{!}{\input{pictures/BellTestEMproof.tikz}}
\end{equation}
Marginalising over all outputs leaves us with $\trace{}\circ \rho$, which equals $1$ (independently of the measurement context $\underline{m}$) because $\rho$ is normalised. Hence the state $(\zeta_{\,\underline{m}})_{\underline{m} \in \mathcal{M}}$ is also an $R$-distribution as desired, completing our proof.
\qed

\section{Mermin's Original Argument}
\label{section_merminOriginalArg}

\subsection*{The parity argument} 

\noindent In the original \cite{Mermin1990}, Mermin considers a 3-qubit GHZ state in the computational basis, the basis of eigenstates for the single-qubit Pauli $Z$ observable, together with the following four joint measurements\footnote{Where $X_j$ and $Y_j$ are the single-qubit Pauli $X$ and $Y$ observables on qubit $j$, for $j=1,2,3$.}:
\begin{enumerate}[\indent(a)]
	\item the GHZ state is measured in the observable $X_1 \tensor X_2 \tensor X_3$;
	\item the GHZ state is measured in the observable $\;Y_1 \tensor \;Y_2 \tensor X_3$;
	\item the GHZ state is measured in the observable $\;Y_1 \tensor X_2 \tensor \;Y_3$;
	\item the GHZ state is measured in the observable $X_1 \tensor \;Y_2 \tensor \;Y_3$.
\end{enumerate}
We will denote the eigenstates of the Pauli $Z$ observable by $\ket{z_0},\ket{z_1}$, the eigenstates of the Pauli $X$ observable by $\ket{\pm} := \frac{1}{\sqrt{2}}(\ket{z_0} \pm \ket{z_1})$ and the eigenstates of the Pauli $Y$ observable by $\ket{\pm i} := \frac{1}{\sqrt{2}}(\ket{z_0} \pm i \ket{z_1})$. Mermin's argument is a parity argument, where measurement outcomes are valued in the abelian group $\integersMod{2} = \{0,1\}$ according to the following bijections:
\begin{enumerate}[\indent(i)]
	\item for the $X$ observable, $\ket{+} \mapsto 0$ and $\ket{-} \mapsto 1$
	\item for the $Y$ observable, $\ket{+i} \mapsto 0$ and $\ket{-i} \mapsto 1$
\end{enumerate}

\noindent The argument then proceeds as follows. While the joint measurement outcomes are probabilistic, the $\integersMod{2}$ sum of the three outcomes turns out to be deterministic, yielding the following system of equations ($\oplus$ here denotes the sum in $\integersMod{2}$):
\begin{equation}
\label{eqn_MerminSystemZ2Equations}
\begin{cases}
X_1 \oplus X_2 \oplus X_3 &= 0 \\
\;Y_1 \oplus \;Y_2 \oplus X_3 &= 1 \\
\;Y_1 \oplus X_2 \oplus \;Y_3 &= 1 \\
X_1 \oplus \;Y_2 \oplus \;Y_3 &= 1 
\end{cases}
\end{equation}
If there was a non-contextual assignment of outcomes for all measurements ($X_1,X_2,X_3,Y_1,Y_2$ and $Y_3$), i.e. if there existed a non-contextual hidden variable model, then System \ref{eqn_MerminSystemZ2Equations} would have a solution in $\integersMod{2}$, and in particular it would have to be consistent. However, the sum of the left hand sides yields $0$ in $\integersMod{2}$:
\begin{equation}
	\label{eqn_MerminSystemZ2EquationsLHSSum}
	2 X_1 \oplus 2 X_2 \oplus ... \oplus 2 Y_3 = 0 X_1 \oplus ... \oplus 0 Y_3 = 0
\end{equation}
while the sum of the right hand sides yields $0 \oplus 1 \oplus 1 \oplus 1 = 3 = 1$ in $\integersMod{2}$. This shows the system to be inconsistent. Equivalently, one could observe that the sum of the LHS from Equation \ref{eqn_MerminSystemZ2EquationsLHSSum} can be written as $2(Y_1 \oplus Y_2 \oplus Y_3)$, and that inconsistency of the system is witnessed by the fact that the equation $2 y = 1$ has no solution in~$\integersMod{2}$. 

The first point of view, where contextuality is witnessed by an inconsistent system where each equation individually admits a solution, is behind the generalisation of Mermin's argument to All-vs-Nothing arguments, presented in \cite{Abramsky2015}. The second point of view, where contextuality is witnessed by the single unsatisfiable equation $2 y = 1$, will inspire the generalisation presented in this work.

\subsection*{The role of phases}

To understand the role played by the equation $2 y = 1$ in the original Mermin argument, we need to take a step back. First of all, we observe that the Pauli $Y$ measurement can be equivalently obtained as a Pauli $X$ measurement preceded by an appropriate unitary. A single-qubit \textbf{phase gate}, in the computational basis (the Pauli $Z$ observable), is a unitary transformation in the following form:
\begin{equation}
	\label{eqn_Z2PhaseGates}
	\phasegate{\alpha} := \left(
	\begin{array}{cc}
		1 & 0 \\
		0 & e^{i \alpha}
	\end{array}
	\right)
\end{equation}
where we eliminated global phases by setting the first diagonal element to 1. Measuring in the single-qubit $Y$ observable is equivalent to first applying the single-qubit phase gate $\phasegate{\frac{\pi}{2}}$, and then measuring in the Pauli $X$ observable.

Because they pairwise commute, phase gates come with a natural abelian group structure given by composition, resulting in an isomorphism $\alpha \mapsto P_\alpha$ between them and the abelian group\footnote{The abelian group $\reals/(2\pi\integers)$ is isomorphic to the circle group $S^1$. We prefer the former because of its additive notation, as opposed to the traditionally multiplicative notation of the latter (which is a subgroup of the non-zero multiplicative complex numbers $\complexs^\times$).} $\reals/(2\pi\integers)$. Of all the phase gates, $\phasegate{0}$ (the identity element of the group) and $\phasegate{\pi}$ stand out because of their well-defined action on the (unnormalised) eigenstates of the Pauli $X$ observable:
\begin{align}
	\label{eqn_PzeroPiAction}
	\phasegate{0} & = \ket{\pm} \mapsto \ket{\pm} \nonumber\\
	\phasegate{\pi} & = \ket{\pm} \mapsto \ket{\mp} 
\end{align}
If we see $\ket{\pm}$ as the subgroup\footnote{Corresponding to $\{\pm 1\} < S^1$ in the circle group.} $\{0,\pi\} < \reals / (2\pi\integers)$, then Equation \ref{eqn_PzeroPiAction} looks a lot like the regular action of $\{0,\pi\}$ on itself. This is not a coincidence. Each phase gate $\phasegate{\alpha}$ can be (faithfully) associated the unique \textbf{phase state} $\ket{\alpha} := \ket{z_0} + e^{i \alpha} \ket{z_1}$ obtained from its diagonal, and these phase states can be abstractly characterised in terms of the Pauli $Z$ observable, with no reference to the phase gates they came from (\cf~section \ref{section_phaseGroup}). The phase states inherit the abelian group structure of the phase gates, and their regular action coincides with the action of the group of phase gates on them. In particular, the phase gates $\phasegate{0}$ and $\phasegate{\pi}$ have orthogonal eigenstates of the Pauli $X$ observable as their associated phase states $\ket{0}$ and $\ket{\pi}$, which coincide with $\sqrt{2}\ket{+}$ and $\sqrt{2}\ket{-}$ respectively: this endows the outcomes of Pauli $X$ measurements with the natural $\integersMod{2}$ abelian group structure arising\footnote{Natural because there is a unique isomorphism $\integersMod{2} \isom \{0,\pi\}$.} from the inclusion $\{0,\pi\} < \reals/(2\pi\integers)$. We will henceforth refer to the group of phase states as the \textbf{group of $Z$-phase states}, and to the subgroup $\{0,\pi\}$ as the \textbf{subgroup of $X$-classical states}; the latter will also be used to label the corresponding measurement outcomes.

In order to pave the way to our generalisation, we now proceed to show how Mermin's original argument can be re-constructed from the following statement: 
\begin{displayquote}
	\textit{the equation $2 y = \pi$ has no solution in the subgroup $\{0,\pi\}$ of $X$-classical states, but a solution\footnote{Corresponding to $y = e^{i\frac{\pi}{2}} = +i$ in the circle group $S^1$.} $y = \frac{\pi}{2}$ can be found in the larger group $\reals/(2\pi\integers)$ of $Z$-phase states.}
\end{displayquote}
We begin by observing that tripartite qubit GHZ state used in Mermin's argument has a special property when it comes to the application of phase gates followed by measurements in the Pauli $X$ observable.
\begin{lemC}[\cite{Coecke2012c}]\label{lemma_qubitGHZphaseSum} If $\alpha_j \in \reals/(2\pi\integers)$, denote by $X_j^{\alpha_j}$ the measurement outcome on qubit $j$ obtained by first applying phase gate $\phasegate{\alpha_j}$, and then measuring in the Pauli $X$ observable. If $\alpha_1 \oplus \alpha_2 \oplus \alpha_3 = \modclass{0\text{ or }\pi}{2\pi}$, then $X_1^{\alpha_1} \oplus X_2^{\alpha_2} \oplus X_3^{\alpha_3} = \modclass{0 \text{ or }\pi}{2\pi}$ respectively.
\end{lemC}
\noindent Now consider System \ref{eqn_MerminSystemZ2Equations} again, with values on the the RHS now obtained by applying Lemma~\ref{lemma_qubitGHZphaseSum} to $X_j := X_j^{0}$ and $Y_j := X_j^{\frac{\pi}{2}}$ (and valued in $\{0,\pi\}$ instead of the original $\integersMod{2}$):
\begin{equation}
	\label{eqn_MerminSystemZ2EquationsPM}
	\begin{cases}
	X^{0}_1\; \oplus X^{0}_2\; \oplus X^{0}_3\; &= 0 \text{, the control}\\
	X^{\frac{\pi}{2}}_1 \oplus X^{\frac{\pi}{2}}_2 \oplus X^{0}_3\; &= \pi \text{, the first variation}\\
	X^{\frac{\pi}{2}}_1 \oplus X^{0}_2\; \oplus X^{\frac{\pi}{2}}_3 &= \pi \text{, the second variation}\\
	X^{0}_1\; \oplus X^{\frac{\pi}{2}}_2 \oplus X^{\frac{\pi}{2}}_3 &= \pi \text{, the third variation}
	\end{cases}
\end{equation}
There are two complementary parts to the Mermin non-locality argument: (i) System \ref{eqn_MerminSystemZ2EquationsPM} above must be inconsistent, to rule out the existence of a non-contextual hidden variable model, and (ii) joint measurements yielding the individual equations must be possible (in quantum theory). For the first part, inconsistency of the system is witnessed by the fact that the equation $2 y = \pi$ has no solution in the subgroup of $X$-classical states. For the second part, notice that only measurements in the $Y$ observable contribute to the sum for each equation, as measurements in the $X$ observable are associated with the group unit $0$ of the group of $Z$-phase states. As a consequence, the existence of measurements implementing each individual equation reduces to the existence of a $Z$-phase state $\ket{y}$ satisfying equation $2 y = \pi$: the $Y$ observable is chosen exactly because $y = \pi/2$ gives one such $Z$-phase state.

The following steps summarise the skeleton of the argument, and open the way to our generalisation: 
\begin{enumerate}[\indent 1.]
	\item consider a non-degenerate observable, call it $Z$, on an arbitrary quantum system;
	\item consider another non-degenerate observable, call it $X$, such that the $X$-classical states are a subgroup (call it $K$) of the abelian group of $Z$-phase states (call it $P$);
	\item consider an equation in the following form, generalising $2 y = \pi$:
	\begin{equation}\label{eqn_generalisedInconsistencyEquation}
		n_1 y_1 \oplus ... \oplus n_M y_M = a
	\end{equation}
	(here $a \in K$, $n_1,...,n_M$ are integers\footnote{This is a general equation in abelian groups, seen equivalently as $\integers$-modules.}, and $\oplus$ is the group addition in $P$);
	\item construct an appropriate system of equations, generalising System \ref{eqn_MerminSystemZ2EquationsPM}, with inconsistency witnessed by non-existence of solutions for  Equation \ref{eqn_generalisedInconsistencyEquation} in $K$, and consistency of the individual equations witnessed by the existence of solutions in $P$;
	\item a measurement scenario can be implemented if and only if a solution exists in $P$;
	\item the measurement scenario is contextual if and only if no solutions exist in $K$.
\end{enumerate} 

\noindent To give a first example of how such an appropriate system of equations might be constructed, we consider the simple generalisation of the argument from a $3$-partite to an $N$-partite GHZ state, for appropriate values of $N \geq 2$. Our requirements are as follows: 
\begin{enumerate}[\indent(i)]
	\item we want the phases in the control to sum to $0$, and hence we will take them all to be $0$ (i.e. measurements in the $X$ observable), just as in the original argument; 
	\item we also want the phases in each variation to sum to $\pi$, and hence we will take two measurements in each variation to be with phase $\pi/2$ (i.e. measurements in the $Y$ observable), and all the other ones to be with phase $0$; 
	\item we want an odd number $V$ of variations, so that the RHSs will sum to $0 \oplus V \pi = \pi$;
	\item we want the LHSs to sum to an even multiple of $X_1^{\frac{\pi}{2}} \oplus ... \oplus X_N^{\frac{\pi}{2}}$;
\end{enumerate}
An appropriate choice is given by the following system of equations, where $V := N$ and all  variations are cyclic permutations of the first one:
\begin{equation}
	\begin{cases}
	X_1^{0}\; \oplus X_2^{0}\; \oplus X_3^{0}\; \oplus\;\;\;\, ...\;\;\;\hspace{0.2mm} \oplus X_{N-1}^{0}\; \oplus X_N^{0}\; &= 0 \text{, the control}\\
	X_1^{\frac{\pi}{2}} \oplus X_2^{\frac{\pi}{2}} \oplus X_3^{0}\; \oplus\;\;\;\, ...\;\;\;\hspace{0.4mm} \oplus X_{N-1}^{0}\; \oplus X_N^{0}\; &= \pi \text{, the $1^{st}$ variation}\\
	X_1^{\frac{\pi}{2}} \oplus X_2^{0}\; \oplus\;\, ...\;\hspace{0.2mm} \oplus X_{N-2}^{0}\; \oplus X_{N-1}^{0}\; \oplus X_N^{\frac{\pi}{2}} &= \pi \text{, the $2^{nd}$ variation}\\
	\hspace{29.6mm}\vdots \\
	X_1^{0}\; \oplus X_2^{\frac{\pi}{2}} \oplus X_3^{\frac{\pi}{2}} \oplus \;\;\;X_4^{0}\;\;\hspace{0.5mm} \oplus \;\;\;\, ...\;\;\;\hspace{0.3mm} \oplus X_{N}^{0}\; &= \pi \text{, the $N^{th}$ variation}\\
	\end{cases}
\end{equation}
As long as $N = \modclass{1}{k}$, where $k=2$ is the exponent\footnote{The smallest positive integer such that $k x = 0$ for all $x \in K$.} of $K$, the RHSs will sum to $\pi$ in $K$. Having \vspace{-1mm} chosen our variations by cyclic permutation also makes for the desired sum of the LHSs, since each $X_j^{\pi/2}$ will be counted exactly twice:
\vspace{3mm}
\begin{center}
\begin{tabular}{ccccc}
$\big(X_1^{0} \oplus ... \oplus X_N^{0}\big)$ & $\oplus$ & $2 \cdot \big(X_1^{\frac{\pi}{2}} \oplus ... \oplus X_N^{\frac{\pi}{2}})$ & $\oplus$ & $(N-2)\cdot \big(X_1^{0} \oplus ... \oplus X_N^{0})$ \\
control & & $X_j^{\frac{\pi}{2}}$s from the variations & &  $X_j^{0}$s from the variations\\ 
\end{tabular}
\end{center}
\vspace{3mm}
Writing $x$ for $X_1^{0} \oplus ... \oplus X_N^{0}$ and $y$ for $X_1^{\frac{\pi}{2}} \oplus ... \oplus X_N^{\frac{\pi}{2}}$, the sum above can be rearranged to take the form $(N-1) x \oplus 2 y$, which is equal to $2 y$ in $K$ (since $(N-1) = \modclass{0}{k}$)\footnote{In this specific case, it is also true that $2 = \modclass{0}{k}$, but this is not key to the argument.}. Hence summing all the LHSs and RHSs leaves us with the equation $2 y = \pi$, which we know to be unsatisfiable in $K$.

\section{Strong Complementarity}
\label{section_strongCompl}

Mermin's parity argument is fundamentally group-theoretic, and it depends almost entirely on the special relationship between the Pauli $Z$ and Pauli $X$ observables. Fixing the eigenstates of the Pauli $Z$ observable as the computational basis, the requirement that the $X$-classical states are $Z$-phase states is satisfied by the Pauli $X$ observable, but also by the Pauli $Y$: in fact, the $Z$-phase states are exactly the \textbf{unbiased states} for the Pauli $Z$ observables, the states lying on the equator of the Bloch sphere, and hence any observable \textbf{complementary}, or \textbf{mutually unbiased}, to Pauli $Z$ would do the trick; because their eigenstates lie on the equator of the Bloch sphere, we will refer to observables complementary to Pauli $Z$ as \textbf{equatorial observables}. Definition \ref{def_complementarity} gives an algebraic/diagrammatic presentation of complementarity using Hopf's Law, and Lemma \ref{lem_complementarityUnbiased} shows that observables which are complementary observables under this definition are always mutually unbiased. A more general result relating complementarity and mutual unbias in $\dagger$-SMCs will be given by Theorem \ref{thm_characterisingCandSC} in the next section.

\begin{defi}\label{def_complementarity}
Two $\dagger$-qSFAs $\hbox{\input{symbols/ZbwdotSym.tex}}\!\!$ and $\hbox{\input{symbols/DdotSym.tex}}\!\!$ on the same object $\SpaceH$ of a $\dagger$-SMC are said to be \textbf{complementary} if they satisfy the following \textbf{Hopf's Law}:
\begin{equation}\label{eqn_HopfsLaw}
\input{pictures/HopfsLaw.tikz}
\end{equation}
where the \textbf{antipode} $\hbox{\input{symbols/antipodeSym.tex}}\!\!: \SpaceH \rightarrow \SpaceH$ is the unitary defined as follows, which we require to be self-adjoint (or equivalently self-inverse) as part of the definition of complementarity:
\begin{equation}\label{eqn_antipode}
\input{pictures/antipode.tikz}
\end{equation}
\end{defi}
\begin{defi}\label{def_unbiasedStates}
Let $\hbox{\input{symbols/YbwdotSym.tex}}\!\!$ be a $\dagger$-qSFA in a dagger compact category. Then a state $u$ is a \textbf{$\hbox{\input{symbols/YbwdotSym.tex}}\!\!$-unbiased} state if the following holds:
\begin{equation}\label{eqn_YunbiasedState}
\input{pictures/YunbiasedState.tikz}
\end{equation}
Equation \ref{eqn_YunbiasedState} can be unfolded into the following more general definition, which holds in an arbitrary $\dagger$-SMCs:
\begin{equation}
	\input{pictures/ZphaseStateDefExplained.tikz}
\end{equation}
\end{defi}
In $\fdHilbCategory$, the $\hbox{\input{symbols/YbwdotSym.tex}}\!\!$-unbiased states are those which, once normalised, yield the uniform distribution upon measurement in the $\hbox{\input{symbols/YbwdotSym.tex}}\!\!$ observable. 
\begin{lem}\label{lem_complementarityUnbiased}
Consider a complementary pair of $\dagger$-qSFAs $\hbox{\input{symbols/ZbwdotSym.tex}}\!\!$ and $\hbox{\input{symbols/DdotSym.tex}}\!\!$. The $\hbox{\input{symbols/DdotSym.tex}}\!\!$-classical states are $\hbox{\input{symbols/ZbwdotSym.tex}}\!\!$-unbiased, and the $\hbox{\input{symbols/ZbwdotSym.tex}}\!\!$-classical states are $\hbox{\input{symbols/DdotSym.tex}}\!\!$-unbiased.
\end{lem}
\proof We prove that a $\hbox{\input{symbols/DdotSym.tex}}\!\!$-classical state $\goodchi$ is $\hbox{\input{symbols/ZbwdotSym.tex}}\!\!$-unbiased: 
\begin{equation}\label{eqn_HopfLawUnbiasedStatesProof}
\resizebox{\textwidth}{!}{\input{pictures/HopfLawUnbiasedStatesProof.tikz}}
\end{equation}
The first equality is by $\hbox{\input{symbols/DdotSym.tex}}\!\!$-classicality (delete condition), the second equality is Hopf's law (together with the self-adjoint requirement for the antipode), the third equality is again by $\hbox{\input{symbols/DdotSym.tex}}\!\!$-classicality (copy condition for the bottom state, transpose condition for the top state), the last equality is by Frobenius law and unit law for $\hbox{\input{symbols/ZbwdotSym.tex}}\!\!$. The proof for $\hbox{\input{symbols/ZbwdotSym.tex}}\!\!$-classical states is the same, with colours swapped.\qed

Complementarity is not sufficient for Mermin's argument: Lemma \ref{lemma_qubitGHZphaseSum} only holds if we measure the GHZ state in the Pauli $X$ observable, not in any other equatorial observable. The algebraic relationship between the Pauli $X$, $Y$ and $Z$ observables is vividly captured by the ZX calculus \cite{Coecke2011}: there, the special property relating the Pauli $Z$ and $X$ observables is axiomatised under the name of \textbf{strong complementarity}, to distinguish it from the complementarity of Pauli $Z$ and any other equatorial observable (such as Pauli $Y$). Strong complementarity is behind the proof of Lemma \ref{lemma_qubitGHZphaseSum}, which lies at core of the fully diagrammatic treatment of Mermin's original argument appearing in \cite{Coecke2012c}. 

Definition \ref{def_strongComplementarity} gives an algebraic/diagrammatic presentation of strong complementarity, while Theorem \ref{thm_SCHilb} provides an exact correspondence in finite-dimensional Hilbert spaces between complementarity and the representation theory of finite abelian groups. A more general characterisation of strong complementarity in $\dagger$-SMCs will be given by Theorem \ref{thm_characterisingCandSC} in the next section.

\begin{defi}\label{def_strongComplementarity}
Two $\dagger$-qSFAs $\hbox{\input{symbols/ZbwdotSym.tex}}\!\!$ and $\hbox{\input{symbols/DdotSym.tex}}\!\!$ on the same object $\SpaceH$ of a $\dagger$-SMC are said to be \textbf{strongly complementary} if they are complementary and furthermore satisfy the following equations\footnote{The empty diagram on the RHS of the top right equation is the scalar 1.}:
\begin{equation}\label{eqns_SC}
\resizebox{\textwidth}{!}{\input{pictures/strongComplementarityAlt.tikz}}
\end{equation}
\end{defi}
\begin{rmrk}
Technically speaking, the central equations of both rows are not necessary: indeed, they do not usually feature in the literature. The central equation on the top row of \ref{eqns_SC} is in fact a consequence of Hopf's law and the other two equations of the top row:
\begin{equation}
\resizebox{\textwidth}{!}{\input{pictures/HopfsLawClassicalAdjunctionsConsequence.tikz}}
\end{equation}
Similarly, the central equation of the top row with colours swapped is a consequence of Hopf's law, the rightmost equation of the top row, and the rightmost equation of the bottom row. The central equation of the bottom row can also be obtained using Hopf's law, self-adjointness of the antipode, and the other equations.
\end{rmrk}

Strong complementarity means that the eigenstates of the Pauli $X$ observable are very specific equatorial states, given by the two multiplicative characters $\goodchi: \integersMod{2} \rightarrow S^1$ of the abelian group $\integersMod{2}$:
\begin{equation}
	\ket{\goodchi} := \goodchi(0)\ket{z_0} + \goodchi(1)\ket{z_1} = 
	\begin{cases}
		\sqrt{2} \ket{+} & \text{if $\goodchi$ is the trivial character } \goodchi(j) := +1\\ 
		\sqrt{2} \ket{-} & \text{if $\goodchi$ is the alternating character } \goodchi(j) := (-1)^j 
	\end{cases}
\end{equation}
This group-theoretic characterisation of strong complementarity fully generalises to arbitrary finite-dimensional Hilbert spaces and finite abelian groups.

\begin{thmC}[\cite{Coecke2012c,Kissinger2012}]\label{thm_SCHilb}\hfill\\
Let $\hbox{\input{symbols/ZbwdotSym.tex}}\!\!$ and $\hbox{\input{symbols/DdotSym.tex}}\!\!$ be a $\dagger$-SCFA and a $\dagger$-qSCFA on the same finite-dimensional Hilbert space $\SpaceH$. Then $\hbox{\input{symbols/ZbwdotSym.tex}}\!\!$ and $\hbox{\input{symbols/DdotSym.tex}}\!\!$ are strongly complementary iff there exists an abelian group $(G,\oplus,0)$ such that $(\!\hbox{\input{symbols/DmultSym.tex}}\!\!,\!\hbox{\input{symbols/DunitSym.tex}}\!\!)$ endows the set of $\hbox{\input{symbols/ZbwdotSym.tex}}\!\!$-classical states with the abelian group structure of $G$, i.e. iff we can label the $\hbox{\input{symbols/ZbwdotSym.tex}}\!\!$-classical states as $(\ket{g})_{g \in G}$ in a way such that:
	\begin{equation}
		\input{pictures/multG.tikz}
	\end{equation}
If $\hbox{\input{symbols/ZbwdotSym.tex}}\!\!$ and $\hbox{\input{symbols/DdotSym.tex}}\!\!$ are strongly complementary, then the $\hbox{\input{symbols/DdotSym.tex}}\!\!$-classical are labelled by the multiplicative characters $\goodchi: G \rightarrow S^1$ of the abelian group $G$, and take the following form:
\begin{equation}\label{eqn_XclassicalStatesMultChars}
	\ket{\goodchi} := \sum_{g \in G} \goodchi(g) \ket{g} 
\end{equation}
Furthermore, $(\!\hbox{\input{symbols/ZbwmultSym.tex}}\!\!,\!\hbox{\input{symbols/ZbwunitSym.tex}}\!\!)$ turns the set of $X$-classical states into the finite abelian group $(G^\wedge, \cdot, \mathbb{1})$ of multiplicative characters\footnote{Multiplicative characters form an abelian group under pointwise multiplication $\big(\chi \cdot \chi'\big)(g) := \chi(g) \cdot \chi'(g)$ and with the trivial character $\mathbb{1} := g \mapsto 1$ as group unit. This is known as the \textbf{Pontryagin dual} $G^\wedge$ of $G$.} of $G$:
	\begin{equation}\label{eqn_XclassicalStatesGroupStructure}
		\input{pictures/multGwedge.tikz}
	\end{equation}
\end{thmC}

The importance of strong complementarity for quantum algorithms \cite{Vicary2012a,Gogioso2017b} mostly lies in the following observation: if the $\hbox{\input{symbols/ZbwdotSym.tex}}\!\!$-classical states are taken to form the computational basis, then the $\hbox{\input{symbols/DdotSym.tex}}\!\!$-classical states form the (unnormalised) Fourier basis, and measuring in the $\hbox{\input{symbols/DdotSym.tex}}\!\!$ observable amounts to performing the quantum Fourier transform. The situation with Mermin-type arguments, however, is different: the relevant facet of strong complementarity will be the special relationship between $\hbox{\input{symbols/DdotSym.tex}}\!\!$-classical points and $\hbox{\input{symbols/ZbwdotSym.tex}}\!\!$-phase states, explored in detail in the coming section.

\begin{rmrk}
Theorem \ref{thm_SCHilb} extend to the case where $\hbox{\input{symbols/DdotSym.tex}}\!\!$ is a $\dagger$-qSFA, i.e. not commutative, as long as the adjective abelian is dropped for the group $G$ (which becomes a generic finite group); it should however be noted that the $\hbox{\input{symbols/DdotSym.tex}}\!\!$-classical states form a basis if and only if $G$ is abelian (if and only if $\hbox{\input{symbols/DdotSym.tex}}\!\!$ is commutative, i.e. a non-degenerate observable). The group $G^\wedge$ of multiplicative characters of a finite group $G$ is always abelian, and Pontryagin duality\footnote{The existence of a (canonical) isomorphism $(G^\wedge)^\wedge \isom G$.} necessarily fails when $G$ is not abelian. Theorem \ref{thm_SCHilb} also extends to the case where $\hbox{\input{symbols/ZbwdotSym.tex}}\!\!$ is a $\dagger$-qSCFA, in which case the $\hbox{\input{symbols/ZbwdotSym.tex}}\!\!$-classical states form an orthogonal basis, rather than an orthonormal one. Most of Theorem \ref{thm_SCHilb} can be further extended to the case where $\hbox{\input{symbols/ZbwdotSym.tex}}\!\!$ is a generic $\dagger$-qSFA: it is still true that $(\!\hbox{\input{symbols/DmultSym.tex}}\!\!,\!\hbox{\input{symbols/DunitSym.tex}}\!\!)$ endows the $\hbox{\input{symbols/ZbwdotSym.tex}}\!\!$-classical states with the structure of a finite group, and that $(\!\hbox{\input{symbols/ZbwmultSym.tex}}\!\!,\!\hbox{\input{symbols/ZbwunitSym.tex}}\!\!)$ endows the $\hbox{\input{symbols/DdotSym.tex}}\!\!$-classical states with the structure of a finite group, but Equation \ref{eqn_XclassicalStatesMultChars} need not hold, and the group structure given by $(\!\hbox{\input{symbols/ZbwmultSym.tex}}\!\!,\!\hbox{\input{symbols/ZbwunitSym.tex}}\!\!)$ need not be that of the group of multiplicative characters. 
\end{rmrk}

\section{The phase group}
\label{section_phaseGroup}

If strong complementarity is the fundamental algebraic property at work in Mermin's argument, phase gates and GHZ states are the operational components key to its implementation. Phase gates arise in the context of quantum-to-classical transitions, where they provide a characterisation, in the spirit of groups and symmetries, of how much information is lost by performing a (demolition) measurement in a non-degenerate observable.
\begin{defi}
Let $\hbox{\input{symbols/ZbwdotSym.tex}}\!\!$ be a $\dagger$-qSFA on an object $\SpaceH$ of a dagger compact category. Then the \textbf{$\hbox{\input{symbols/ZbwdotSym.tex}}\!\!$-phase gates} are the unitaries $U: \SpaceH \rightarrow \SpaceH$ which are annihilated by the measurement:
\begin{equation}\label{eqn_ZphaseGateDef}
	\input{pictures/ZphaseGateDef.tikz}
\end{equation} 
Equation \ref{eqn_ZphaseGateDef} can be unfolded into the following equivalent definition, which extends to an arbitrary $\dagger$-SMC:
\begin{equation}\label{eqn_ZphaseGateDefExplained}
	\input{pictures/ZphaseGateDefExplained.tikz}
\end{equation}
\end{defi}

\begin{rmrk} A simpler algebraic characterisation of phase gates is given by the following two equations, which are equivalent to Equation \ref{eqn_ZphaseGateDefExplained} (because $U$ is assumed to be unitary):
\begin{equation}\label{eqn_ZphaseGateConsequencesComonoid}
	\input{pictures/ZphaseGateConsequencesComonoid.tikz}
\end{equation}
\begin{equation}\label{eqn_ZphaseGateConsequencesMonoid}
	\input{pictures/ZphaseGateConsequencesMonoid.tikz}
\end{equation}
Both equations will play a pivotal role in this section: Equation \ref{eqn_ZphaseGateConsequencesComonoid} will features shortly in Lemma \ref{lem_GHZphaseGates}, the result relating phase gates and GHZ states, while Equation \ref{eqn_ZphaseGateConsequencesMonoid} will feature later on in Theorem \ref{thm_PhaseGatesUnbiasedStates}, the result relating phase gates and unbiased states. 
\end{rmrk}

From Equation \ref{eqn_ZphaseGateDef}, it is not hard to see that $\hbox{\input{symbols/ZbwdotSym.tex}}\!\!$-phase gates form a group: we will refer to this as the \textbf{$\hbox{\input{symbols/ZbwdotSym.tex}}\!\!$-phase group}, and we will denote it by $\phaseGroup{\hbox{\input{symbols/ZbwdotSym.tex}}\!\!}$. If $\hbox{\input{symbols/ZbwdotSym.tex}}\!\!$ is a $\dagger$-SFA on a finite-dimensional Hilbert space $\SpaceH$, associated with a direct sum decomposition $\SpaceH = \oplus_j \SpaceH_j$, then the phase group $\phaseGroup{\hbox{\input{symbols/ZbwdotSym.tex}}\!\!}$ is given by the corresponding direct sum of unitary groups, modulo a global phase: 
\begin{equation}
\phaseGroup{\hbox{\input{symbols/ZbwdotSym.tex}}\!\!} = \Big(\oplus_j U(\SpaceH_j)\Big) \big/ S^1
\end{equation}
In the special case where $\hbox{\input{symbols/ZbwdotSym.tex}}\!\!$ is a $\dagger$-SCFA on $\SpaceH$, i.e. when all $\SpaceH_j$ subspaces are 1-dimensional, the phase group is abelian, the translation group of a torus: 
\begin{equation}
\phaseGroup{\hbox{\input{symbols/ZbwdotSym.tex}}\!\!} = \Big(\oplus_{j=1}^{\dim{\SpaceH}}U(1) \Big) \big/ S^1 \isom T^{\dim{\SpaceH}-1}
\end{equation}  
The connection between abelian phase groups and commutative Frobenius algebras generalises from $\fdHilbCategory$ to arbitrary dagger compact categories. The following result shows that the phase group of a commutative Frobenius algebra is always abelian, while the converse will be proven later on in Corollary \ref{cor_ZphaseGroupAbelianIff} (conditional to the existence of enough unbiased states)
\begin{lem}\label{lem_ZphaseGroupAbelianProof}
Let $\hbox{\input{symbols/ZbwdotSym.tex}}\!\!$ be a $\dagger$-qSFA on an object $\SpaceH$ of a dagger compact category. If $\hbox{\input{symbols/ZbwdotSym.tex}}\!\!$ is commutative, then the $\hbox{\input{symbols/ZbwdotSym.tex}}\!\!$-phase group $\phaseGroup{\hbox{\input{symbols/ZbwdotSym.tex}}\!\!}$ is abelian.
\end{lem}
\proof 
\begin{equation}
\input{pictures/ZphaseGroupAbelianProof.tikz}
\end{equation}
The first equality is by unit law for $\hbox{\input{symbols/ZbwdotSym.tex}}\!\!$; the second equality is by Equation \ref{eqn_ZphaseGateConsequencesComonoid}; the third equality is some topological manipulation; the fourth equality (top right to bottom left) is by commutativity of $\hbox{\input{symbols/ZbwdotSym.tex}}\!\!$; the fifth equality is by Equation \ref{eqn_ZphaseGateConsequencesComonoid}; the sixth equality is commutativity of $\hbox{\input{symbols/ZbwdotSym.tex}}\!\!$; the seventh and last equality is by Equation \ref{eqn_ZphaseGateConsequencesComonoid}, followed by unit law for $\hbox{\input{symbols/ZbwdotSym.tex}}\!\!$. \qed

Having defined the phase group and proven Lemma \ref{lem_ZphaseGroupAbelianProof}, we are now in a position to state the first important result of this section. Lemma \ref{lem_GHZphaseGates} characterises the sates that can be obtained by application of phases gates to a GHZ state: in the context of our generalised Mermin-type arguments, it will play the same role that Lemma \ref{lemma_qubitGHZphaseSum} played in Mermin's original argument.     
\begin{defi}\label{def_GHZ}
If $\hbox{\input{symbols/ZbwdotSym.tex}}\!\!$ is a $\dagger$-qSFA on an object $\SpaceH$ of a dagger compact category, the \textbf{$N$-partite $\hbox{\input{symbols/ZbwdotSym.tex}}\!\!$-GHZ state} is the following state of $\SpaceH^{\otimes N}$:
\begin{equation}
\input{pictures/GHZstate.tikz}
\end{equation}
\end{defi}
\begin{lem}\label{lem_GHZphaseGates}
Let $\hbox{\input{symbols/ZbwdotSym.tex}}\!\!$ be a $\dagger$-qSCFA on an object $\SpaceH$ of a dagger compact category. Then the state obtained by applying $\hbox{\input{symbols/ZbwdotSym.tex}}\!\!$-phase gates $U_1,...,U_N$ to the $N$-partite $\hbox{\input{symbols/ZbwdotSym.tex}}\!\!$-GHZ state only depends on the composition $U_1 \cdot ...\cdot U_N$ of the phase gates:
\begin{equation}
\input{pictures/GHZstatePhaseGates.tikz}
\end{equation}
\end{lem}
\proof Each $\hbox{\input{symbols/ZbwdotSym.tex}}\!\!$-phase gate is pushed down by using Equation \ref{eqn_ZphaseGateConsequencesComonoid} and commutativity of $\hbox{\input{symbols/ZbwdotSym.tex}}\!\!$. Formally, the proof is by induction, with inductive step given by the following equality:
\begin{equation}
\resizebox{\textwidth}{!}{\input{pictures/GHZstatePhaseGatesProof.tikz}}
\end{equation}
\qed

We have remarked before that the phase gates in Mermin's original argument are associated to certain phase states, extracted from their diagonalisation, which are also unbiased states for the relevant observable. As the following Theorem \ref{thm_PhaseGatesUnbiasedStates} shows, the connection between $\hbox{\input{symbols/ZbwdotSym.tex}}\!\!$-phase gates and $\hbox{\input{symbols/ZbwdotSym.tex}}\!\!$-unbiased states holds true in full generality, and as a consequence we will also refer to $\hbox{\input{symbols/ZbwdotSym.tex}}\!\!$-unbiased states as \textbf{$\hbox{\input{symbols/ZbwdotSym.tex}}\!\!$-phase states}. In the case of $\fdHilbCategory$, the decomposition of a $\hbox{\input{symbols/ZbwdotSym.tex}}\!\!$-phase gate $U$ given by Equation \ref{eqn_ZphaseGateZphaseStateStatement} for a $\dagger$-SCFA $\hbox{\input{symbols/ZbwdotSym.tex}}\!\!$ is equivalent to saying that $U$ is diagonal in the orthonormal basis $(\ket{x})_x$ associated with $\hbox{\input{symbols/ZbwdotSym.tex}}\!\!$, and has diagonal encoded by state $\ket{u}$ as $U_{xx} = \braket{x}{u}$.
\begin{thm}\label{thm_PhaseGatesUnbiasedStates}
Let $\hbox{\input{symbols/ZbwdotSym.tex}}\!\!$ be a $\dagger$-qSFA on an object $\SpaceH$ of a dagger compact category. Then the $\hbox{\input{symbols/ZbwdotSym.tex}}\!\!$-phase gates are exactly the maps $\phasegate{u}$ taking the following form for a $\hbox{\input{symbols/ZbwdotSym.tex}}\!\!$-unbiased state $u$:
\begin{equation}\label{eqn_ZphaseGateZphaseStateStatement}
	\input{pictures/ZphaseGateZphaseStateStatement.tikz}
\end{equation}
\end{thm}
\proof First we prove that any phase gate $U$ takes the form above, for some $\hbox{\input{symbols/ZbwdotSym.tex}}\!\!$-unbiased state $u$.
An appropriate state $u$ can then be obtained by unit law for $\hbox{\input{symbols/ZbwdotSym.tex}}\!\!$: 
\begin{equation}
	\input{pictures/ZphaseGateZphaseStateProof1.tikz}
\end{equation}
By using Equation \ref{eqn_ZphaseGateDefExplained}, we can prove that the state we obtained is $\hbox{\input{symbols/ZbwdotSym.tex}}\!\!$-unbiased:
\begin{equation}
	\input{pictures/ZphaseGateZphaseStateProof2.tikz}
\end{equation}
Then we prove that any $U$ in the form above with $u$ a $\hbox{\input{symbols/ZbwdotSym.tex}}\!\!$-unbiased state is a unitary:
\begin{equation}
	\input{pictures/ZphaseStateZphaseGateUnitaryProof.tikz}
\end{equation}
Finally, we prove that any unitary $U$ in the form above with $u$ a $\hbox{\input{symbols/ZbwdotSym.tex}}\!\!$-unbiased state is a $\hbox{\input{symbols/ZbwdotSym.tex}}\!\!$-phase gate:
\begin{equation}
	\input{pictures/ZphaseStateZphaseGateProof.tikz}
\end{equation}
\qed
\noindent Because of the correspondence above, we will adopt a uniform notation for phase gates and phase states, known in the literature as \textbf{decorated spider} notation \cite{Coecke2011,Coecke2016a}:
\begin{equation}
	\input{pictures/ZphaseGatesStatesDecoratedSpiders.tikz}
\end{equation}

\begin{cor}\label{cor_GHZphaseStates}
Let $\hbox{\input{symbols/ZbwdotSym.tex}}\!\!$ be a $\dagger$-qSCFA on an object $\SpaceH$ of a dagger compact category. Then the state obtained by applying $\hbox{\input{symbols/ZbwdotSym.tex}}\!\!$-phase gates $\phasegate{u_1},...,\phasegate{u_N}$ to the $N$-partite $\hbox{\input{symbols/ZbwdotSym.tex}}\!\!$-GHZ state takes the following form in terms of the corresponding $\hbox{\input{symbols/ZbwdotSym.tex}}\!\!$-phase states ${u_1},...,{u_N}$:
\begin{equation}
\input{pictures/GHZstatePhaseStates.tikz}
\end{equation}
That is, the states that can be obtained by applying $\hbox{\input{symbols/ZbwdotSym.tex}}\!\!$-phase gates to the $N$-partite $\hbox{\input{symbols/ZbwdotSym.tex}}\!\!$-GHZ state are exactly those obtained by comultiplying $N$-times some $\hbox{\input{symbols/ZbwdotSym.tex}}\!\!$-unbiased state $u$ (specifically, above we have $u = u_1 \cdot ... \cdot u_N$, and all $\hbox{\input{symbols/ZbwdotSym.tex}}\!\!$-unbiased states can be obtained this way). 
\end{cor}
\proof From Lemma \ref{lem_GHZphaseGates}, by re-writing each $\hbox{\input{symbols/ZbwdotSym.tex}}\!\!$-phase gate in terms of the corresponding $\hbox{\input{symbols/ZbwdotSym.tex}}\!\!$-phase state using Theorem \ref{thm_PhaseGatesUnbiasedStates}, and then using associativity to group the $\hbox{\input{symbols/ZbwdotSym.tex}}\!\!$-phase states.
\qed

The group structure of phase gates transfers to unbiased states via the correspondence given by Theorem \ref{thm_PhaseGatesUnbiasedStates}. Albeit not surprising, this result plays an important role in our generalisation of Mermin-type arguments, where it connects the operational side of phase gates and GHZ states to the algebraic side of strong complementarity (in the group-theoretic characterisation given by Theorem \ref{thm_SCHilb} and Theorem \ref{thm_characterisingCandSC} below).
\begin{lem}
Let $\hbox{\input{symbols/ZbwdotSym.tex}}\!\!$ be a $\dagger$-qSFA on an object $\SpaceH$ of a dagger compact category. Then $(\!\hbox{\input{symbols/ZbwmultSym.tex}}\!\!,\!\hbox{\input{symbols/ZbwunitSym.tex}}\!\!)$ endows the set of $\hbox{\input{symbols/ZbwdotSym.tex}}\!\!$-unbiased states with the structure of $\phaseGroup{\hbox{\input{symbols/ZbwdotSym.tex}}\!\!}$.
\end{lem}
\proof The $\hbox{\input{symbols/ZbwdotSym.tex}}\!\!$-phase gate corresponding to the $\hbox{\input{symbols/ZbwdotSym.tex}}\!\!$-unbiased state $\!\hbox{\input{symbols/ZbwunitSym.tex}}\!\!$ is the identity, the unit of $\phaseGroup{\hbox{\input{symbols/ZbwdotSym.tex}}\!\!}$, so all we need to show is that composition of phase gates is the same as multiplication under $\!\hbox{\input{symbols/ZbwmultSym.tex}}\!\!$ of the corresponding $\hbox{\input{symbols/ZbwdotSym.tex}}\!\!$-unbiased states:
\begin{equation}
	\input{pictures/ZphaseGroupProof.tikz}
\end{equation}
\qed

As a bonus, the correspondence between the $\hbox{\input{symbols/ZbwdotSym.tex}}\!\!$-phase group and the group structure on $\hbox{\input{symbols/ZbwdotSym.tex}}\!\!$-unbiased states can be used to prove a converse to Lemma \ref{lem_ZphaseGroupAbelianProof}.
\begin{cor}\label{cor_ZphaseGroupAbelianIff}
Let $\hbox{\input{symbols/ZbwdotSym.tex}}\!\!$ be a $\dagger$-qSFA on an object $\SpaceH$ of a dagger compact category, and assume that $\hbox{\input{symbols/ZbwdotSym.tex}}\!\!$ has \textbf{enough unbiased states}\footnote{I.e. that two morphisms $F,G: \SpaceH \rightarrow \SpaceK$ are equal whenever $F\circ {u} = G \circ {u}$ for all $\hbox{\input{symbols/ZbwdotSym.tex}}\!\!$-unbiased states $u$.}. Then $\hbox{\input{symbols/ZbwdotSym.tex}}\!\!$ is commutative iff $\phaseGroup{\hbox{\input{symbols/ZbwdotSym.tex}}\!\!}$ is abelian.
\end{cor}
\proof We already know from Lemma \ref{lem_ZphaseGroupAbelianProof} that if $\hbox{\input{symbols/ZbwdotSym.tex}}\!\!$ is commutative then the $\hbox{\input{symbols/ZbwdotSym.tex}}\!\!$-phase group $\phaseGroup{\hbox{\input{symbols/ZbwdotSym.tex}}\!\!}$ must be abelian. Conversely, if $\phaseGroup{\hbox{\input{symbols/ZbwdotSym.tex}}\!\!}$ is abelian then so is the group structure induced by $(\!\hbox{\input{symbols/ZbwmultSym.tex}}\!\!,\!\hbox{\input{symbols/ZbwunitSym.tex}}\!\!)$ on the $\hbox{\input{symbols/ZbwdotSym.tex}}\!\!$-unbiased states: in particular, this means that $\!\hbox{\input{symbols/ZbwmultSym.tex}}\!\!$ is commutative whenever it is applied to $\hbox{\input{symbols/ZbwdotSym.tex}}\!\!$-unbiased states, and the existence of enough unbiased states allows us to conclude that $\hbox{\input{symbols/ZbwdotSym.tex}}\!\!$ is always commutative.
\qed

With Theorem \ref{thm_PhaseGatesUnbiasedStates} we have proven a general correspondence between phase gates and unbiased states, while with Lemma \ref{lem_GHZphaseGates} and Corollary \ref{cor_GHZphaseStates} we have characterised the states that can be obtained by applying phase gates to GHZ states. Phase gates and the GHZ state for the Pauli $Z$ observable are the key operational ingredients for Mermin's original argument. However, just as important is the special algebraic standing of those phase gates derived from the eigenstates of the Pauli $X$ observable (an observable strongly complementary to Pauli $Z$), as opposed to the phase gates derived from other equatorial states (the eigenstates of observables complementary to Pauli $Z$). 

The last result of this section, Theorem \ref{thm_characterisingCandSC}, provides a general characterisation of complementarity and strong complementarity in terms of the relation between classical states of one observable and unbiased states of the other. Together with Theorem \ref{thm_PhaseGatesUnbiasedStates} and \ref{cor_GHZphaseStates}, it will form the basis for the formulation of our generalised Mermin-type arguments in the next section.

\begin{thm}\label{thm_characterisingCandSC}\hfill\\
Let $\hbox{\input{symbols/ZbwdotSym.tex}}\!\!$ and $\hbox{\input{symbols/DdotSym.tex}}\!\!$ be $\dagger$-qSFAs on an object $\SpaceH$ of a $\dagger$-SMC. The following implications always hold:
\begin{enumerate}[(i)]
\item if $\hbox{\input{symbols/ZbwdotSym.tex}}\!\!$ and $\hbox{\input{symbols/DdotSym.tex}}\!\!$ are complementary, then the $\hbox{\input{symbols/DdotSym.tex}}\!\!$-classical states form a subset of the $\hbox{\input{symbols/ZbwdotSym.tex}}\!\!$-unbiased states; 
\item if $\hbox{\input{symbols/ZbwdotSym.tex}}\!\!$ and $\hbox{\input{symbols/DdotSym.tex}}\!\!$ are strongly complementary, then the $\hbox{\input{symbols/DdotSym.tex}}\!\!$-classical states form a subgroup of the $\hbox{\input{symbols/ZbwdotSym.tex}}\!\!$-unbiased states. 
\end{enumerate}
The converse implications hold if $\hbox{\input{symbols/DdotSym.tex}}\!\!$ has enough classical states: 
\begin{enumerate}[(i)]
\setcounter{enumii}{2}
\item if the $\hbox{\input{symbols/DdotSym.tex}}\!\!$-classical states form a subset of the $\hbox{\input{symbols/ZbwdotSym.tex}}\!\!$-unbiased states, then $\hbox{\input{symbols/ZbwdotSym.tex}}\!\!$ and $\hbox{\input{symbols/DdotSym.tex}}\!\!$ are complementary; 
\item if the $\hbox{\input{symbols/DdotSym.tex}}\!\!$-classical states form a subgroup of the $\hbox{\input{symbols/ZbwdotSym.tex}}\!\!$-unbiased states, then $\hbox{\input{symbols/ZbwdotSym.tex}}\!\!$ and $\hbox{\input{symbols/DdotSym.tex}}\!\!$ are strongly complementary.
\end{enumerate}
Note that the existence of enough $\hbox{\input{symbols/DdotSym.tex}}\!\!$-classical states implies the existence of enough $\hbox{\input{symbols/ZbwdotSym.tex}}\!\!$-unbiased states when the former are a subset/subgroup of the latter.
\end{thm}
\proof Implication (i) is the statement of Lemma \ref{lem_complementarityUnbiased}: in its proof, Equation \ref{eqn_HopfLawUnbiasedStatesProof} is shown that Hopf's law is equivalent to the defining equation of a $\hbox{\input{symbols/ZbwdotSym.tex}}\!\!$-unbiased state when it applied to $\hbox{\input{symbols/DdotSym.tex}}\!\!$-classical state. 
\begin{equation}\label{eqn_HopfLawUnbiasedStatesProof2}
	\resizebox{\textwidth}{!}{\input{pictures/HopfLawUnbiasedStatesProof.tikz}}
\end{equation}
Implication (ii) follows by applying the four defining equations of strong complementarity, together with Hopf's law, to $\hbox{\input{symbols/DdotSym.tex}}\!\!$-classical states. The top row in \ref{eqns_SC} holds if and only if the unit $\!\hbox{\input{symbols/ZbwunitSym.tex}}\!\!$ is a $\hbox{\input{symbols/DdotSym.tex}}\!\!$-classical state: it is coherently copied, transposed and deleted by $\hbox{\input{symbols/DdotSym.tex}}\!\!$.
\begin{equation}\label{eqn_strongComplementarityAltTopRow}
	\input{pictures/strongComplementarityAltTopRow.tikz}
\end{equation}
The bottom row in \ref{eqns_SC} holds applied to two $\hbox{\input{symbols/DdotSym.tex}}\!\!$-classical states if and only if the multiplication under $\!\hbox{\input{symbols/ZbwmultSym.tex}}\!\!$ of two $\hbox{\input{symbols/DdotSym.tex}}\!\!$-classical states is a $\hbox{\input{symbols/DdotSym.tex}}\!\!$-classical state.
\begin{equation}\label{eqn_strongComplementarityAltBottomRowApplied}
	\resizebox{\textwidth}{!}{\input{pictures/strongComplementarityAltBottomRowApplied.tikz}}
\end{equation}
Conditional to $(\hbox{\input{symbols/ZbwdotSym.tex}}\!\!,\!\hbox{\input{symbols/ZbwunitSym.tex}}\!\!)$ endowing the $\hbox{\input{symbols/DdotSym.tex}}\!\!$-classical states with the structure of a monoid, Hopf's law applied to a $\hbox{\input{symbols/DdotSym.tex}}\!\!$-classical state is equivalent to the antipode acting as group inverse on $\hbox{\input{symbols/DdotSym.tex}}\!\!$-classical states. 
\begin{equation}\label{eqn_HopfLawApplied}
	\input{pictures/HopfLawApplied.tikz}
\end{equation}
Implications (iii) and (iv) follow the same lines as implications (i) and (ii). Under the assumptions of (iii) we can conclude that Equation \ref{eqn_HopfLawUnbiasedStatesProof2} holds, and under the assumption of (iv) we can conclude that Equations \ref{eqn_strongComplementarityAltTopRow}, \ref{eqn_strongComplementarityAltBottomRowApplied} and \ref{eqn_HopfLawApplied} hold: from the existence of enough $\hbox{\input{symbols/DdotSym.tex}}\!\!$-classical states, we can conclude that the laws of complementarity and strong complementarity hold as desired.
\qed

\section{Generalised Mermin-type Arguments}
\label{section_generalisedMerminArg}

Armed with the necessary results relating the classical and unbiased states of strongly complementary observables, we are now in a position to formulate our generalised Mermin-type arguments. To do so, we first review the ingredients of Mermin's original parity argument for qubit GHZ states:
\begin{enumerate}[(a)]
\item a 3-partite qubit GHZ state for the Pauli $Z$ observable;
\item the abelian group $\phaseGroup{Z} \isom \reals/(2\pi\integers)$ of phase states for the Pauli $Z$ observable;
\item the finite subgroup $\{0,\pi\} \isom \integersMod{2}$ given by the eigenstates of the Pauli $X$ observable;
\item an equation $2 x = 1$ with no solution in the subgroup $\{0,\pi\}$ given by the Pauli $X$ eigenstates, but with a solution $\pi/2$ in the group $\reals/(2\pi\integers)$ of Pauli $Z$ phase states;
\item measurements in the Pauli $X$ observable.
\end{enumerate}
Similarly, our generalised Mermin-type arguments will involve the following ingredients:
\begin{enumerate}[(a)]
\item an $N$-partite GHZ state for a $\dagger$-qSCFA $\hbox{\input{symbols/ZbwdotSym.tex}}\!\!$;
\item the abelian group $(\phaseGroup{\hbox{\input{symbols/ZbwdotSym.tex}}\!\!}, \oplus, 0)$ of $\hbox{\input{symbols/ZbwdotSym.tex}}\!\!$-phase states\footnote{Isomorphic, by Theorem \ref{thm_PhaseGatesUnbiasedStates}, to the $\hbox{\input{symbols/ZbwdotSym.tex}}\!\!$-phase group, which we will denote by $(\phaseGroup{\hbox{\input{symbols/ZbwdotSym.tex}}\!\!}, \cdot, \id{})$.}; 
\item the subgroup $(\classicalStates{\hbox{\input{symbols/DdotSym.tex}}\!\!},\oplus,0)$, assumed to be finite, of $\hbox{\input{symbols/DdotSym.tex}}\!\!$-classical states for a $\dagger$-qSFA $\hbox{\input{symbols/DdotSym.tex}}\!\!$ which is strongly complementary to $\hbox{\input{symbols/ZbwdotSym.tex}}\!\!$;
\item a finite system of $\integers$-module equations, together with a solution in the group $\phaseGroup{\hbox{\input{symbols/ZbwdotSym.tex}}\!\!}$;
\item measurements in the $\hbox{\input{symbols/DdotSym.tex}}\!\!$ observable.
\end{enumerate}
The non-existence of a solution in the subgroup $\classicalStates{\hbox{\input{symbols/DdotSym.tex}}\!\!}$ of $\hbox{\input{symbols/DdotSym.tex}}\!\!$-classical states is not part of our generalised setup: it will be explicitly characterised as the necessary and sufficient condition for contextuality. Also, $N$ will not be a free parameter, being instead determined by the exponent of the finite abelian group $\classicalStates{\hbox{\input{symbols/DdotSym.tex}}\!\!}$. 

\begin{defi}
Consider an $R$-probabilistic CP* Category $\CPStarCategory{\CategoryC}$. A \textbf{generalised Mermin-type argument}~in~$\CPStarCategory{\CategoryC}$ is specified by the following data:
\begin{enumerate}[(i)]
	\item a strongly complementary pair $(\hbox{\input{symbols/ZbwdotSym.tex}}\!\!,\hbox{\input{symbols/DdotSym.tex}}\!\!)$ of a canonical $\dagger$-qSCFA $\hbox{\input{symbols/ZbwdotSym.tex}}\!\!$ and a canonical $\dagger$-SCFA~$\hbox{\input{symbols/DdotSym.tex}}\!\!$ on some object $\SpaceH$ of $\CategoryC$, such that $\hbox{\input{symbols/DdotSym.tex}}\!\!$ has enough classical states; we furthermore assume that the set $\classicalStates{\hbox{\input{symbols/DdotSym.tex}}\!\!}$ of $\hbox{\input{symbols/DdotSym.tex}}\!\!$-classical states is finite\footnote{This, together with commutativity of $\hbox{\input{symbols/ZbwdotSym.tex}}\!\!$, means that $(\classicalStates{\hbox{\input{symbols/DdotSym.tex}}\!\!},\oplus,0)$ is a finite abelian group.}, and that $|\classicalStates{\hbox{\input{symbols/DdotSym.tex}}\!\!}|$ is invertible as an element of the semiring $R$ of scalars of $\CategoryC$;
	\item a finite system of $\integers$-module equations\footnote{I.e. equations with integer coefficients  $n_r^s \in \integers$ and valued in abelian groups (aka $\integers$-modules).} in the following form, with $a^1,...,a^S \in \classicalStates{\hbox{\input{symbols/DdotSym.tex}}\!\!}$:
	\begin{equation}\label{eqn_system}
	\mathcal{S} = \begin{cases}
		\bigoplus_{r=1}^{M} n^1_r \, y_r = a^1 \\
		\hspace{5mm}\vdots\\
		\bigoplus_{r=1}^{M} n^S_r \, y_r = a^S 
	\end{cases}
	\end{equation}  
	\item a given solution $(y_r := \beta_r)_{r=1}^M$ in the abelian group $\phaseGroup{\hbox{\input{symbols/ZbwdotSym.tex}}\!\!}$ of $\hbox{\input{symbols/ZbwdotSym.tex}}\!\!$-phase states;
	\item a positive integer $N$ such that $N \geq \sum_{r=1}^M n_r^s$ for all $s=1,...,S$, and satisfying $\gcd(N,\exp[\classicalStates{\hbox{\input{symbols/DdotSym.tex}}\!\!}]) = 1$, where $\exp[\classicalStates{\hbox{\input{symbols/DdotSym.tex}}\!\!}]$ is the exponent\footnote{The smallest positive integer $e$ such that $e \cdot g = 0$ for all $g \in \classicalStates{\hbox{\input{symbols/DdotSym.tex}}\!\!}$.} of $\classicalStates{\hbox{\input{symbols/DdotSym.tex}}\!\!}$.
\end{enumerate}
Therefore a generalised Mermin-type argument is specified by a quintuple $(\hbox{\input{symbols/ZbwdotSym.tex}}\!\!,\hbox{\input{symbols/DdotSym.tex}}\!\!, \mathcal{S}, \beta, N)$. 
\end{defi} 

The quintuple $(\hbox{\input{symbols/ZbwdotSym.tex}}\!\!,\hbox{\input{symbols/DdotSym.tex}}\!\!, \mathcal{S}, \beta, N)$ contains all the algebraic and operational ingredients we need to formulate a measurement scenario, which sees $N$ no-signalling parties sharing an $N$-partied $\hbox{\input{symbols/ZbwdotSym.tex}}\!\!$-GHZ state. Each party makes a measurement choice $m_j \in \{0,1,...,M\}$, applies the phase gate $\phasegate{\beta_{m_j}}$ to her system, and then measures it in the $\hbox{\input{symbols/DdotSym.tex}}\!\!$ observable (i.e. measurement outcomes are valued in the set $\classicalStates{\hbox{\input{symbols/DdotSym.tex}}\!\!}$ of $\hbox{\input{symbols/DdotSym.tex}}\!\!$-classical states). 

Not all combinations of measurement choices are needed for the argument, and the measurement contexts will be determined by System \ref{eqn_system}. We begin by zero-padding the system as follows, so that exactly $N$ phase states are involved in each equation:
\begin{equation}\label{eqn_systemAlgExt}
\begin{cases}
	n^0_0 \,y_0 \oplus \;\;0 \,y_1 ... \oplus \;\;\;\,0 \,y_M = 0 \\
	n^1_0 \,y_0 \oplus n^1_1 \,y_1 ... \oplus n^1_M \,y_M = a^1 \\
	\hspace{10mm}\vdots\\
	n^S_0 \,y_0 \hspace{-0.5mm} \oplus n^S_1 \,y_1 ...\hspace{-0.5mm} \oplus n^S_M \,y_M = a^S 
\end{cases}
\end{equation}  
where we have defined $a^0 := 0$, $n_0^s := N - \sum_{r=1}^M n_r^s$ for all $s=1,...,S$, $n_0^0 := N$ and $n_r^0 := 0$ for all $r=1,...,M$; we will also extend the given solution by setting $\beta_0 := 0$. The first equation in System \ref{eqn_systemAlgExt} (which we will refer to by the special value $s=0$ of the parameter $s$) will contribute to a single measurement context, the \textbf{control}; each further equation (i.e. for each value $s=1,....,S$ of the parameter $s$) will give rise to $N$ measurement contexts, the \textbf{variations}, for a total of $1+S \cdot N$ measurement contexts involved in the scenario.

In the control, all parties choose $m_j^0 = 0$, i.e. perform no phase gate before measuring. They obtain the following global state (where $1 / |\classicalStates{\hbox{\input{symbols/DdotSym.tex}}\!\!}|^{N-1}$ is the normalisation factor required to obtain a $R$-distribution):
\begin{equation}\label{eqn_MerminControl}
	\input{pictures/MerminControl.tikz}
\end{equation}
The first variation for each value $s=1,...,S$ is specified by the corresponding equation in System \ref{eqn_systemAlgExt}: the first $n^s_0$ parties choose $m_j^s=0$, the next $n^s_1$ parties choose $m_j^s = 1$, the next $n^s_2$ parties choose $m_j^s = 2$ and so on, until the last $n^s_M$ parties choose $m_j^s = M$:
\begin{equation}\label{eqn_measurementChoices}
	m_j^s := \text{the largest $m \in \{0,...,M\}$ such that }  j \geq \sum_{r=0}^{m-1} n^s_r
\end{equation}
They obtain the following global state, where the equality results from an application of Corollary \ref{cor_GHZphaseStates}, using the relevant equation from System \ref{eqn_systemAlgExt}:
\begin{equation}\label{eqn_MerminVariation1}
	\input{pictures/MerminVariation1.tikz}
\end{equation}
For each fixed value of $s$, the next $N-1$ variations are cyclic permutations of the first: the measurement choice for the $j^{th}$ party at the $k^{th}$ variation of a given $s$ is $m_{j+(k-1)}^s$, where the sum $j+(k-1)$ is taken modulo $N$: 
\begin{equation}\label{system_variations}
\begin{array}{c|ccccc}
	\text{Parties:} & 1 & 2 & ... & N-1 & N \\
	\hline
	\text{ $1^{st}$ variation for } s & m_1^s & m_2^s & ... & m_{N-1}^s & m_N^s \\
	\text{ $2^{nd}$ variation for } s & m_2^s & m_3^s & ... & m_N^s & m_1^s \\
	\text{ $3^{rd}$ variation for } s & m_3^s & m_4^s & ... & m_1^s & m_2^s \\
	\vdots & \vdots & \vdots &  & \vdots & \vdots \\
	\text{ $N^{th}$ variation for } s  & m_N^s & m_1^s & ... & m_{N-2}^s & m_{N-1}^s
\end{array}
\end{equation} 
Because $\hbox{\input{symbols/ZbwdotSym.tex}}\!\!$ is commutative, the global state obtained is the same as that for the first variation for that value of $s$ (shown on the RHS of Equation \ref{eqn_MerminVariation1}).

By using strong complementarity and Theorem \ref{thm_characterisingCandSC}, we rewrite the global state obtained by the $N$ parties in the control and variations, obtaining an explicit $R$-distribution over the set $\classicalStates{\hbox{\input{symbols/DdotSym.tex}}\!\!}^N$ of joint measurement outcomes (from now on, the parameter $s$ can take any value in $\{0,1,...,S\}$, unless otherwise specified). 
\begin{lem}
\begin{equation}\label{eqn_MerminVariationDistrib}
	\input{pictures/MerminVariationDistrib.tikz}
\end{equation}
\end{lem}
\proof 
Strong complementarity can be used to swap $\hbox{\input{symbols/ZbwdotSym.tex}}\!\!$ and $\hbox{\input{symbols/DdotSym.tex}}\!\!$, as shown in Corollary 4.1 of \cite{Coecke2012c}, and then $a^s$ can be pushed through because it is a $\hbox{\input{symbols/DdotSym.tex}}\!\!$-classical state (we have left normalisation aside, and we use $a^0:= 0$ to treat control and variations uniformly): 
\begin{equation}\label{eqn_MerminVariationSCDistrib1}
	\input{pictures/MerminVariationSCDistrib1.tikz}
\end{equation}
Using fact that $\hbox{\input{symbols/DdotSym.tex}}\!\!$ has enough classical states, and recalling from Theorem \ref{thm_characterisingCandSC} that $(\!\hbox{\input{symbols/ZbwmultSym.tex}}\!\!,\!\hbox{\input{symbols/ZbwunitSym.tex}}\!\!)$ acts as the group multiplication of $\classicalStates{\hbox{\input{symbols/DdotSym.tex}}\!\!}$ when restricted to the $\hbox{\input{symbols/DdotSym.tex}}\!\!$-classical states, we can further decompose the state on the RHS of Equation \ref{eqn_MerminVariationSCDistrib1} into an $R$-distribution over the set $\classicalStates{\hbox{\input{symbols/DdotSym.tex}}\!\!}^N$:
\begin{equation}\label{eqn_MerminVariationSCDistrib2}
	\input{pictures/MerminVariationSCDistrib2.tikz}
\end{equation}
\qed

The joint outcome of measurements for the control is uniformly distributed over the subgroup $H_0 \normalSubgroup \classicalStates{\hbox{\input{symbols/DdotSym.tex}}\!\!}^N$ specified by $H_0 := \suchthat{(g_1,...,g_N)}{g_1 \oplus ... \oplus g_N = 0}$, while the joint outcome of any of the $N$ variations for each specific value of $s$ is uniformly distributed over the coset $H_{a_s} := (a^s,0,...,0) \oplus H_0$. For each $s,s' \in \{0,1,...,S\}$, the cosets $H_{a_s}$ and $H_{a_{s'}}$ are disjoint whenever $a^s \neq a^{s'}$. All in all, we get the following empirical model for the generalised Mermin-type argument:
\begin{align}
\mathbb{P}[(g_1,...,g_N) | \text{control}] &= 
	\begin{cases}
		\dfrac{1}{|\classicalStates{\hbox{\input{symbols/DdotSym.tex}}\!\!}|^{N-1}} & \text{ if } g_1 \oplus ... \oplus g_N = 0 \\
		\hfill 0 \hfill & \text{ otherwise }
	\end{cases}\label{eqn_empiricalModelControl}\\
\mathbb{P}[(g_1,...,g_N) | \text{$k^{th}$ variation for $s$}] &= 
	\begin{cases}
		\dfrac{1}{|\classicalStates{\hbox{\input{symbols/DdotSym.tex}}\!\!}|^{N-1}} & \text{ if } g_1 \oplus ... \oplus g_N = a^s \\
		\hfill 0 \hfill & \text{ otherwise }
	\end{cases}\label{eqn_empiricalModelVariations}
\end{align} 

One of the catchy features of Mermin's original argument is that it is entirely deterministic: instead of relying on the violation of some probabilistic inequality, the proof of contextuality shows that the existence of a local hidden variable (LHV) model leads would lead to existence of solutions to an unsatisfiable parity equation (i.e. one which doesn't admit solutions in the finite abelian group $\integersMod{2}$). The proof of contextuality for our generalised Mermin-type arguments goes by similar lines, showing that the existence of a LHV model is equivalent to System \ref{eqn_system} admitting solutions in the finite abelian group $\classicalStates{\hbox{\input{symbols/DdotSym.tex}}\!\!}$. 

\begin{thm}\label{thm_contextuality}
Consider an $R$-probabilistic CP* category $\CPStarCategory{\CategoryC}$, and let $(\hbox{\input{symbols/ZbwdotSym.tex}}\!\!,\hbox{\input{symbols/DdotSym.tex}}\!\!, \mathcal{S}, \beta, N)$ be a generalised Mermin-type argument in it. If the associated empirical model is contextual, then the system $\mathcal{S}$ admits no solution in the finite abelian group $\classicalStates{\hbox{\input{symbols/DdotSym.tex}}\!\!}$. Conversely, if the system $\mathcal{S}$ admits no solution in $\classicalStates{\hbox{\input{symbols/DdotSym.tex}}\!\!}$ and $R$ is a positive semiring, then the empirical model is contextual.
\end{thm}
\proof 

The proof comes in two parts: ($\Rightarrow$) we show that any solution in $\classicalStates{\hbox{\input{symbols/DdotSym.tex}}\!\!}$ can be turned into a LHV model; ($\Leftarrow$) we show that, as long as $R$ is a positive semiring, any LHV model can be turned into a solution in $\classicalStates{\hbox{\input{symbols/DdotSym.tex}}\!\!}$. 

\noindent \textbf{Proof of ($\Rightarrow$).} Assume that the system $\mathcal{S}$ (in the form of System \ref{eqn_system}) admits a solution $(y_r := b_r)_{r=1}^{M}$, and define $b_0 := 0$. A LHV model can be obtained as follows:
\begin{enumerate}[(i)]
	\item the uniform $R$-distribution on $H_0 \normalSubgroup \classicalStates{\hbox{\input{symbols/DdotSym.tex}}\!\!}^N$ is taken as a shared classical state amongst the $N$ parties:
		\begin{equation}\label{eqn_MerminLHVDistrib1}
			\input{pictures/MerminLHVDistrib1.tikz}
		\end{equation}
	\item upon measurement choice $m_j \in \{0,1,...,M\}$ for the $j^{th}$ party, a translation by $b_{m_j}$ in the group $\classicalStates{\hbox{\input{symbols/DdotSym.tex}}\!\!}$ is applied to the respective classical subsystem, independently of the measurement choices of the other parties:
		\begin{equation}\label{eqn_MerminLHVDistrib2}
			\input{pictures/MerminLHVDistrib2.tikz}
		\end{equation}
\end{enumerate} 
All we need to show is that the procedure above produces the same $R$-distributions on $\classicalStates{\hbox{\input{symbols/DdotSym.tex}}\!\!}^N$ as those given by the empirical model of Equations \ref{eqn_empiricalModelControl} and \ref{eqn_empiricalModelVariations}. To do so, we simply observe that the global state obtained with the procedure above is the same as the global states obtained in the control \ref{eqn_MerminControl} and in the variations \ref{eqn_MerminVariation1} (which we treat uniformly by considering $s=0,1,...,S$), because $b_0,b_1,...,b_N$ satisfy the same equations satisfied by the phases $\beta_0,\beta_1,...,\beta_N$:
\begin{equation}\label{eqn_MerminLHVDistrib}
	\input{pictures/MerminLHVDistrib.tikz}
\end{equation}

\noindent \textbf{Proof of ($\Leftarrow$).} Now assume that $R$ is a positive semiring, and that the scenario admits a LHV model:
\begin{enumerate}[(i)]
	\item there is a some finite set $\Lambda$, the set of values for the hidden variable, coming with an $R$-distribution $p : \Lambda \rightarrow R$;
	\item for each possible measurement choice $r=0,1,...,M$ that each party $i=1,...,N$ can make, there is a family $(c_{r}^{i,\lambda})_{\lambda \in \Lambda}$ of $\hbox{\input{symbols/DdotSym.tex}}\!\!$-classical states, the deterministic local outcomes for each value of the hidden variable;
	\item for each measurement context (either $s=0$, $k=1$ for the control, or $(s,k)\in \{1,...,S\}\times\{1,...,N\}$ for the $N \cdot S$ variations), a definite $\hbox{\input{symbols/DdotSym.tex}}\!\!$-classical outcome $d_{s,k}^{i,\lambda}$ is obtained by each party $i=1,...,N$ at each definite value $\lambda \in \Lambda$ of the hidden variable:
		\begin{equation}
			d_{s,k}^{i,\lambda} := c_{m_{i+(k-1)}^s}^{i,\lambda}
		\end{equation}
	\item if these definite $\hbox{\input{symbols/DdotSym.tex}}\!\!$-classical global states are weighted based on the $R$-distribution $p$ on $\Lambda$, one obtains the same $R$-distribution on joint measurement outcomes that would be expected from the measurement context:
		\begin{equation}\label{eqn_MerminLHV}
			\input{pictures/MerminLHV.tikz}
		\end{equation} 
\end{enumerate}
Given a LHV model, we can sum up all $N$ outcomes of each side of Equation \ref{eqn_MerminLHV} in $(\classicalStates{\hbox{\input{symbols/DdotSym.tex}}\!\!},\oplus,0)$ to obtain an equation between $R$-distribution over $\classicalStates{\hbox{\input{symbols/DdotSym.tex}}\!\!}$:
\begin{equation}\label{eqn_MerminLHVsummed}
	\input{pictures/MerminLHVsummed.tikz}
\end{equation}
The last equation used the fact that $\hbox{\input{symbols/DdotSym.tex}}\!\!$ was chosen to be special\footnote{The special $\hbox{\input{symbols/DdotSym.tex}}\!\!$ could have been replaced by a more general $\dagger$-qSCFA, but at the price of an additional normalisation factor in all global states.}, and hence the normalisation factor for the $\dagger$-qSCFA $\hbox{\input{symbols/ZbwdotSym.tex}}\!\!$ is $|\classicalStates{\hbox{\input{symbols/DdotSym.tex}}\!\!}|$ (because $\hbox{\input{symbols/DdotSym.tex}}\!\!$ has enough classical states)\footnote{The normalisation factor $|\classicalStates{\hbox{\input{symbols/DdotSym.tex}}\!\!}|$ refers to two wires: each additional wire is an additional copy of $|\classicalStates{\hbox{\input{symbols/DdotSym.tex}}\!\!}|$, for a total of $|\classicalStates{\hbox{\input{symbols/DdotSym.tex}}\!\!}|^{N-1}$ in the $N$-wire case here.}. Equation \ref{eqn_MerminLHVsummed} can be turned into the following conditions on the LHV: 
\begin{equation}
	\sum_{\lambda \text{ s.t. }\bigoplus_{i=1}^N d_{s,k}^{i,\lambda} = a^s} \hspace{-0.75cm} p(\lambda) = 1 \hspace{2cm}
	\sum_{\lambda \text{ s.t. }\bigoplus_{i=1}^N d_{s,k}^{i,\lambda} \neq a^s} \hspace{-0.75cm} p(\lambda) = 0
\end{equation}
Because $R$ is a positive semiring, $p(\lambda) = 0$ for any $\lambda$ such that $\oplus_{i=1}^N d_{s,k}^{i,\lambda} \neq a^s$ for some $s$. Conversely, picking any $\lambda_+$ such that $p(\lambda_+) > 0$ (and at least one such $\lambda_+$ exists, because $p$ is an $R$-distribution) yields a family $(d_{s,k}^{i,\lambda_+})_{s,k,i}$ such that $\oplus_{i=1}^N d_{s,k}^{i,\lambda_+} = a^s$ for all $s$ and $k$. For the control ($s=0$ and $k=1$), we obtain the following equation: 
\begin{equation}\label{eqn_LHVstatesSummedUpControl}
	\oplus_{i=1}^N c_{0}^{i,\lambda_+} = 0
\end{equation}
For each variation $(s,k) \in \{1,...,S\}\times\{1,...,N\}$, we obtain the following equation:
\begin{equation}
	\oplus_{i=1}^N c_{m_{i+(k-1)}^s}^{i,\lambda_+} = a^s
\end{equation}
If $c_r^{i,\lambda_+}$ was independent of the party $i$ for all $r=1,...,M$, this equation would yield a solution to system $\mathcal{S}$ in the form of $b_r := c_r^{i,\lambda_+}$ for any $i$; unfortunately, this need not be the case. This is where our cyclic definition of the $N$ variations for each value of $s$ comes into play. For each fixed value of $s$, we add up the $N$ equations for $k=1,...,N$:
\begin{equation}\label{eqn_LHVstatesSummedUp}
	\oplus_{k=1}^{N}\oplus_{i=1}^N c_{m_{i+(k-1)}^s}^{i,\lambda_+} = N a^s
\end{equation}
Because $\gcd(N,\exp[\classicalStates{\hbox{\input{symbols/DdotSym.tex}}\!\!}]) = 1$, we can take the inverse of $N$ modulo $\exp[\classicalStates{\hbox{\input{symbols/DdotSym.tex}}\!\!}]$, and the equation above has solutions if and only if the equation below does:
\begin{equation}\label{eqn_LHVstatesSummedUpEquiv}
	\oplus_{k=1}^{N}\oplus_{i=1}^N N^{-1} c_{m_{i+(k-1)}^s}^{i,\lambda_+} = a^s
\end{equation}
Now refer to the Table \ref{system_variations} defining the $N$ variations for $s$, as well as to Equation \ref{eqn_measurementChoices} which defines the measurement choices,. The LHS of Equation \ref{eqn_LHVstatesSummedUp} is a sum by rows of the $N^2$ measurement choices in Table \ref{system_variations}: each $r=0,1,...,M$ appears $n_r^s$ times in each row, but the changing value of $i$ along each row stops us from turning it into a solution to system $\mathcal{S}$. However, we can switch the summations in Equation \ref{eqn_LHVstatesSummedUp} to obtain a sum by columns of the table, where each $r=0,1,...,M$ still appears $n_r^s$ times in each column (by the cyclic definition), but now $i$ is constant along each column:
\begin{equation}\label{eqn_LHVstatesSummedUp2}
	\oplus_{i=1}^N \oplus_{k=1}^{N} c_{m_{i+(k-1)}^s}^{i,\lambda_+} = \oplus_{i=1}^N \oplus_{r=0}^M n_r^s c_r^{i,\lambda_+}
\end{equation}
We can then sum up all $(c_r^{i,\lambda_+})_{i=1}^N$ for each $r=0,1,...,M$, and use Equation \ref{eqn_LHVstatesSummedUp} (together with Equation \ref{eqn_LHVstatesSummedUpControl} to cancel out the contribution from $r=0$) to finally obtain the desired solution $(b_r)_{r=1}^M$ to system $\mathcal{S}$:
\begin{equation}
	\input{pictures/MerminLHVsolution.tikz}
\end{equation}
\qed

\section{Quantum realisability}
\label{section_quantumRealisab}

In quantum theory, i.e. in the probabilistic CP* category $\CPStarCategory{\fdHilbCategory}$, many of the requirements of generalised Mermin-type arguments are automatically satisfied: canonical $\dagger$-SCFAs in $\CPMCategory{\fdHilbCategory}$ (i.e. $\dagger$-SCFAs in $\fdHilbCategory$) always have enough classical states (and finitely many so), the semiring $\reals^+$ of scalars is positive, and any non-zero integer is invertible in it. Hence, only strong complementarity is required in point (i) of the definition of generalised Mermin-type arguments, and Theorem \ref{thm_contextuality} establishes an unconditional equivalence between contextuality of a generalised Mermin-type argument $(\hbox{\input{symbols/ZbwdotSym.tex}}\!\!,\hbox{\input{symbols/DdotSym.tex}}\!\!, \mathcal{S}, \beta, N)$ and the existence of solutions to system $\mathcal{S}$ in the finite abelian group $\classicalStates{\hbox{\input{symbols/DdotSym.tex}}\!\!}$ of $\hbox{\input{symbols/DdotSym.tex}}\!\!$-classical states. 

The remarks above show that the correspondence between systems of equations in finite abelian groups and generalised Mermin-type arguments is particularly tight in the case of quantum theory, but an important question remains unanswered: which systems of $\integers$-module equations lead to arguments which can be realised in quantum theory? As it turns out, all of them (but an obvious caveat applies).

\begin{thm}\label{thm_quantumRealisability}
Let $(K,\oplus,0)$ be a finite abelian group, and $\mathcal{S}$ be a finite system of $\integers$-module equations in the following form, with $a^1,...,a^S \in K$:
\begin{equation}\label{eqn_systemQuantumRealisability}
\mathcal{S} = \begin{cases}
	\bigoplus_{r=1}^{M} n^1_r \, y_r = a^1 \\
	\hspace{5mm}\vdots\\
	\bigoplus_{r=1}^{M} n^S_r \, y_r = a^S 
\end{cases}
\end{equation}  
Assume that the system is \textbf{consistent} in the following sense, where by $\underline{n}^s \in \integers^{M}$ we denoted the row vectors of System \ref{eqn_systemQuantumRealisability}:
\begin{equation}
	\bigoplus_{s=1}^{S} c_s \cdot \underline{n}^{s} =_{\integers^M} \underline{0} \implies \bigoplus_{s=1}^{S} c_s \cdot a^{s} =_{K} 0,
\end{equation}
Then for every $|K|$-dimensional quantum system $\SpaceH$ and every $\dagger$-qSCFA $\hbox{\input{symbols/ZbwdotSym.tex}}\!\!$ on $\SpaceH$ with normalisation factor $|K|$, there exists a generalised Mermin-type argument $(\hbox{\input{symbols/ZbwdotSym.tex}}\!\!,\hbox{\input{symbols/DdotSym.tex}}\!\!, \mathcal{S}, \beta, N)$ corresponding to System \ref{eqn_systemQuantumRealisability}, i.e. we can always find:
\begin{enumerate}[(i)]
	\item a $\dagger$-SCFA $\hbox{\input{symbols/DdotSym.tex}}\!\!$, strongly complementary to $\hbox{\input{symbols/ZbwdotSym.tex}}\!\!$, such that $(\classicalStates{\hbox{\input{symbols/DdotSym.tex}}\!\!},\!\hbox{\input{symbols/ZbwmultSym.tex}}\!\!,\!\hbox{\input{symbols/ZbwunitSym.tex}}\!\!) \isom (K,\oplus,0)$; 
	\item a solution $(y_r := \beta_r)_{r=1}^M$ to $\mathcal{S}$ in $\phaseGroup{\hbox{\input{symbols/ZbwdotSym.tex}}\!\!} \isom T^{|K|-1}$;
	\item a positive integer $N$ (infinitely many, in fact) such that $N \geq \sum_{r=1}^{M} n_r^s$ for all $s=1,...,S$, and such that $\gcd(N,\exp[\classicalStates{\hbox{\input{symbols/DdotSym.tex}}\!\!}]) = 1$. 
\end{enumerate}
\end{thm}
\proof 
Point (iii) is trivial: there are infinitely many $N$ such that $\gcd(N,\exp[\classicalStates{\hbox{\input{symbols/DdotSym.tex}}\!\!}]) = 1$, and hence we can always find one such that $N \geq \sum_{r=1}^{M} n_r^s$ for all $s=1,...,S$. Point (i) is more interesting, and relies on Theorem \ref{thm_SCHilb} and Pontryagin duality for finite abelian groups. Point (ii) is perhaps the most interesting, and relies on the possibility of solving consistent systems of $\integers$-module equations in the torus $T^{|K|-1}$.

\noindent \textbf{Proof of point (i).} Because $\hbox{\input{symbols/ZbwdotSym.tex}}\!\!$ is a $\dagger$-qSCFA with normalisation factor $|K|$ on a $|K|$-dimensional Hilbert space $\SpaceH$, it is associated with a basis of $|K|$ vectors, each having norm $\sqrt{|K|}$. Label the basis vectors by the $|K|$ multiplicative characters $\goodchi \in K^\wedge$ of the finite abelian group $K$, and construct an orthonormal basis by using the multiplicative characters $\tau \in (K^\wedge)^\wedge$ of the finite abelian group $K^\wedge$:
\begin{equation}\label{eqn_doubleDualBasis}
	\ket{\tau} := \frac{1}{|K|} \sum_{\goodchi \in K^\wedge} \tau(\goodchi) \ket{\goodchi}
\end{equation}
By Pontryagin duality, there is a canonical isomorphism $(K^\wedge)^\wedge \isom K$, so that the new orthonormal basis given by Equation \ref{eqn_doubleDualBasis} is canonically labelled by elements of $K$. Consider the $\dagger$-SCFA $\hbox{\input{symbols/DdotSym.tex}}\!\!$ associated to the orthonormal basis thus defined to obtain the desired $(\classicalStates{\hbox{\input{symbols/DdotSym.tex}}\!\!},\!\hbox{\input{symbols/ZbwmultSym.tex}}\!\!,\!\hbox{\input{symbols/ZbwunitSym.tex}}\!\!) \isom (K,\oplus,0)$.

\noindent \textbf{Proof of point (ii).} The phase group $\phaseGroup{\hbox{\input{symbols/ZbwdotSym.tex}}\!\!}$ for a canonical $\dagger$-qSCFA on a $|K|$-dimensional Hilbert space in $\CPMCategory{\fdHilbCategory}$ is isomorphic to the $(|K|-1)$-dimensional torus, an abelian Lie group. To find a solution $(y_r := \beta_r)_{r=1}^M$ to System \ref{eqn_systemQuantumRealisability}, we will show that one can always find solutions to arbitrary consistent systems of $\integers$-module equations in a torus.

While all $K$-valued systems with solutions in some super-group of $K$ must necessarily be consistent, the converse is not true in general: given a super-group $P$ of $K$ there may be consistent systems with no solutions in $P$. Certainly if $P$ is finite then at least one such system exists (because of the finite exponent), and certainly if $P=\rationals^d$ then no such system exists; in fact, every divisible torsion-free abelian group $P$ is canonically a $\rationals$-vector space, and thus every consistent system of $\integers$-modules equations  (and, in fact, of $\rationals$-vector space equations) valued in a divisible torsion-free abelian group $P$ has solutions in $P$ (e.g. by Gaussian elimination over the field $\rationals$). Unfortunately, while tori are divisible, they are not torsion-free, and in particular not $\rationals$-vector spaces: as a consequence, the reasoning above does not apply. 

However, a more general argument can be used to show that any consistent system of equations can be solved in any divisible abelian group, regardless of whether the group is torsion-free or not \cite{fuchs2015abelian} (although uniqueness of solution need not hold for systems with linearly independent row vectors). As tori are divisible abelian groups, all consistent systems of $\integers$-module equations can be solved in them, and in particular we can find our solution $(y_r := \beta_r)_{r=1}^M$ to System \ref{eqn_systemQuantumRealisability}.
\qed

\newcommand{\eqnIndex}[1]{\operatorname{index}(#1)}
\newcommand{\RlinearTheory}[2]{\mathbb{T}_{#1}(#2)}
\newcommand{\AvN}[2]{\operatorname{AvN}_{#1,#2}}
\newcommand{\AvNring}[1]{\operatorname{AvN}_{#1}}

\section{All-vs-Nothing Arguments}
\label{section_AvN}

Strong contextuality can be reformulated directly in terms of the supports of the distributions. The supports of the global sections, i.e. the $d \in \presheafOfDistributions{\mathbb{B}}{\mathcal{X}}$ satisfying Equation \ref{eqn_StrongContextualityCondition}, form a (possibly empty) lattice, and thus a probabilistic empirical model is strongly contextual iff the following set is empty:
\begin{equation}
	\supportSubpresheaf{\mathcal{X}} := \suchthat{s \in \sheafOfEvents{\mathcal{X}}}{ \restrict{s}{C} \in \support{\zeta_C} \text{ for all } C \in \mathcal{M}}
\end{equation}
For a possibilistic (no-signalling) empirical model $(\zeta_C)_{C \in \mathcal{M}}$, we can define \cite{Abramsky2015} a \textbf{support subpresheaf} $\supportSubpresheafSym \subseteq \sheafOfEventsSym$ by setting:
\begin{equation}
	\supportSubpresheaf{U} := \suchthat{s \in \sheafOfEvents{U}}{ \restrict{s}{C \cap U} \in \support{\restrict{\zeta_C}{U \cap C}} \text{ for all } C \in \mathcal{M}}
\end{equation}
Then a possibilistic empirical model is strongly contextual if and only if $\supportSubpresheaf{\mathcal{X}} = \emptyset$.

The fundamental observation behind the \textbf{All-vs-Nothing arguments} of \cite{Abramsky2015} is that contextuality of Mermin's original argument follows from the existence of the system of $\integersMod{2}$ equations which has no global solution (corresponding to $\supportSubpresheaf{\mathcal{X}} = \emptyset$ in the sheaf-theoretic framework for contextuality \cite{Abramsky2011}), but where each equation admits a solution (i.e. we have $\supportSubpresheaf{C} \neq \emptyset$ for the measurement context $C$ associated to each equation). In this section we summarise the basic framework of All-vs-Nothing arguments from \cite{Abramsky2015}, taking the liberty of slightly generalising the definitions therein from rings to modules over rings.

Let $\mathcal{R}$ be a commutative ring with unit: we will denote by $+$ the addition in the ring $\mathcal{R}$, and by $\oplus$ the addition in $\mathcal{R}$-modules. The ring $\mathcal{R}$ should not be confused with the semiring $R$ over which the distributions are taken (i.e. the semiring of scalars of the enriched CPM category the arguments take place in). If $G$ is some $\mathcal{R}$-module, we will define an \textbf{$\mathcal{R}$-linear equation valued in $G$} to be a triple $\phi = (C,n,b)$ where:
\begin{enumerate}[leftmargin=8mm]
	\item[(i)] $C$ is some finite set, and we define $\eqnIndex{\phi} := C$;
	\item[(ii)] $n: C \rightarrow \mathcal{R}$ is any function;
	\item[(iii)] $b \in G$ is a given element of $G$.
\end{enumerate}
If $\phi = (C,n,b)$ is an $\mathcal{R}$-linear equation valued in $G$, we will say that a function $s: C \rightarrow G$ (henceforth an \textbf{assignment}) \textbf{satisfies} $\phi$, written $s \models \phi$, if and only if the following equation holds in $G$:
\begin{equation}
	\bigoplus_{m \in C} n_m s_m = b
\end{equation}
where we denoted $n_m := n(m)$ and $s_m := s(m)$. Any set $W$ of assignments $C \rightarrow G$ can be associated a corresponding set $\RlinearTheory{\mathcal{R}}{W}$ of satisfied equations, which is itself an $\mathcal{R}$-module\footnote{This gives rise to some interesting results on affine closures, see \cite{Abramsky2015}.}:
\begin{equation}
	\label{eqn_RlinearTheory}
	\RlinearTheory{\mathcal{R}}{W} := \suchthat{\phi}{s \models \phi \text{ for all }s \in W}
\end{equation} 

Let $(\zeta_C)_{C \in \mathcal{M}}$ be a possibilistic empirical model for a measurement scenario $(\sheafOfEventsSym,\mathcal{M})$, such that all measurements have the same $\mathcal{R}$-module $G$ as their set of outcomes (for example we had $G = \integersMod{2}$, a $\integers$-module, for Mermin's original argument). Let $\supportSubpresheafSym \subseteq \sheafOfEventsSym$ be the support subpresheaf for the empirical model and define its \textbf{$\textbf{R}$-linear theory} to be:
\begin{equation}
	\RlinearTheory{\mathcal{R}}{\supportSubpresheafSym} := \bigcup_{C \in \mathcal{M}}  \RlinearTheory{\mathcal{R}}{\supportSubpresheaf{C}}
\end{equation} 
We say that a possibilistic empirical model is \textbf{All-vs-Nothing} with respect to ring $\mathcal{R}$ and $\mathcal{R}$-module $G$, written $\AvN{\mathcal{R}}{G}$, iff the $\mathcal{R}$-linear theory admits no solution in $G$, i.e. iff there exists no global assignment $s: \mathcal{X} \rightarrow G$ such that:
\begin{equation}
	\label{eqn_AvNDefinition}
	\restrict{s}{C} \models \phi \text{ for all } C \in \mathcal{M} \text{ and all } \phi \in \RlinearTheory{\mathcal{R}}{\supportSubpresheaf{C}}
\end{equation}
To connect back with the notation in \cite{Abramsky2015}, we will simply write $\AvNring{\mathcal{R}}$ for $\AvN{\mathcal{R}}{\mathcal{R}}$. 

A straightforward generalisation (from rings to modules) of a result by \cite{Abramsky2015} proves that any possibilistic empirical model which is $\AvN{\mathcal{R}}{G}$ for some ring $\mathcal{R}$ and some $\mathcal{R}$-module $G$ is strongly contextual: if the model weren't strongly contextual, then there would be some global section $s \in \supportSubpresheaf{\mathcal{X}}$, and this would imply $\restrict{s}{C} \in \supportSubpresheaf{C}$ for all $C \in \mathcal{M}$, which in turn would prove that global assignment $s$ satisfies Equation \ref{eqn_AvNDefinition} (by appealing to Equation \ref{eqn_RlinearTheory}). 

A result by \cite{Abramsky2011} shows that a probabilistic empirical model is strongly contextual if and only if it is maximally contextual, i.e. if and only if it lies on a face of the no-signalling polytope with no local vertices. As a consequence, showing that our generalised Mermin-type arguments are $\AvN{\mathcal{R}}{G}$ is a particularly neat way of proving that they are maximally contextual, a highly desirable property for the device-independent security of the quantum-classical secret sharing protocol which we will present in the next section.

\begin{thm}\label{thm_AvNMermin}
Consider a $R$-probabilistic CP* category $\CPStarCategory{\CategoryC}$, where $R$ is a positive semiring, and let $(\hbox{\input{symbols/ZbwdotSym.tex}}\!\!,\hbox{\input{symbols/DdotSym.tex}}\!\!, \mathcal{S}, \beta, N)$ be a generalised Mermin-type argument in it. If the associated empirical model is contextual, then it is $\AvN{\integers}{K}$.
\end{thm}
\proof
The associated probabilistic empirical model is given by Equations \ref{eqn_empiricalModelControl} and \ref{eqn_empiricalModelVariations}: the only scalars appearing are $0$ and the invertible $\frac{1}{|\classicalStates{\hbox{\input{symbols/DdotSym.tex}}\!\!}|}$, which are (necessarily) sent to $0$ and $1$ respectively in the passage to the possibilistic empirical model. The possibilistic empirical model is as follows:
\begin{align}
\mathbb{P}[(g_1,...,g_N) | \text{control}] &= 
	\begin{cases}
		1 & \text{ if } g_1 \oplus ... \oplus g_N = 0 \\
		0 & \text{ otherwise }
	\end{cases}\label{eqn_empiricalModelControlBool}\\
\mathbb{P}[(g_1,...,g_N) | \text{$k^{th}$ variation for $s$}] &= 
	\begin{cases}
		1 & \text{ if } g_1 \oplus ... \oplus g_N = a^s \\
		0 & \text{ otherwise }
	\end{cases}\label{eqn_empiricalModelVariationsBool}
\end{align} 
The possibilistic empirical model has the following support subpresheaf $\supportSubpresheafSym \subseteq \sheafOfEventsSym$:
\begin{align}
	\supportSubpresheaf{\text{control}} &= \suchthat{(c^{i}_{m^0_{i}})_{i=1}^N \in K^N}{\oplus_{i=1}^N c^{i}_{m^0_{i}} =_K 0} \\
	\supportSubpresheaf{\text{$k^{th}$ variation for $s$}} &= \suchthat{(c^{i}_{m^s_{i+(k-1)}})_{i=1}^N \in K^N}{\oplus_{i=1}^N c^{i}_{m^s_{i+(k-1)}} =_K a^s} 
\end{align} 
Amongst the (many) equations in $\RlinearTheory{\integers}{\supportSubpresheafSym}$ we can find the following $1+N\cdot S$ equations:
\begin{align}
	\bigoplus_{m} s_m &= 0 \text{, satisfied by all } s \in \supportSubpresheaf{\text{control}}\\
	\bigoplus_{m} s_m &= a^s \text{, satisfied by all } s \in \supportSubpresheaf{\text{$k^{th}$ variation for $s$}} 
\end{align} 
Any global assignment satisfying all equations in $\RlinearTheory{\integers}{\supportSubpresheafSym}$ would in particular satisfy the $1+N\cdot S$ equations above, and hence provide a solution in $K$ to the system $\mathcal{S}$. By Theorem \ref{thm_contextuality}, if the empirical model is contextual then no such solution exists: hence no global assignment satisfying all equations in $\RlinearTheory{\integers}{\supportSubpresheafSym}$ can exist, proving that the model is in particular $\AvN{\integers}{K}$.
\qed

\begin{cor} The generalised Mermin-type arguments provide an infinite family of quantum realisable $\AvN{\integers}{K}$ empirical models, indexed by all finite abelian groups $K$ and all finite consistent systems $\mathbb{S}$ of $\integers$-module equations valued in $K$ which admit no solution in $K$. Furthermore, all $\AvN{\integers}{K}$ arguments for some fixed $K$ are equivalently $\AvN{\integersMod{n}}{K}$ for any positive integer $n$ divisible by the exponent of $K$: as a consequence, there are generalised Mermin-type arguments providing quantum realisable $\AvNring{\integersMod{n}}$ models for positive integers $n \geq 2$.
\end{cor}
\proof 
The first part is a straightforward consequence of Theorems \ref{thm_contextuality}, \ref{thm_quantumRealisability} and \ref{thm_AvNMermin}. The second part is a consequence of the fact that any $\integers$-module equation valued in a finite abelian group $K$ is equivalent to a $\integersMod{\exp[K]}$-module equation (by taking remainders modulo $\exp[K]$ of all coefficients), and to a $\integersMod{n}$-module equation for any $n$ divisible by the exponent $\exp[K]$ (taking reminders modulo $n$ of coefficients). The last part is the special case where we consider the finite abelian group $K=\integersMod{n}$ as a module over the ring $\mathcal{R} = \integersMod{n}$.
\qed

One open question about All-vs-Nothing arguments asks whether all quantum realisable $\AvNring{\integers}$ models are in fact $\AvNring{\integersMod{2}}$. The following result answers the question negatively, showing that the infinite family of $\AvNring{\integers}$ models provided by the previous corollary form a non-collapsing hierarchy of $\AvNring{\integersMod{p}}$ models for all $n \geq 2$.

\begin{thm} 
For each $n \geq 2$, there is a quantum realisable $\AvNring{\integersMod{n}}$ (and hence also $\AvN{\integers}{\integersMod{n}}$) empirical model which is not $\AvN{\integersMod{m}}{K'}$ for any $m \geq 2$ coprime with $n$ and any non-trivial abelian group $K'$ with exponent dividing $m$; in particular, it is not $\AvNring{\integersMod{m}}$.
\end{thm}
\proof
The next Section fully works out the example of $K := \integersMod{n}$ with the system $\mathcal{S}$ consisting of a single $\integers$-module equation $t y = 1$. If we pick a $t \in \{2,...,n-1\}$ which divides $n$, the equation cannot be satisfied for $K = \integersMod{n}$, giving rise to a model which is both $\AvN{\integers}{\integersMod{n}}$ and $\AvNring{\integersMod{n}}$ (because the equation can be replaced by an equivalent $\integersMod{n}$-module equation). 
Now consider some $m$ coprime with $n$, and some abelian group $K'$ with exponent dividing $m$. Then the equation has solutions in $K'$, giving rise to a model which is not $\AvN{\integers}{K'}$ nor $\AvN{\integersMod{m}}{K'}$ (nor $\AvNring{\integersMod{m}}$, in the case $K' := \integersMod{m}$). Indeed, we must have $K' \isom \prod_{l=1}^{L} \integersMod{p_l^{e_l}}$ for some primes $p_l$ not dividing $n$ and some exponents $e_l \geq 1$, and the equation has solutions in $\integersMod{p_l^{e_l}}$ for all $l$ (because $t$ has the same prime factors of $n$, and hence no $p_l$ can divide $t$). 
\qed

\section{A fully worked-out example}
\label{section_example}

In this Section, we fully work out a generalised Mermin-type argument, for the group $K := \integersMod{d}$ and the system $\mathcal{S}$ consisting of a single $\integers$-module equation $t y = 1$ (i.e. we have $S=M=1$); we take $d \geq 2$ and $t \in \{1,...,d-1\}$. This can equivalently be seen as a $\integersMod{d}$-module equation $t y = \modclass{1}{d}$. Our presentation goes through four distinct Subsections, covering all aspects from abstract definition of the argument to concrete realisation in quantum theory. In the first Subsection, we present the measurement scenario and empirical model, by writing down the table of measurement choices (for the measurement scenario) and the table of outcome probabilities (for the empirical model). In the second Subsection, we characterise those cases in which the empirical model admits a local hidden variable model, which we describe in full detail. In the third Subsection, we write down the equations turning the empirical model into an All-vs-Nothing argument, in those cases in which a local hidden variable model is ruled out. In the fourth and final Subsection, we give an explicit quantum realisation in terms of GHZ states and phase gates on qudits (i.e. $d$-dimensional quantum systems). When restricted to the special case $d=2$ and $t = 1$, the setup we describe coincides with the one originally proposed by Mermin \cite{Mermin1990}, as long as we take the observable $\hbox{\input{symbols/ZbwdotSym.tex}}\!\!$ in the quantum realisation to correspond to Pauli Z, and the observable $\hbox{\input{symbols/DdotSym.tex}}\!\!$ to correspond to Pauli X (recall that performing the Pauli Z phase gate $P_{e^{i\frac{\pi}{2}}}$ and then measuring in Pauli X is the same as measuring in Pauli Y).

\subsection{Measurement scenario and empirical model.} 
Firstly, the exponent of $\integersMod{d}$ is $d$, and we fix a number of parties $N$ such that $\gcd(N,d) = 1$ (e.g. $N := d+1$). Each party $i=1,...,N$ can make a measurement choice $m_i$ in the set $\{0,1\}$, and the measurement contexts take the following form: in the control, all parties make measurement choice $0$, while the variations take the form of $N$ cyclic permutations, each one featuring $N-t$ contiguous parties making measurement choice $0$ and $t$ parties making measurement choice $1$. This is summarised by the table below:
\begin{equation}
\begin{array}{c|ccccccccc}
	\text{Party:} & 1 & 2 & ... & N-t-1 & N-t & N-t+1 &... & N-1 & N \\
	\hline
	\text{ control} 				& 0 & 0 & ... & 0 & 0 & 0 & ...  & 0 & 0 \\
	\text{ $1^{st}$ variation} 		& 0 & 0 & ... & 0 & 0 & 1 & ...  & 1 & 1 \\
	\text{ $2^{nd}$ variation} 		& 0 & 0 & ... & 0 & 1 & 1 & ...  & 1 & 0 \\
	\text{ $3^{rd}$ variation} 		& 0 & 0 & ... & 1 & 1 & 1 & ...  & 0 & 0 \\
	\vdots & \vdots & \vdots &  & \vdots & \vdots & \vdots & & \vdots & \vdots \\
	\text{ $N^{th}$ variation}  & 1 & 0 & ... & 0 & 0 & 0 & ...  & 1 & 1
\end{array}
\end{equation}
The joint measurement outcomes $(g_1,...,g_N)$ for the $N$ parties are valued in $\integersMod{d}^N$, and the generalised Mermin-type argument is associated with the following empirical model:
\begin{equation}\label{explicitEmpiricalModel}
\begin{array}{c||c|c|c}
	 & g_1\oplus...\oplus g_N = 0  & g_1 \oplus ... \oplus g_N = 1 & g_1 \oplus ... \oplus g_N \neq 0,1\\
	\hline
	\text{ control} 				& \frac{1}{d^{N-1}} & 0 & 0 \\
	\text{ $1^{st}$ variation} 		& 0 & \frac{1}{d^{N-1}} & 0 \\
	\text{ $2^{nd}$ variation} 		& 0 & \frac{1}{d^{N-1}} & 0 \\
	\text{ $3^{rd}$ variation} 		& 0 & \frac{1}{d^{N-1}} & 0 \\
	\vdots & \vdots & \vdots & \vdots  \\
	\text{ $N^{th}$ variation}  & 0 & \frac{1}{d^{N-1}} & 0
\end{array}
\end{equation}

\subsection{Local hidden variable models.}
When $t$ and $d$ are coprime, the equation $t y = \modclass{1}{d}$ has a (unique) solution $y :=  \modclass{t^{-1}}{d}$, and a local hidden variable model for the empirical model \ref{explicitEmpiricalModel} can be obtained as follows. Consider the set $\Lambda$ of all the $(g_1,...,g_N) \in \integersMod{d}^N$ such that $g_1\oplus...\oplus g_N = 0$, together with the uniform probability distribution $p: \Lambda \rightarrow \reals^+$ on $\Lambda$ (i.e. $p(g_1,...,g_N) = \frac{1}{d^{N-1}}$). Also, consider deterministic local outcomes for each fixed value $\underline{g} \in \Lambda$ of the hidden variable such that, upon measurement choice $m_i$ for party $i$, the measurement outcome is $g_i$ whenever $m_i = 0$ and $g_i \oplus t^{-1}$ whenever $m_i = 1$. In the control, all parties $i=1,...,N$ will choose $m_i = 0$, and the joint measurement outcome will be uniformly distributed over the subgroup $\Lambda \subset \integersMod{d}^N$. In any variation, $t$ parties will choose $m_i = 1$ and $N-t$ parties will choose $m_i = 0$, and the joint measurement outcome will be uniformly distributed over the coset  $(1,0,...,0) \oplus \Lambda \subset \integersMod{d}^N$ (using the fact that $t \cdot t^{-1} = 1$ in $\integersMod{d}$). Hence this really defines a local hidden variable model for the empirical model \ref{explicitEmpiricalModel} associated with the generalised Mermin-type argument.

\subsection{All-vs-Nothing arguments.}
When $t$ and $d$ are not coprime, the equation $t y = \modclass{1}{d}$ cannot have solutions in $K = \integersMod{d}$ (by a standard argument from number theory). The possibilistic empirical model associated with the argument has the following support subpresheaf $\supportSubpresheafSym \subseteq \sheafOfEventsSym$ (only the control and the first three variations are explicitly shown here, to exemplify the pattern):
\begin{align}
	\supportSubpresheaf{\text{control}} &= \text{ the set of all } (g^{1}_{0},g^{2}_{0},...,g^{N-t-1}_{0},g^{N-t}_{0},g^{N-t+1}_{0},...,g^{N-1}_{0},g^{N}_{0}) \in \integersMod{d}^N\nonumber\\
	& \text{ such that } \bigoplus_{i=1}^{N} g^{i}_{0} = 0 \label{AvNequationControl}
\end{align}
\begin{align}
	\supportSubpresheaf{\text{$1^{st}$ var'n}} &= \text{ the set of all } (g^{1}_{0},g^{2}_{0},...,g^{N-t-1}_{0},g^{N-t}_{0},g^{N-t+1}_{1},...,g^{N-1}_{1},g^{N}_{1}) \in \integersMod{d}^N\nonumber\\
	& \text{ such that } \Big(\bigoplus_{i=1}^{N-t} g^{i}_{0}\Big) \oplus \Big(\bigoplus_{i=N-t+1}^{N} g^{i}_{1}\Big)  = 1 \label{AvNequationVar1}
\end{align}
\begin{align}
	\supportSubpresheaf{\text{$2^{nd}$ var'n}} &= \text{ the set of all } (g^{1}_{0},g^{2}_{0},...,g^{N-t-1}_{0},g^{N-t}_{1},g^{N-t+1}_{1},...,g^{N-1}_{1},g^{N}_{0}) \in \integersMod{d}^N\nonumber\\
	& \text{ such that } \Big(g^{N}_{0} \oplus \bigoplus_{i=1}^{N-t-1} g^{i}_{0}\Big) \oplus \Big(\bigoplus_{i=N-t}^{N-1} g^{i}_{1}\Big)  = 1 \label{AvNequationVar2}
\end{align}
\begin{align}
	\supportSubpresheaf{\text{$3^{rd}$ var'n}} &= \text{ the set of all } (g^{1}_{0},g^{2}_{0},...,g^{N-t-1}_{1},g^{N-t}_{1},g^{N-t+1}_{1},...,g^{N-1}_{0},g^{N}_{0}) \in \integersMod{d}^N\nonumber\\
	& \text{ such that } \Big(g^{N-1}_{0} \oplus g^{N}_{0} \oplus \bigoplus_{i=1}^{N-t-2} g^{i}_{0}\Big) \oplus \Big(\bigoplus_{i=N-t-1}^{N-2} g^{i}_{1}\Big)  = 1 \label{AvNequationVar3}
\end{align}
Amongst the (many) equations in $\RlinearTheory{\integers}{\supportSubpresheafSym}$ we can find the $N+1$ equations equations above, one for the control (Equation \ref{AvNequationControl}) and $N$ for the variations (Equations \ref{AvNequationVar1}, \ref{AvNequationVar2} and \ref{AvNequationVar3}, corresponding to the first three variations, exemplify the pattern). Any global assignment $(g^{i}_{r})^{i=1,...,N}_{r=0,1}$ which satisfies all equations in  $\RlinearTheory{\integers}{\supportSubpresheafSym}$ would in particular satisfy those $N+1$ equations. However, adding up the $N$ equations corresponding to the variations yields, after a bit of rearranging, the following equation:
\begin{equation}
(N-t)\Big(\bigoplus_{i=1}^{N} g^{i}_{0}\Big) \oplus t\Big(\bigoplus_{i=1}^{N} g^{i}_{1}\Big)  = N 
\end{equation}
Taking this together with the equation $\bigoplus_{i=1}^{N} g^{i}_{0} = 0$ associated with the control then results in the following equation (using $\gcd(N,d) = 1$ to obtain the inverse $N^{-1}$ modulo $d$):
\begin{equation}
t\Big(\bigoplus_{i=1}^{N} N^{-1}g^{i}_{1}\Big) = 1
\end{equation}
But this means that setting $y := \bigoplus_{i=1}^{N} g^{i}_{1}$ would yield a solution to the equation $t y = 1$ in $\integersMod{d}$, which we assumed not to exist. Hence we cannot have any global assignment satisfying all equations in $\RlinearTheory{\integers}{\supportSubpresheafSym}$, and the model is both $\AvN{\integers}{\integersMod{d}}$ and $\AvNring{\integersMod{d}}$.

\subsection{Quantum realisation.}
Now consider the $d$-dimensional quantum system $\complexs[\integersMod{d}]$. Pick $\hbox{\input{symbols/DdotSym.tex}}\!\!$ to be the special commutative $\dagger$-Frobenius algebra associated with the orthonormal basis $\ket{g}_{g \in \integersMod{d}}$, and $\hbox{\input{symbols/ZbwdotSym.tex}}\!\!$ to be the quasi-special commutative $\dagger$-Frobenius algebra associated with the orthogonal basis $\ket{\goodchi}_{\goodchi \in \integersMod{d}^\wedge}$ defined as follows in terms of the multiplicative characters $\goodchi: \integersMod{d} \rightarrow S^1$ of the finite abelian group $\integersMod{d}$:
\begin{equation}
\ket{\goodchi} := \sum_{g \in \integersMod{d}} \goodchi(g) \ket{g}
\end{equation}
Let $H_0^N$ be the subgroup of $\integersMod{d}^N$ defined by  $H_0 := \suchthat{\underline{g} \in \integersMod{d}^N}{g_1 \oplus ... \oplus g_N = 0}$. Then the (normalised) $N$-party GHZ state $\ket{GHZ_{\hbox{\input{symbols/ZbwdotSym.tex}}\!\!}}$ in the observable $\hbox{\input{symbols/ZbwdotSym.tex}}\!\!$ takes the following form:
\begin{equation}
\ket{GHZ_{\hbox{\input{symbols/ZbwdotSym.tex}}\!\!}}  := \frac{1}{d^{N-1}}\sum_{\underline{g} \in H_0} \ket{g_1} \otimes... \otimes \ket{g_N}
\end{equation}
Let $P_{\alpha_1},...,P_{\alpha_N}$ be phase gates for the $\hbox{\input{symbols/ZbwdotSym.tex}}\!\!$ observable, where each $\alpha_i$ is a function $\integersMod{d}^\wedge \rightarrow S^1$ (not necessarily a group homomorphism): writing the phase gates explicitly we get $P_{\alpha_i} := \frac{1}{d} \sum_{\goodchi \in \integersMod{d}^\wedge} \alpha_i(\goodchi) \, \ket{\goodchi}\bra{\goodchi}$. Applying the $N$ phase gates  $P_{\alpha_1},...,P_{\alpha_N}$ to the $N$ subsystems of the GHZ state is the same as applying their composition  $P_{\alpha}:=P_{\alpha_1}\circ...\circ P_{\alpha_N}$ (where $\alpha := \alpha_1 \cdot ... \cdot \alpha_N$) to a single subsystem.

Amongst the many functions $\alpha : \integersMod{d}^\wedge \rightarrow S^1$ are the group homomorphisms, which form the group $\big(\integersMod{d}^\wedge \big)^\wedge$ under pointwise product. By Pontryagin duality, there is a canonical isomorphism $\integersMod{d} \isom \big(\integersMod{d}^\wedge \big)^\wedge$, given explicitly by $h \mapsto (\goodchi \mapsto \goodchi(h))$, and we can consider phase gates $P_{h}$ corresponding to all $h \in \integersMod{d}$. The phase gate $P_{h}$ acts as $P_{h} \ket{g} = \ket{g \oplus h}$ on the basis $\ket{g}_{g \in \integersMod{d}}$, and hence applying $P_{h}$ to a single subsystem of the GHZ state results in the following state, involving the coset $H_h = \suchthat{\underline{g} \in \integersMod{d}^N}{g_1 \oplus ... \oplus g_N = h}$:
\begin{equation}
\Big(P_{h} \otimes \id{} \otimes... \otimes \id{}\Big) \ket{GHZ_{\hbox{\input{symbols/ZbwdotSym.tex}}\!\!}} = \frac{1}{d^{N-1}}\sum_{\underline{g} \in H_h^N} \ket{g_1} \otimes... \otimes \ket{g_N}
\end{equation}
We wish to find a solution to $t y = 1$ in the group of phase gates for $\hbox{\input{symbols/ZbwdotSym.tex}}\!\!$, i.e. we want some $\beta :  \integersMod{d}^\wedge \rightarrow S^1$ such that $(P_{\beta})^t = P_1 $. To do so, first note that the multiplicative characters $\goodchi: \integersMod{d} \rightarrow S^1$ can equivalently be formulated in terms of elements $k \in \integersMod{d}$, e.g. by considering $\goodchi_{k} := g \mapsto e^{i 2\pi \frac{k \cdot g}{d}}$, and then rewriting the equation $(P_{\beta})^t = P_1 $ as follows: 
\begin{equation}
 \frac{1}{d}\sum_{k \in \integersMod{d}} \beta(\goodchi_k)^t \ket{\goodchi_k}\bra{\goodchi_k} = \frac{1}{d}\sum_{k \in \integersMod{d}} e^{i 2\pi \frac{k}{d}} \ket{\goodchi_k}\bra{\goodchi_k}
\end{equation}
It is now easy to see that a solution to $(P_{\beta})^t = P_1 $ is given by $\beta := \goodchi_k \mapsto e^{i 2\pi \frac{1}{t}\frac{k}{d}}$. 

\begin{rmrk}
The explicit construction presented here for the case of cyclic groups $\integersMod{d}$ in ordinary quantum theory is related to the recent work of Ref. \cite{Zukowski2013} (as well as previous work by Refs. \cite{Lee2006,Cerf2002,Kaszlikowski2002,Zukowski1999}), and the relationship between the dimension $d$ of the quantum systems and the allowed numbers $N$ of parties (i.e. $\gcd(N,d) = 1$) is the same here and in Ref.~\cite{Zukowski2013}. Even when restricted to the special case of quantum theory, however, the work presented here is a significant generalisation of the work of Ref.~\cite{Zukowski2013}: in the latter, the authors focus on a specific family of $M=2, S=1$ systems with values in a finite cyclic group $\integersMod{d}$; in this work, we provide necessary and sufficient algebraic conditions for arbitrary systems and arbitrary finite abelian groups. 
\end{rmrk}

\section{Quantum-classical Secret Sharing}
\label{section_QSS}

In contrast to other information security protocols, classical secret sharing comes with the intrinsic assumption that some participants cannot, to some extent, be trusted. A \textit{dealer} is interested in sharing some \textit{secret} with a number of \textit{players}, with the caveat that the secret be revealed to the players only when all players agree to cooperate\footnote{More in general, a minimum number of cooperating players can be specified.}. Integrity and availability of communications is guaranteed by the existence of authenticated classical channels between dealer and players, and the protocol is only concerned with confidentiality, defined as the impossibility of recovering the secret unless all players cooperate.

The quantum-classical scheme of Hillery, Bu\v{z}ek and Berthiaume \cite{Hillery1999} introduces a new layer of security to secret sharing, employing entangled states and non-commuting observables to detect eavesdropping. The HBB scheme is based on the same measurement contexts of Mermin's original parity argument: a dealer and $N-1$ players share $N$ qubits in a GHZ state (with respect to the computational basis associated with the Pauli $Z$ observable), and randomly choose to measure their qubit in either of the mutually unbiased Pauli $X$ or Pauli $Y$ observables. It can be shown \cite{Zamdzhiev2012} that confidentiality is an immediate consequence of strong complementarity of the Pauli $Z$ and $X$ observables, while eavesdropping detection follows from mutual unbias of the Pauli $X$ and $Y$ observables. 

We extend the HBB scheme from Mermin's original parity argument to our generalised Mermin-type arguments, and we use our result on contextuality to provide a number of device-independent security guarantees. For the remainder of this section, we will consider a generalised Mermin-type argument $(\hbox{\input{symbols/ZbwdotSym.tex}}\!\!,\hbox{\input{symbols/DdotSym.tex}}\!\!, \mathcal{S}, \beta, N)$, on an object $\SpaceH$ of a $R$-probabilistic CP* category $\CPStarCategory{\CategoryC}$, where $R$ is a positive semiring.

Consider a \textbf{dealer}, call her Alice, who wishes to share a \textbf{secret} with $N'$ \textit{players}, where $2 \leq N' < N$. As the owner of the secret, Alice is always a trusted party, the \textit{only} trusted party in the protocol. The secret is assumed to take the form of a string of elements of $\classicalStates{\hbox{\input{symbols/DdotSym.tex}}\!\!}$, the \textbf{plaintext} (at most one element of $\classicalStates{\hbox{\input{symbols/DdotSym.tex}}\!\!}$, the \textbf{round plaintext}, transmitted for each round of the protocol). We wish to ensure that the plaintext can be decoded from the information Alice sends, the \textbf{cyphertext}, if and only if all players agree to cooperate (by which we mean that they all reveal their secret keys to some party in possession of the cyphertext). Alice and the players are given $N$ devices (one per player, and $N-N'$ for Alice): at each round $w$, each device $B_j$ is fed an \textbf{input} $m_j^w \in \{0,1,...,M\}$ and returns an \textbf{output} $g_j^w \in \classicalStates{\hbox{\input{symbols/DdotSym.tex}}\!\!}$ (we also refer to the outputs $g_1^w,...,g_{N'}^w$ as the \textbf{secret keys} of the players for round $w$).
We furthermore assume the following \textbf{security conditions} to hold. 
\begin{enumerate}
	\item[(i)] Alice and the players share an authenticated classical channel, ensuring integrity and availability of all classical communications involved in the protocol.
	\item[(iia)] Alice and the players are in possession of $N$ secure independent classical sources of randomness, to generate independent inputs at each round which are uniformly distributed in $\{0,1,...,M\}$.
	\item[(iib)] Alice is in possession of a secure classical source of randomness, independent from all other, to decide which rounds will be \textbf{secret rounds} (with probability $(1-\tau)>0$) and which rounds will be \textbf{test rounds} (with probability $\tau > 0$).
	\item[(iii)] During step 2 of the protocol below, no signalling is possible between distinct parties/devices\footnote{This can be achieved, for example, by ensuring the devices are operated in conditions controlled by Alice (trusted laboratories, synchronized time-stamp servers, etc).}.
	\item[(iv)] We will assume that in step 3 Alice is communicated the measurement choices faithfully\footnote{This can be achieved by entrusting the laboratory setup with the communication of the random measurement choices to Alice, the player and the device.}
\end{enumerate}
Because tampering can only be determined after the protocol has ended and the entirety (or an otherwise significant portion) of the plaintext has been transmitted, we distinguish between the \textbf{plaintext}, the data that can be decoded using the secret keys, and the actual \textbf{secret} that Alice wants the players to share. Before the protocol begins, Alice will obtain the plaintext by encrypting the secret with a secure symmetric encryption protocol\footnote{If the secret is in the form of a string of elements of $\classicalStates{\hbox{\input{symbols/DdotSym.tex}}\!\!}$, the natural choice for this protocol, then the plaintext can be obtained by generating a string of uniformly random $k^w$ elements of $\classicalStates{\hbox{\input{symbols/DdotSym.tex}}\!\!}$, obtaining the round plaintext $p^w$ from the corresponding \inlineQuote{round secret} $q^w$ as $p^w = q^w \oplus k^w$. Once the string of random elements is broadcast, upon successful completion of the protocol, the secret can be recovered from the decoded plaintext as $q^w = p^w \ominus k^w$.}, using a freshly generated ephemeral key which she will broadcast only if the protocol is successful. If the protocol fails, the random key will not be broadcast and the secret will be unrecoverable even if the plaintext is decoded. 

\begin{figure}
	\input{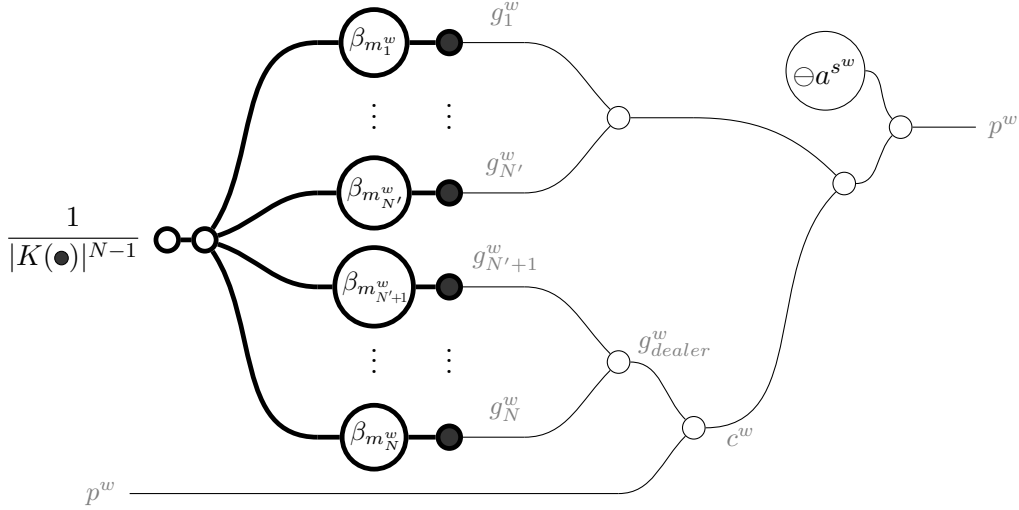}
	\caption{Graphical presentation of a noiseless, trusted implementation. 
	\label{fig_trustedProtocol}}
\end{figure}

The quantum-classical secret sharing protocol then proceeds as follows for each round $w=1,...,W$, until the entire secret has been transmitted. An individual round for a noiseless, trusted implementation is presented in Figure \ref{fig_trustedProtocol}. Throughout the protocol, Alice keeps a count of occurrences of joint outputs $g_1,...,g_N$ conditional to each joint input $m_1,...,m_N$ that she observes in test rounds.
\begin{enumerate}
	\item[1.] Alice and the players share $N$ subsystems of a state $\rho$: each player has an individual subsystem and Alice keeps the remaining $N-N'$ subsystems. In a noiseless, trusted implementation, $\rho$ is the $N$-partite $\hbox{\input{symbols/ZbwdotSym.tex}}\!\!$-GHZ state. For the purposes of a device-independent security analysis, $\rho$ can be potentially any state.
	\item[2.] Alice and the players each sample their classical source of randomness and obtain inputs $m_1^w,...,m_N^w$ which are passed to the devices $B_1,...,B_N$ and result in outputs $g_1^w,...,g_{N'}^w \in \classicalStates{\hbox{\input{symbols/DdotSym.tex}}\!\!}$ for the players (the secret keys for the round) and $g_{N'+1}^w,...,g_{N}^w \in \classicalStates{\hbox{\input{symbols/DdotSym.tex}}\!\!}$ for Alice. In a noiseless, trusted implementation, $B_j$ with input $m_j^w$ applies the phase gate $\phasegate{\beta_{m_j^w}}$ to the subsystem $j$ and then measures it in the $\hbox{\input{symbols/DdotSym.tex}}\!\!$ observable.
	\item[3.] The inputs for the players are communicated to Alice. She checks that $m_1^w,...,m_N^w$ define a valid \textbf{measurement context} (either the control ($s=0$) or a variation for some $s=1,...,S$).
	\item[4.] Alice samples her source of randomness to decide whether the round will be a test round or a secret round.
	\item[4a.] If the round is a test round, Alice requests all players to communicate their secret keys, and she increases the occurrence count for joint output $(g_1^w,...,g_N^w)$ conditional to joint input $m_1^w,...,m_N^w$.
	\item[4b.] If the round is a secret round, Alice computes $g_{dealer}^w := \bigoplus_{j=N'+1}^N g_j^w$ and broadcasts the \textbf{round ciphertext} $c^w:= p^w \oplus g_{dealer}^w$ to the players, where the \textbf{round plaintext} $p^w$ is the next element of the plaintext to be sent. She also broadcasts the relevant value $s^w \in \{0,1,...,S\}$ she obtained from the joint inputs $m_1^w,...,m_N^w$.
	\item[5.] Anyone in possession of the round ciphertext $c^w$ and all secret keys $g_1^w,...,g_{N'}^w$ can obtain the round plaintext $p^w$ by computing $p^w = (c^w\, \oplus g_1^w \oplus ... \oplus g_{N'}^w)\ominus a^{s^w}$, where $s^w$ is the value broadcast in Step 4. 
\end{enumerate}
The chosen generalised Mermin-type argument determines the following \textbf{promised conditional distribution} $\mathbb{P}_{promised}\big[\,\underline{g}\,\big\vert\, \underline{m}\,\big]$, the one which Alice and the players expect to observe (asymptotically) in a trusted noiseless implementation (we use the more compact notation $\underline{g} := (g_1,...,g_N)$ for the joint output and $\underline{m} := (m_1,...,m_N)$ for the joint input): 
\begin{equation}\label{Ppromised}
\mathbb{P}_{promised}\big[\,\underline{g}\,\big\vert\, \underline{m}\,\big] = 
\begin{cases}
	\dfrac{1}{|\classicalStates{\hbox{\input{symbols/DdotSym.tex}}\!\!}|^{N-1}} &\text{ if } g_1 \oplus ... \oplus g_N = \beta{m_1} \oplus ... \oplus \beta{m_N} \\
	0 &\text{ otherwise}
\end{cases}
\end{equation}
At the end of the protocol, Alice normalises her joint output counts for each joint input to obtain the \textbf{observed conditional distribution} $\mathbb{P}_{observed}\big[\,\underline{g}\,\big\vert\, \underline{m}\,\big]$ (which need not be no-signalling). She then computes the \textbf{noise parameter} $\epsilon$ as follows: 
\begin{equation}
\epsilon := 1 - |\classicalStates{\hbox{\input{symbols/DdotSym.tex}}\!\!}|^{N-1} \min \Big\{ \mathbb{P}_{observed}\big[\,\underline{g}\,\big\vert\, \underline{m}\,\big] \Big\vert g_1 \oplus ... \oplus g_N = \beta_{m_1} \oplus ... \oplus \beta_{m_N} \Big\} 
\end{equation}
The error parameter as defined above is the smallest $\epsilon \in [0,1]$ such that the observed conditional distribution can be decomposed as the following convex combination of  promised conditional distribution and some \textbf{noise conditional distribution} $\mathbb{P}_{noise}\big[\,\underline{g}\,\big\vert\, \underline{m}\,\big]$:
\begin{equation}
\mathbb{P}_{observed}\big[\,\underline{g}\,\big\vert\, \underline{m}\,\big] = (1 - \epsilon) \; \mathbb{P}_{promised}\big[\,\underline{g}\,\big\vert\, \underline{m}\,\big] + \epsilon \;\mathbb{P}_{noise}\big[\,\underline{g}\,\big\vert\, \underline{m}\,\big]
\end{equation} 
Before a run of the protocol begins, Alice sets a maximum $\epsilon_{max}$ that she is going to accept for the noise parameter. Alice chooses as low an $\epsilon_{max}$ as possible compatibly with the specifications of the device provider (and any other beliefs she might have) on the amount of noise she should expect from the devices and states in the absence of any tampering from Eve. At the end of the protocol run, Alice compares the noise parameter $\epsilon$ she computed with the maximum $\epsilon_{max}$ she decided to accept: if $\epsilon \leq \epsilon_{max}$, she declares the protocol run a success and broadcasts the ephemeral key she used to encode the secret into the plaintext; if $\epsilon > \epsilon_{max}$, she declares the protocol run a failure and she destroys the ephemeral key, rendering the secret unrecoverable even if the plaintext is at some point obtained by the players or by Eve.

The HBB quantum-classical secret sharing protocol comes with two security guarantees: (i) ignorance about any one secret key for a round denies knowledge about the plaintext for that round; (ii) successful, undetected eavesdropping has low probability. It can be shown \cite{Zamdzhiev2012} that in a noiseless and trusted implementation the first guarantee follows abstractly from strong complementarity of the Pauli $Z$ and $X$ observables, and the proof straightforwardly transfers to the strongly complementary pairs $(\hbox{\input{symbols/ZbwdotSym.tex}}\!\!,\hbox{\input{symbols/DdotSym.tex}}\!\!)$ appearing in our generalised protocol. Instead of treating eavesdropping directly, we will present a more general, device-independent proof of security, based solely on contextuality of the generalised Mermin-type argument used by the protocol. 

Works on device-independent security (such as \cite{Barrett2005,Vazirani2014} on quantum key distribution) usually posit Eve to be an adversary who can arbitrarily tamper with the shared state and measurement devices, and is only bound in her attempts by the physical theory under consideration\footnote{Eve is often assumed to be bound by the laws of quantum theory, but sometimes super-quantum attackers are also considered, bound only by causality and no-signalling.} and by the security conditions explicitly enforced by the protocol (including no-signalling). Examples of things that the Eve is allowed to do include:
\begin{enumerate}[(i)]
\item the measurement outcomes broadcast at a test round can reveal to Eve information about measurement outcomes in previous secret rounds;
\item Eve can keep a subsystem of the shared state to herself, which she can optimally measure, once all inputs and test round outputs have been broadcast, to obtain information about the secret keys.
\end{enumerate}
Our choice of a device-independent setting comes from the more modest desire to show that the security guarantees follow from contextuality of the generalised Mermin-type argument, regardless of the specific implementation; as a consequence, we will be content with a more restricted model of attack. We assume that Alice and the players might be provided with noisy or imperfect states and devices, which might give Eve a variety of security loopholes to exploit. However, we assume that the device provider shows no malice: 
\begin{enumerate}[(i)]
\item the devices are memoryless and operate independently at each round;
\item the states used at different rounds are independent and identical; 
\item the states are not entangled with any additional system.
\end{enumerate}
However, Eve might possess classical information about the states which is unavailable to the players (such as information leaked through noise or side channels, information acquired via eavesdropping, etc).

\begin{figure}
	\input{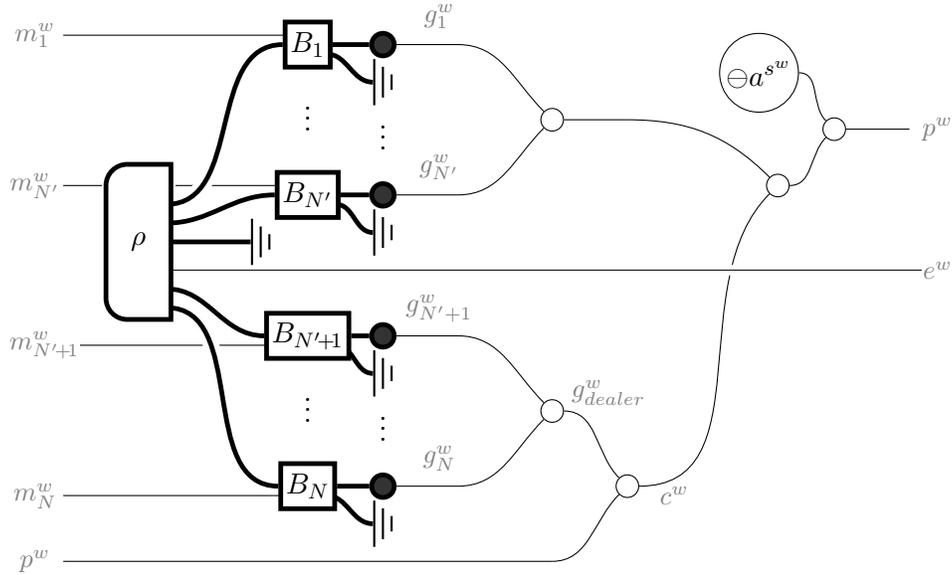}
	\caption{Graphical presentation of a generic, untrusted implementation at a single round of the protocol. Eve might have some classical information $e^w$ about the states which is unknown to Alice and the players. The classical side of the protocol is entirely in the hands of Alice and the players, and proceeds as in the trusted noiseless case.
	\label{fig_untrustedProtocol}}
\end{figure}

Although not fully general, this setup subsumes a variety of more specialised security scenarios that are of interest in classical and quantum cryptography:
\begin{enumerate}[(i)]
\item Real-world implementations are unavoidably noisy, and one should consider any noise as a potential source of cryptophthora. Our setup allows for the possibility that both the shared state and the measurement devices be noisy, with no dependence on a specific model of noise; it also allows for the possibility that what looks like random noise to Alice and the players might actually carry side-channel information to Eve.
\item Eavesdropping detection is a typical desideratum in quantum cryptography, where Eve intercepts the local state of a player\footnote{In our secret sharing protocol, a single player's secret key is all that Eve needs to break confidentiality, as we may freely assume that the remaining players are colluding and asked Eve to help them.}, measures it in some basis to obtain classical information, and forwards the resulting collapsed state to the player. Our setup allows for the possibility of eavesdropping\footnote{However, it does not cover a more advanced attack in which Eve sends through a subsystem of an entangled state, keeping the rest of the state to herself and measuring it in the future to obtain more information about the player's outcome.}: the classical information that Eve possesses about the state can be used to model the information she acquired by eavesdropping. Our security proof then has eavesdropping detection as a special case of protocol failure.
\end{enumerate} 

\noindent Figure \ref{fig_untrustedProtocol} displays a single round $w$ of the protocol in a generic, untrusted implementation. An $N$-partite state $\rho$ is shared between Alice and the players at a given round of the protocol, with no additional subsystem accessible to Eve (who might however be in possession of classical information $e^w$ about it). The measurement devices $B_1,...,B_N$ operate independently at each round, with no memory or shared resource other than the state $\rho$. At each round $w$, device $B_j$ takes measurement choice $m_j^w$ as a classical input and returns measurement outcome $g_j^w$ as a classical output. The rest of the protocol is entirely in the hands of Alice and the players, and proceeds as in the trusted noiseless case.

Our first result shows that lack of contextuality implies the existence of a scenario in which a perfect undetectable attack may take place. In fact, the scenario is not particularly remote: it might well happen happen that the device provider inadvertently chose phases states $\beta_1,...,\beta_M$ which happen to be $\hbox{\input{symbols/DdotSym.tex}}\!\!$-classical states (maybe she did not notice, maybe she was tricked by Eve into choosing them), and that the GHZ state decoheres (spontaneously or with a malicious helping hand) in the $\hbox{\input{symbols/DdotSym.tex}}\!\!$ observable. In that case, Alice and the player will notice nothing wrong with their protocol, and Eve will obtain the entirety of the secret all by herself.

\begin{thm}
Consider a quantum-classical secret sharing protocol based on a generalised Mermin-type argument $(\hbox{\input{symbols/ZbwdotSym.tex}}\!\!,\hbox{\input{symbols/DdotSym.tex}}\!\!, \mathcal{S}, \beta, N)$, in an $R$-probabilistic CP* category $\CPStarCategory{\CategoryC}$, where $R$ is a positive semiring. If the associated empirical model is non-contextual, then there are shared states $\rho$ and measurement devices $B_1,...,B_N$ such that test rounds will succeed with certainty, and Eve will always know all the secret keys.
\end{thm}
\proof
By Theorem \ref{thm_contextuality}, if the empirical model is non-contextual then there exists a solution $(y_r := b_r)_{r=1}^M$ in $\classicalStates{\hbox{\input{symbols/DdotSym.tex}}\!\!}$ to the system $\mathcal{S}$ (which we take to be in the form of System \ref{eqn_system}). For each round $w$, Eve samples a random variable uniformly distributed over the following set:
\begin{equation}
\suchthat{(h_1^w,...,h_N^w) \in \classicalStates{\hbox{\input{symbols/DdotSym.tex}}\!\!}^N}{h_1^w \oplus ... \oplus h_N^w = 0}
\end{equation}
Now assume that the separable pure state $\rho^w = \ket{h_1} \otimes ... \otimes \ket{h_N}$ is given in input to the measurement devices $B_1,...,B_N$, which are designed so that $B_j$ returns $g_j^w := h_j^w \oplus b_{m_j^w}$ upon measurement choice $m_j^w$ (i.e. applies a phase $b_{m_j^w}$ which happens to be $\hbox{\input{symbols/DdotSym.tex}}\!\!$-classical). Once the measurement $(m_j^w)_{j=1}^{N'}$ choices for the players are broadcast, Eve can compute all the secret keys $(g_j^w)_{j=1}^{N'}$. Furthermore, since $(b_r)_{r=1}^{M}$ is a solution to $\mathcal{S}$, the measurement outcomes obtained from this setup will have the same distribution as the ones from a noiseless trusted implementation, and all test rounds will succeed with certainty.
\qed

Our second result is restricted to probabilistic theories, i.e. distributively $\CMonCategory$-enriched CPM categories having $\reals^{+}$ as their semiring of scalars. Consider the no-signalling polytope associated with the measurement scenario of a contextual generalised Mermin-type argument $(\hbox{\input{symbols/ZbwdotSym.tex}}\!\!,\hbox{\input{symbols/DdotSym.tex}}\!\!, \mathcal{S}, \beta, N)$, and let $F$ be the face of the polytope specified by the support of the empirical model (the one defined by Equation \ref{Ppromised}). For each vertex $v \in F$ of that face, corresponding to empirical model $\mathbb{P}_{v}\big[\,\underline{g}\,\big\vert\,\underline{m}\,\big]$, let $H_v$ be the average entropy across all measurement contexts:
\begin{equation}
H_v := \frac{1}{1+N \cdot S} \sum_{\underline{m} \in \mathcal{M}} H\Big[\, \mathbb{P}_{v}\big[\,\emptyArg\,\big\vert\,\underline{m}\,\big] \,\Big]
\end{equation}
Let $H_{promised}^{(min)} := \min_{v \in F} H_v$ be the minimum average entropy across all vertices of the face: because the generalised Mermin-type argument is strongly contextual, the face cannot contain any local vertices, and hence the minimum average entropy $H_{promised}^{(min)}$ is always strictly positive; a tighter estimation of this quantity is left to future work. Call $\eta := \Big(1-\dfrac{H_{promised}^{(min)}}{|\classicalStates{\hbox{\input{symbols/DdotSym.tex}}\!\!}|^{N-1}}\Big) \in [0,1)$ the \textbf{information leakage fraction} for the face: it is the maximum fraction of plaintexts that Eve can expect to decipher when the empirical model she sees lies on face $F$.

We will now show that protocols based on contextual generalised Mermin-type arguments always provide a certain amount of security: for observed noise parameter $\epsilon$ small enough, the maximum expected fraction of plaintexts that Eve can expect to decipher is sharply peaked somewhere between $\eta$ and $c \cdot \epsilon$, where $c$ is some constant depending on the geometry of the no-signalling polytope. In one extreme, we may have $\eta = 0$, i.e. all empirical model on the face carry the same maximal amount of entropy. In this case, Eve's chances of learning some parts of the secret rely entirely on the noise parameter $\epsilon$: in her best case scenario, she observes a deterministic empirical model for some fraction $\epsilon$ of rounds, in which case she can gain complete knowledge about the round plaintext. In the other extreme, we have $\eta \gg \epsilon$, i.e. there are empirical models on the face $F$ which might lead to more leakage of plaintext information than any number of deterministic model which might be lurking in the noise $\epsilon$. In this case, Eve's best bet might just be to exploit the empirical models on the face $F$ itself.

\begin{thm}
Consider a quantum-classical secret sharing protocol based on a generalised Mermin-type argument $(\hbox{\input{symbols/ZbwdotSym.tex}}\!\!,\hbox{\input{symbols/DdotSym.tex}}\!\!, \mathcal{S}, \beta, N)$, in a probabilistic CP* category $\CPStarCategory{\CategoryC}$ (with $\reals^{+}$ as its positive semiring of scalars). Consider a run of the protocol with a large number $W$ of rounds, of which $P$ secret rounds and $T$ test rounds (with $P \rightarrow (1-\tau)W$ and $T \rightarrow \tau W$ almost certainly as $W \rightarrow \infty$). Let $\epsilon$ be the noise parameter observed by Alice at the end (a random variable), and let $P_{Eve}$ be maximum number of round plaintexts that Eve expects to successfully decipher (another random variable). Then the maximum fraction of plaintexts $P_{Eve}/P$ that Eve expects to successfully decipher is sharply peaked around some value between $\eta$ and $O(\epsilon)$, with variance bounded above by $O(\frac{\tau(1-\tau)}{W})$ almost certainly for $W \rightarrow \infty$ (where the big-$O$ notation hides a constant depending on the geometry of the polytope alone).
\end{thm}
\proof 
As part of this proof, a number of different conditional distributions will be considered: 
\begin{enumerate}[(i)]
\item the no-signalling conditional distribution $\mathbb{P}_{true}(e)\big[\,\underline{g}\,\big\vert\, \underline{m}\,\big]$ determined by $\rho$ and the devices $B_1,...,B_N$ conditional to Eve obtaining information $e$ (this is the conditional distribution as seen from Eve's vantage point);
\item the no-signalling conditional distribution $\mathbb{P}_{true}\big[\,\underline{g}\,\big\vert\, \underline{m}\,\big] := \sum_{e}\mathbb{P}[e] \cdot \mathbb{P}_{true}(e)\big[\,\underline{g}\,\big\vert\, \underline{m}\,\big]$ determined by $\rho$ and the devices $B_1,...,B_N$, averaged over Eve's information (this is the \textbf{true conditional distribution} as seen from Alice's vantage point, which her tests will estimate);
\item the no-signalling conditional distribution $\mathbb{P}_{promised}\big[\,\underline{g}\,\big\vert\, \underline{m}\,\big]$ derived from the generalised Mermin-type argument (this is what Alice would expect to estimate in the absence of any noise or tampering);
\item the conditional distribution $\mathbb{P}_{observed}\big[\,\underline{g}\,\big\vert\, \underline{m}\,\big]$ estimated by Alice.
\end{enumerate}
Alice's estimate of the true conditional distribution $\mathbb{P}_{true}\big[\,\underline{g}\,\big\vert\, \underline{m}\,\big]$ can be modelled by considering the vector-valued random variables $\underline{X}^w := \big(X_{(\underline{g},\underline{m})}^w\big)$ for all test rounds $w$, where $X_{(\underline{g},\underline{m})}^w$ is the real-valued random variable defined as follows (note that $\underline{g}^w$ is a random element of $\classicalStates{\hbox{\input{symbols/DdotSym.tex}}\!\!}^N$, and $\underline{m}^w$ is a uniformly random element of the set of $1+NS$ measurement contexts):
\begin{equation}
X_{(\underline{g},\underline{m})}^w = 
\begin{cases}
1 & \text{ if } \underline{g} = \underline{g}^w \text{ and } \underline{m} = \underline{m}^w\\
0 & \text{ otherwise}
\end{cases}
\end{equation} 
The vector $\underline{X}^w$ takes the value $1$ over the joint input/joint output pair  recorded by Alice for round $w$, and $0$ everywhere else: Alice's estimate of the true conditional distribution is then obtained from the average random variable $\frac{1}{T}\!\!\sum\limits_{w \text{ test}}\!\! \underline{X}^w$. By the central limit theorem, Alice's estimate $\mathbb{P}_{observed}\big[\,\underline{g}\,\big\vert\, \underline{m}\,\big]$ will be normally distributed around the true conditional distribution, with variance $O(\frac{1}{T})$; because the noise parameter $\epsilon$ observed by Alice is obtained from this estimate, it will similarly be distributed around the true noise parameter $\epsilon_{true}$ defined below, with variance bounded above by $O(\frac{1}{T})$ (almost certainly for $T \rightarrow \infty$).

We define the \textbf{true noise parameter} $\epsilon_{true}$ to be obtained from the conditional distribution $\mathbb{P}_{true}\big[\,\underline{g}\,\big\vert\, \underline{m}\,\big]$ in the same way that $\epsilon$ is obtained from the conditional distribution $\mathbb{P}_{observed}\big[\,\underline{g}\,\big\vert\, \underline{m}\,\big]$. This means $\epsilon_{true}$ is the largest such that $\mathbb{P}_{true}\big[\,\underline{g}\,\big\vert\, \underline{m}\,\big]$ decomposes as follows, for some conditional distribution $\mathbb{P}_{true,noise}\big[\,\underline{g}\,\big\vert\, \underline{m}\,\big]$:
\begin{equation}
(1-\epsilon_{true}) \, \mathbb{P}_{promised}\big[\,\underline{g}\,\big\vert\, \underline{m}\,\big] + (\epsilon_{true}) \mathbb{P}_{true,noise}\big[\,\underline{g}\,\big\vert\, \underline{m}\,\big]
\end{equation}
For each value $e \in E$ that Eve's information can take, we define the parameter $\xi(e) \in [0,1]$ to be the smallest possible such that the conditional distribution $\mathbb{P}_{true}(e)\big[\,\underline{g}\,\big\vert\, \underline{m}\,\big]$ decomposes as follows:
\begin{equation}
(1-\xi(e))\mathbb{P}_{F}(e)\big[\,\underline{g}\,\big\vert\, \underline{m}\,\big] + \xi(e) \mathbb{P}_{F,noise}(e)\big[\,\underline{g}\,\big\vert\, \underline{m}\,\big]
\end{equation}
for some distribution $\mathbb{P}_{F}(e)\big[\,\underline{g}\,\big\vert\, \underline{m}\,\big]$ lying on the face $F$ and some distribution $\mathbb{P}_{F,noise}(e)\big[\,\underline{g}\,\big\vert\, \underline{m}\,\big]$ lying outside of face $F$. To Eve, in possession of information $e$, the conditional distribution $\mathbb{P}_{true}(e)\big[\,\underline{g}\,\big\vert\, \underline{m}\,\big]$ looks like a biased coin deciding between the two following scenarios:
\begin{enumerate}
\item[(a)] with probability $(1-\xi(e))$, she observes a distribution $\mathbb{P}_{F}(e)\big[\,\underline{g}\,\big\vert\, \underline{m}\,\big]$ lying on face $F$, which means that the fraction of the round plaintext that she expects to learn is bounded above by $\eta$;  
\item[(b)] with probability $\xi(e)$, she observes some other distribution $\mathbb{P}_{F,noise}(e)\big[\,\underline{g}\,\big\vert\, \underline{m}\,\big]$, which in the best case scenario could give her full knowledge of the round plaintext.
\end{enumerate}
Because marginalising over Eve's knowledge\footnote{I.e. taking the convex combination of the conditional distributions $\mathbb{P}_{true}(e)\big[\,\underline{g}\,\big\vert\, \underline{m}\,\big]$ with respect to the probability distribution $\mathbb{P}[e]$ of Eve's side-channel information.} must result in the distribution $\mathbb{P}_{true}\big[\,\underline{g}\,\big\vert\, \underline{m}\,\big]$, the geometry of the polytope implies that the convex combination $\sum_e \mathbb{P}[e] \xi(e)$ must go to zero as $O(\epsilon_{true})$ (i.e. there must be some constant $c > 0$ such that $\sum_e \mathbb{P}[e] \xi(e) \leq c \cdot \epsilon_{true}$).

It should be noted that the information $e$ obtained by Eve is random to Eve herself: sometimes she will obtain information giving her better guessing probability, sometimes she will obtain information giving her worse guessing probability.  When the distribution of $e$ is taken into account, the fraction of round plaintexts that Eve can expect to decipher is bounded above by the following value, falling somewhere between $\eta$ and $O(\epsilon_{true})$:
\begin{equation}
\sum_{e} \mathbb{P}[e]\,\Big((1-\xi(e)) \eta + \xi(e)\Big) 
\end{equation}
Again by central limit theorem, the maximum fraction $P_{Eve}/P$ of round plaintexts that Eve expects to successfully decipher is normally distributed around the value above, with variance $O(\frac{1}{P})$ (almost certainly for $P \rightarrow \infty$).

Finally, because $P_{Eve}/P$ is sharply peaked around some value between $\eta$ and $O(\epsilon_{true})$, with variance $O(\frac{1}{P})$, and because $\epsilon$ is sharply peaked around $\epsilon_{true}$, with variance bounded above by $O(\frac{1}{T})$, we can conclude that $P_{Eve}/P$ is sharply peaked around some value between $\eta$ and $O(\epsilon)$ , with variance bounded above by $O(\frac{1}{T}+\frac{1}{P})$ (which tends to $O(\frac{1}{\tau (1-\tau) W})$ almost certainly as $W \rightarrow \infty$).
\qed

\section*{Conclusions and future work}

\subsection*{Conclusions}

Using phase groups and strongly complementary observables, we have fully generalised Mermin-type non-locality arguments, and we have provided the exact group-theoretic conditions required for non-locality to arise. In this sense, our results complete the line of enquiry on the connection between phase groups and non-locality started in Refs. \cite{Coecke2010a,Coecke2012c}. We have shown that all generalised Mermin-type argument can be realised in quantum mechanics, using GHZ states and appropriate sets of phase gates. Furthermore, the abstract, diagrammatic nature of our proofs makes our results immediately applicable to a much larger class of quantum-like theories, such as real quantum theory, relational quantum theory, hyperbolic quantum theory, p-adic quantum theory, and modal quantum theory \cite{Gogioso2017FQT}.

We have proceeded to investigate the empirical models arising from our generalised arguments, using the sheaf-theoretic framework for non-locality and contextuality \cite{Abramsky2011}. We have shown the models to provide a new infinite family of inequivalent quantum-realisable All-vs-Nothing arguments \cite{Abramsky2015}, and we have concluded that the hierarchy of quantum-realisable All-vs-Nothing arguments over rings of modular integers does not collapse. As a corollary, we have established that generalised Mermin-type arguments give rise to strong contextuality whenever they are contextual in the first place. 

Our generalisations have found practical application in the formulation of an extension of the quantum-classical secret sharing scheme of Hillery, Bu\v{z}ek and Berthiaume \cite{Hillery1999}, which was originally based on Mermin's non-locality argument for qubit GHZ states \cite{Mermin1990}. We have provided a diagrammatic description of the scheme in its most general form, and we have used our results on strong contextuality to provide some device-independent security guarantees (which apply to the original HBB scheme as a special case).

\subsection*{Future work}

In the process of pursuing the connection between phase groups and non-locality, this work  has left several questions unanswered. 

Firstly, our generalised arguments are restricted to finite abelian group algebras, while it is known that strongly complementary pairs extend to all finite group algebras (where one Frobenius algebra is commutative, and one is potentially non-commutative), and even further to certain finite-dimensional compact quantum groups (where both Frobenius algebras are potentially non-commutative). From a quantum perspective, this corresponds to using possibly degenerate observables in place of the non-degenerate ones employed in this work. In the future, we will be interested in investigating the impact that the introduction of degenerate observables---in the form of non-abelian group algebras and compact quantum groups---might have on the connection between non-locality and phase groups that was established here in the abelian group case. 

Secondly, we have shown that our generalised Mermin-type arguments are All-vs-Nothing, but the converse need not hold in general: there are many examples of All-vs-Nothing arguments which do not come in the format of Mermin-type arguments \cite{Abramsky2015}. In the future, we will be interested in fleshing out an exact description of which All-vs-Nothing are equivalent to / arise as deformation of Mermin-type arguments, and in exploring whether the techniques used in this work can be applied in a more general context.

Finally, the model of attack we used in the security proof for our quantum-classical secret sharing scheme is somewhat more restrictive than the gold standard employed in device-independent quantum cryptography. In the future, we will be interested in obtaining a more complete proof of security, covering broader modes of attack (such as memory and side-channel attacks). Also, our current proof relies on some rather lax parameters, which we expect to be tightened as part of upcoming work.

\section*{Acknowledgements}
The authors would like to thank Samson Abramsky, Bob Coecke, Aleks Kissinger and Rui Soares Barbosa for comments, suggestions and useful discussions, as well as Sukrita Chatterji and Nicol\`o Chiappori for their support. The authors would especially like to thank Samson Abramsky for pointing out a mistake in a previous version of this work. Funding from EPSRC (OUCL/2013/SG) and Trinity College (Williams Scholarship) for the first author is gratefully acknowledged.

\input{biblio.bbl}

\end{document}

%% file: packages.tex
\usepackage{mathtools} 
\usepackage{amssymb} 
\usepackage{bbold} 
\usepackage{amsthm} 

\usepackage{microtype} 
\usepackage{csquotes} 

\usepackage{hyperref} 

\usepackage{graphicx} 
\usepackage{xcolor} 
\usepackage{tikz} 
\usepackage{circuitikz} 
\usetikzlibrary{
	arrows,
	shapes,
	decorations,
	intersections,
	backgrounds,
	positioning,
	circuits.ee.IEC
	}

%% file: macros.tex
	\newcommand{\emptyArg}{\,\underline{\hspace{6px}}\,} 
	\newcommand{\inlineQuote}[1]{\textquotedblleft #1\textquotedblright} 
	\newcommand{\goodchi}{\protect\raisebox{2pt}{$\chi$}} 
	\newcommand{\goodrho}{\protect\raisebox{2pt}{$\rho$}} 
	\newcommand{\goodvdots}{\protect\raisebox{7pt}{\vdots}} 
	\newcommand{\verteq}{\rotatebox{90}{$\,=$}}

	\newcommand{\Powerset}[1]{\mathcal{P}(#1)} 
	\newcommand{\naturals}{\mathbb{N}} 
	\newcommand{\integers}{\mathbb{Z}} 
	\newcommand{\rationals}{\mathbb{Q}} 
	\newcommand{\reals}{\mathbb{R}} 
	\newcommand{\complexs}{\mathbb{C}} 
	\newcommand{\integersMod}[1]{\mathbb{Z}_{#1}} 
	\newcommand{\modclass}[2]{#1 \; (\text{mod } #2)} 
	\newcommand{\restrict}[2]{\left. #1 \right\vert_{#2}} 
	\newcommand{\support}[1]{\operatorname{supp}#1} 
	\newcommand{\inject}{\hookrightarrow} 
	
	\newcommand{\normalSubgroup}{\trianglelefteq}

	\newcommand{\suchthat}[2]{\left\{#1 \: \middle\vert \: #2\right\}} 

		\newcommand{\ket}[1]{\vert #1 \rangle} 
		\newcommand{\bra}[1]{\langle #1 \vert} 
		\newcommand{\braket}[2]{\langle #1 \vert #2 \rangle} 
		
		\newcommand{\Trace}[1]{\operatorname{Tr}[#1]} 
		\newcommand{\decohSym}{\operatorname{dec}} 
		\newcommand{\decoh}[1]{\decohSym_{#1}} 
		\newcommand{\CPMdoubled}[1]{\textbf{double}\left[#1\right]}

		\newcommand{\SpaceH}{\mathcal{H}} 
		
		\newcommand{\SpaceK}{\mathcal{K}}

		\newcommand{\isom}{\cong} 
		\newcommand{\id}[1]{id_{#1}} 

		\newcommand{\tensor}{\otimes} 
		\newcommand{\tensorUnit}{I} 

		\newcommand{\Hom}[3]{\operatorname{Hom}_{\,#1}\left[#2,#3\right]} 

		\newcommand{\SetCategory}{\operatorname{Set}} 

		\newcommand{\CMonCategory}{\operatorname{CMon}} 
		\newcommand{\RMatCategory}[1]{#1\operatorname{-Mat}} 

		\newcommand{\fdHilbCategory}{\operatorname{fdHilb}} 
		\newcommand{\fSetCategory}{\operatorname{fSet}} 

		\newcommand{\CategoryC}{\mathcal{C}}
		\newcommand{\CategoryD}{\mathcal{D}}

		\newcommand{\OpCategory}[1]{#1^{\operatorname{op}}} 
		\newcommand{\CPMCategory}[1]{\operatorname{CPM}[#1]} 
		\newcommand{\CPStarCategory}[1]{\operatorname{CP}^\ast[#1]} 

	\newcommand{\classicalStates}[1]{K(#1)} 
	\newcommand{\phaseGroup}[1]{P(#1)} 
	\newcommand{\phasegate}[1]{P_{#1}}
	
	\newcommand{\CPstates}[1]{\mathcal{L}[#1]}

	\newcommand{\sheafOfEventsSym}{\mathcal{E}} 
	\newcommand{\sheafOfEvents}[1]{\sheafOfEventsSym[#1]} 
	\newcommand{\restrictionMap}[2]{\operatorname{res}^{#1}_{#2}} 
	\newcommand{\distributionFunctorSym}[1]{\mathcal{D}_{#1}} 
	\newcommand{\distributionFunctor}[2]{\distributionFunctorSym{#1}[#2]} 
	\newcommand{\presheafOfDistributionsSym}[1]{\distributionFunctorSym{#1}\sheafOfEventsSym} 
	\newcommand{\presheafOfDistributions}[2]{\presheafOfDistributionsSym{#1}[#2]} 
	\newcommand{\supportSubpresheafSym}{\mathbb{S}} 
	\newcommand{\supportSubpresheaf}[1]{\supportSubpresheafSym[#1]} 

	\newcommand{\Xcolour}{Red}
	
	\newcommand{\Zcolour}{YellowGreen}

	\newcommand{\Xaltcolour}{Purple}
	
	\newcommand{\Zaltcolour}{Cyan}
	
	\newcommand{\Dcolour}{black!80}

	\newcommand{\Xbwcolour}{black!80}
	\newcommand{\XbwdotSym}{\hbox{\input{symbols/DdotSym.tex}}\!\!} 

	\newcommand{\Zbwcolour}{white}
	\newcommand{\ZbwdotSym}{\hbox{\input{symbols/ZbwdotSym.tex}}\!\!} 

	\newcommand{\Ybwcolour}{black!20}
	\newcommand{\YbwdotSym}{\hbox{\input{symbols/YbwdotSym.tex}}\!\!} 

	\newcommand{\trace}[1]{\hbox{\input{symbols/traceSym.tex}}\!_{#1}} 

	\tikzset{
	  rectangle with rounded corners north west/.initial=4pt,
	  rectangle with rounded corners south west/.initial=4pt,
	  rectangle with rounded corners north east/.initial=4pt,
	  rectangle with rounded corners south east/.initial=4pt,
	}
	\makeatletter
	\pgfdeclareshape{rectangle with rounded corners}{
	  \inheritsavedanchors[from=rectangle] 
	  \inheritanchorborder[from=rectangle]
	  \inheritanchor[from=rectangle]{center}
	  \inheritanchor[from=rectangle]{north}
	  \inheritanchor[from=rectangle]{south}
	  \inheritanchor[from=rectangle]{west}
	  \inheritanchor[from=rectangle]{east}
	  \inheritanchor[from=rectangle]{north east}
	  \inheritanchor[from=rectangle]{south east}
	  \inheritanchor[from=rectangle]{north west}
	  \inheritanchor[from=rectangle]{south west}
	  \backgroundpath{
	    \southwest \pgf@xa=\pgf@x \pgf@ya=\pgf@y
	    \northeast \pgf@xb=\pgf@x \pgf@yb=\pgf@y
	    \pgfkeysgetvalue{/tikz/rectangle with rounded corners north west}{\pgf@rectc}
	    \pgfsetcornersarced{\pgfpoint{\pgf@rectc}{\pgf@rectc}}
	    \pgfpathmoveto{\pgfpoint{\pgf@xa}{\pgf@ya}}
	    \pgfpathlineto{\pgfpoint{\pgf@xa}{\pgf@yb}}
	    \pgfkeysgetvalue{/tikz/rectangle with rounded corners north east}{\pgf@rectc}
	    \pgfsetcornersarced{\pgfpoint{\pgf@rectc}{\pgf@rectc}}
	    \pgfpathlineto{\pgfpoint{\pgf@xb}{\pgf@yb}}
	    \pgfkeysgetvalue{/tikz/rectangle with rounded corners south east}{\pgf@rectc}
	    \pgfsetcornersarced{\pgfpoint{\pgf@rectc}{\pgf@rectc}}
	    \pgfpathlineto{\pgfpoint{\pgf@xb}{\pgf@ya}}
	    \pgfkeysgetvalue{/tikz/rectangle with rounded corners south west}{\pgf@rectc}
	    \pgfsetcornersarced{\pgfpoint{\pgf@rectc}{\pgf@rectc}}
	    \pgfpathclose
	 }
	}
	\makeatother

	\tikzset{->-/.style={decoration={markings,mark=at position #1 with {\arrow{>}}},postaction={decorate}}}
	\tikzset{-<-/.style={decoration={markings,mark=at position #1 with {\arrow{<}}},postaction={decorate}}}

	\tikzstyle{every picture}=[baseline=-0.25em,scale=0.5]
	\pgfdeclarelayer{edgelayer}
	\pgfdeclarelayer{nodelayer}
	\pgfsetlayers{edgelayer,nodelayer,main}

	\tikzstyle{box} = [draw,shape=rectangle,inner sep=2pt,minimum height=6mm,minimum width=6mm,fill=white] 
	\tikzstyle{boxlarge} = [draw,shape=rectangle,inner sep=2pt,minimum height=1.5cm,minimum width=8mm,fill=white] 
	\tikzstyle{boxLarge} = [draw,shape=rectangle,inner sep=2pt,minimum height=2cm,minimum width=10mm,fill=white] 
	\tikzstyle{boxsmall} = [draw,shape=rectangle,inner sep=2pt,minimum height=3mm,minimum width=3mm,fill=white] 
	\tikzstyle{dot} = [inner sep=0mm,minimum width=3mm,minimum height=3mm,draw,shape=circle,text depth=-0.1mm]
	\tikzstyle{Zbwdot} = [dot, fill=\Zbwcolour]
	\tikzstyle{Xbwdot} = [dot, fill=\Xbwcolour]
	\tikzstyle{Ybwdot} = [dot, fill=\Ybwcolour]
	\tikzstyle{antipode} = [boxsmall] 

	\tikzstyle{state} = [draw, rectangle with rounded corners,
	  rectangle with rounded corners north west=8pt,
	  rectangle with rounded corners south west=8pt,
	  rectangle with rounded corners north east=0pt,
	  rectangle with rounded corners south east=0pt,
	,inner sep=2pt,minimum height=6mm,minimum width=6mm,fill=white]
	\tikzstyle{statelarge} = [draw, rectangle with rounded corners,
	  rectangle with rounded corners north west=8pt,
	  rectangle with rounded corners south west=8pt,
	  rectangle with rounded corners north east=0pt,
	  rectangle with rounded corners south east=0pt,
	,inner sep=2pt,minimum height=1.5cm,minimum width=8mm,fill=white]
	\tikzstyle{stateLarge} = [draw, rectangle with rounded corners,
	  rectangle with rounded corners north west=8pt,
	  rectangle with rounded corners south west=8pt,
	  rectangle with rounded corners north east=0pt,
	  rectangle with rounded corners south east=0pt,
	,inner sep=2pt,minimum height=2cm,minimum width=8mm,fill=white]
	\tikzstyle{effect} = [draw, rectangle with rounded corners,
	  rectangle with rounded corners north west=0pt,
	  rectangle with rounded corners south west=0pt,
	  rectangle with rounded corners north east=8pt,
	  rectangle with rounded corners south east=8pt,
	,inner sep=2pt,minimum height=6mm,minimum width=6mm,fill=white]
	\tikzstyle{scalar}=[diamond,draw,inner sep=0.1mm,font=\small,fill=white]

	\tikzstyle{cdnode}=[fill=white]
	\tikzstyle{labelnode}=[fill=white]
	\tikzstyle{tightlabelnode}=[fill=white,inner sep = 0.1mm]
	\tikzstyle{none}=[inner sep=0pt]
	\tikzstyle{whiteline}=[-, line width=4pt, draw=white]

	\tikzstyle{trace}=[circuit ee IEC,thick,ground,scale=2.5]
	\tikzstyle{cotrace}=[circuit ee IEC,thick,ground,rotate=180,scale=2.5]
	\tikzstyle{upground}=[circuit ee IEC,thick,ground,rotate=90,scale=2.5]
	\tikzstyle{downground}=[circuit ee IEC,thick,ground,rotate=-90,scale=2.5]
	
	\tikzstyle{doubled} = [line width=1.8pt] 


%% file: biblio.bbl
\newcommand{\etalchar}[1]{$^{#1}$}